\documentclass[12pt,preprint,letterpaper]{aastex}

\pdfoutput=1

\usepackage{amssymb,amsmath}
\usepackage{multirow}
\usepackage{xspace}
\usepackage{dcolumn}
\usepackage{color}

\definecolor{urlcolor}{rgb}{0,0,0}
\definecolor{filecolor}{rgb}{0,0,0}
\definecolor{linkcolor}{rgb}{0,0,0}
\definecolor{anchorcolor}{rgb}{0,0,0}
\definecolor{citecolor}{rgb}{0,0,0}
\definecolor{menucolor}{rgb}{0,0,0}
\usepackage[pdftex,
        colorlinks=true, % color (true) or boxes (false)
        bookmarks=true,
        bookmarksopen=true,
        urlcolor=urlcolor,       % \href{...}{...} external (URL)
        filecolor=filecolor,     % \href{...} local file
        linkcolor=linkcolor,       % \ref{...} and \pageref{...}
        anchorcolor=anchorcolor,
        citecolor=citecolor,
        menucolor=menucolor,
        pdftitle={Astrometry.net: Blind astrometric calibration of arbitrary astronomical images},
        pdfauthor={Dustin Lang, David W. Hogg, Keir Mierle, Michael Blanton, and Sam Roweis},
        pdfsubject={astronomy; astrometry; astrometry.net; geometric hashing; computer vision; object recognition},
        pdfproducer={pdfLaTeX},
        letterpaper
]{hyperref}

\newcommand{\numberparagraphs}{}
\newcommand{\nonumberparagraphs}{}

\newcommand{\filesuffix}[2]{#1}

\newcommand{\bigorsmallfig}[2]{#2}

\newcommand{\doctype}{article\xspace}
\newcommand{\thesisonly}[1]{}
\newcommand{\notthesisonly}[1]{#1}
\renewcommand{\comment}[1]{}

\newlength{\figunit}
\newcommand{\unit}[1]{\ensuremath{\mathrm{#1}}}
\renewcommand{\arcmin}{\unit{arcmin}}
\renewcommand{\mag}{\unit{mag}}

\newcommand{\degrees}{\unit{degrees}}
\renewcommand{\deg}{\unit{degree}}
\newcommand{\RA}{\unit{RA}}
\newcommand{\Dec}{\unit{Dec}}
\newcommand{\milliseconds}{\unit{ms}}

\newcommand{\captionpart}[1]{\textbf{#1}}

\newcommand{\metadata}{meta-data\xspace}

\newcommand{\an}{\textsl{Astrometry.net}\xspace}

\newcommand{\kdtree}{kd-tree\xspace}

\newcommand{\xref}[2]{\mbox{#1~\ref{#2}}\xspace}

\newcommand{\secref}[1]{\xref{section}{#1}}

\newcommand{\Fig}{Figure\xspace}
\newcommand{\fig}{figure\xspace}
\newcommand{\figs}{figures\xspace}
\newcommand{\Figs}{Figures\xspace}
\newcommand{\figref}[1]{\xref{\fig}{#1}}
\newcommand{\Figref}[1]{\xref{\Fig}{#1}}

\newcommand{\chapref}[1]{\xref{chapter}{#1}}

\newcommand{\healpix}{HEALPix\xspace}

\newcommand{\squareparens}[1]{\left[ \, #1 \, \right]}
\newcommand{\expect}[1]{\mathbb{E}\squareparens{#1}}

\newcommand{\given}{\,\vert\,}

\newcommand{\ie}{\textit{i.e.}}

\newlength{\gridfigwidth}
\newlength{\quadfigwidth}
\newlength{\densityfigwidth}
\newcommand\tstrut{\rule[-0.5ex]{0pt}{3ex}}
\newcommand{\truepos}{\ensuremath{\mathrm{TP}}}
\newcommand{\falsepos}{\ensuremath{\mathrm{FP}}}
\newcommand{\trueneg}{\ensuremath{\mathrm{TN}}}
\newcommand{\falseneg}{\ensuremath{\mathrm{FN}}}

\newcommand{\starlabel}[1]{\ensuremath{#1}}
\newcommand{\starA}{\starlabel{A}}
\newcommand{\starB}{\starlabel{B}}
\newcommand{\starC}{\starlabel{C}}
\newcommand{\starD}{\starlabel{D}}
\newcommand{\xC}{\ensuremath{x_\starC}}
\newcommand{\yC}{\ensuremath{y_\starC}}
\newcommand{\xD}{\ensuremath{x_\starD}}
\newcommand{\yD}{\ensuremath{y_\starD}}

\newcommand{\fg}{\ensuremath{F}}
\newcommand{\bg}{\ensuremath{B}}
\newcommand{\data}{\ensuremath{D}}

\newcommand{\tableheaderx}[1]{\multicolumn{1}{|c|}{\textbf{#1}}}
\newcommand{\tableheader}[1]{\multicolumn{1}{c|}{\textbf{#1}}}

\newcommand{\percent}{\!\%}

\title{\an: \\
       Blind astrometric calibration of arbitrary astronomical images}
\author{Dustin~Lang\altaffilmark{1,2,3},
        David~W.~Hogg\altaffilmark{4,5},
        Keir~Mierle\altaffilmark{1,6},
        Michael~Blanton\altaffilmark{4},
        Sam~Roweis\altaffilmark{1,6,7}
}
\altaffiltext{1}{Department of Computer Science, University of Toronto,
                 6 King's College Road, Toronto, Ontario, M5S~3G4, Canada}
\altaffiltext{2}{Princeton University Observatory, Princeton, NJ, 08544, USA}
\altaffiltext{3}{to whom correspondence should be addressed: dstn@astro.princeton.edu}
\altaffiltext{4}{Center for Cosmology \& Particle Physics, Department of Physics,
                 New York University, 4 Washington Place, New York, NY, 10003, USA}
\altaffiltext{5}{Max-Planck-Institut f\"ur Astronomie,
                 K\"onigstuhl 17, D-69117, Heidelberg, Germany}
\altaffiltext{6}{Google Inc., Mountain View, CA, 94043, USA}
\altaffiltext{7}{Computer Science Department,
                 New York University, 251 Mercer Street, New York, NY, 10012, USA}

\begin{abstract}
We have built a reliable and robust system that takes as input an
astronomical image, and returns as output the pointing, scale, and
orientation of that image (the astrometric calibration or WCS
information).  The system requires no first guess, and works with the
information in the image pixels alone; that is, the problem is a
generalization of the ``lost in space'' problem in which nothing---not
even the image scale---is known.  After robust source detection is
performed in the input image, asterisms (sets of four or five stars)
are geometrically hashed and compared to pre-indexed hashes to
generate hypotheses about the astrometric calibration.  A hypothesis
is only accepted as true if it passes a Bayesian decision theory test
against a null hypothesis.  With indices built from the USNO-B Catalog
and designed for uniformity of coverage and redundancy, the success
rate is $>99.9~\percent$ for contemporary near-ultraviolet and visual
imaging survey data, with no false positives.  The failure rate is
consistent with the incompleteness of the USNO-B Catalog; augmentation
with indices built from the 2MASS Catalog brings the completeness to
$100~\percent$ with no false positives.  We are using this system to
generate consistent and standards-compliant meta-data for digital and
digitized imaging from plate repositories, automated observatories,
individual scientific investigators, and hobbyists.  This is the first
step in a program of making it possible to trust calibration meta-data
for astronomical data of arbitrary provenance.
\end{abstract}

\keywords{
astrometry
---
catalogs
---
instrumentation: miscellaneous
---
methods: data analysis
---
methods: statistical
---
techniques: image processing
}

\begin{document}
\maketitle

\newlength{\pointonesec}\settowidth{\pointonesec}{$0.1$ s}
\newlength{\acceptable}\settowidth{\acceptable}{Acceptable}
\newlength{\sdsserspc}\settowidth{\sdsserspc}{final}
\newlength{\cw}\settowidth{\cw}{\textbf{Percentage of images recognized}}

\newcommand{\sdssertable}{\begin{tabular}{|r|D{.}{.}{6.0}|D{.}{.}{6.0}|D{.}{.}{3.2}|}
\hline
\multicolumn{1}{|c|}{\textbf{Phase}} & \multicolumn{1}{c|}{\textbf{Images recognized}} & \multicolumn{1}{c|}{\textbf{Unrecognized}} &
\multicolumn{1}{c|}{\textbf{Percent recognized}} \\
\hline
\usnob: \makebox[\sdsserspc][r]{$0.1$ s} & 172,882 & 9,339 & 94.87 \\
\usnob: \makebox[\sdsserspc][r]{$1$ s} & 181,826 & 395 & 99.78 \\
\usnob: \makebox[\sdsserspc][r]{$10$ s} & 182,158 & 63 & 99.97 \\
\usnob: \makebox[\sdsserspc][r]{final} & 182,160 & 61 & 99.97 \\
\twomass & 182,211 & 10 & 99.99 \\
Original images & 182,221 & 0 & 100.00 \\
\hline
\end{tabular}
}

\newcommand{\sdsserradecfig}{\includegraphics[width=2.000000\figunit]{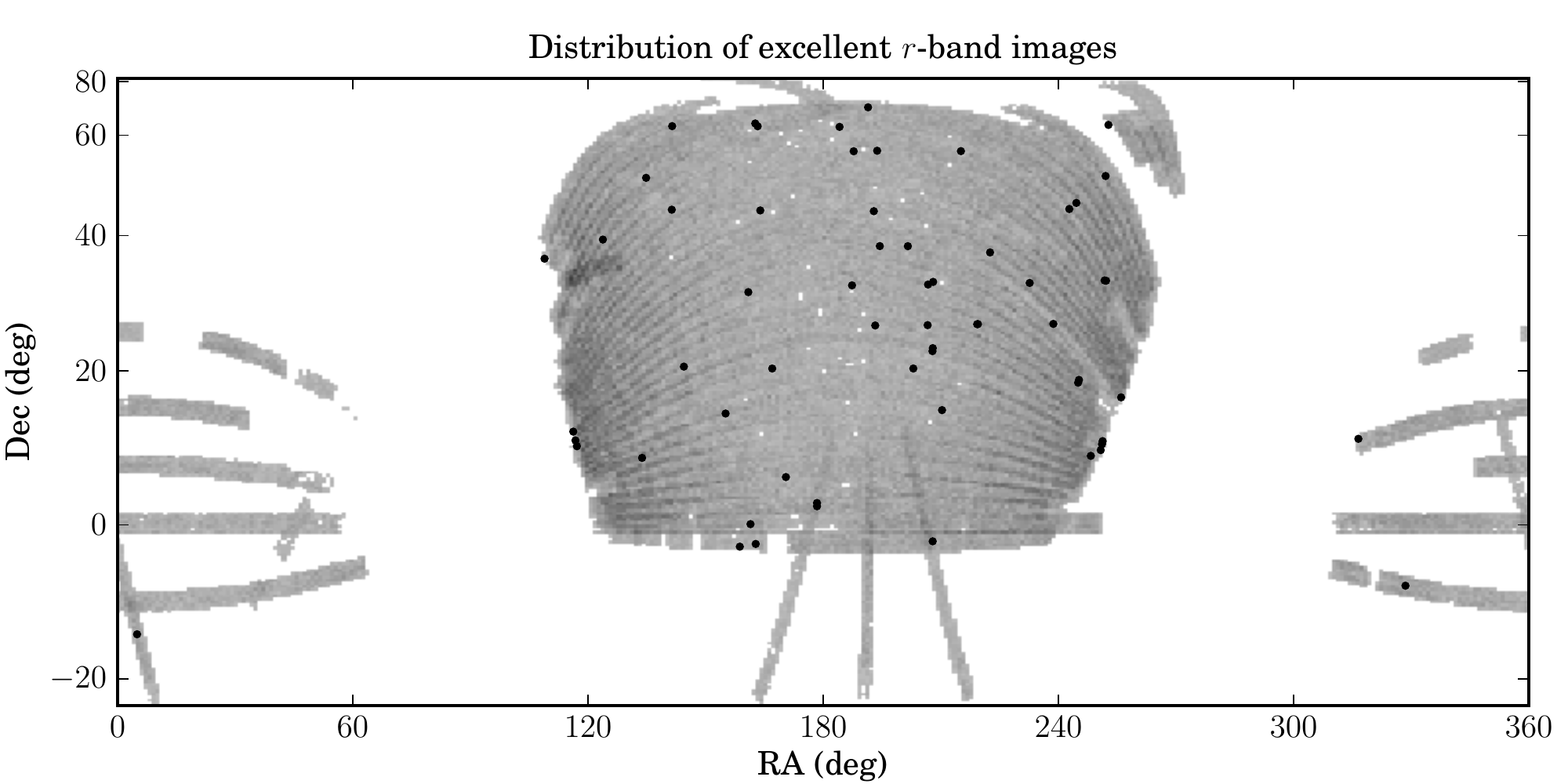}}
\newcommand{\sdssercputimefig}{\includegraphics[width=1.000000\figunit]{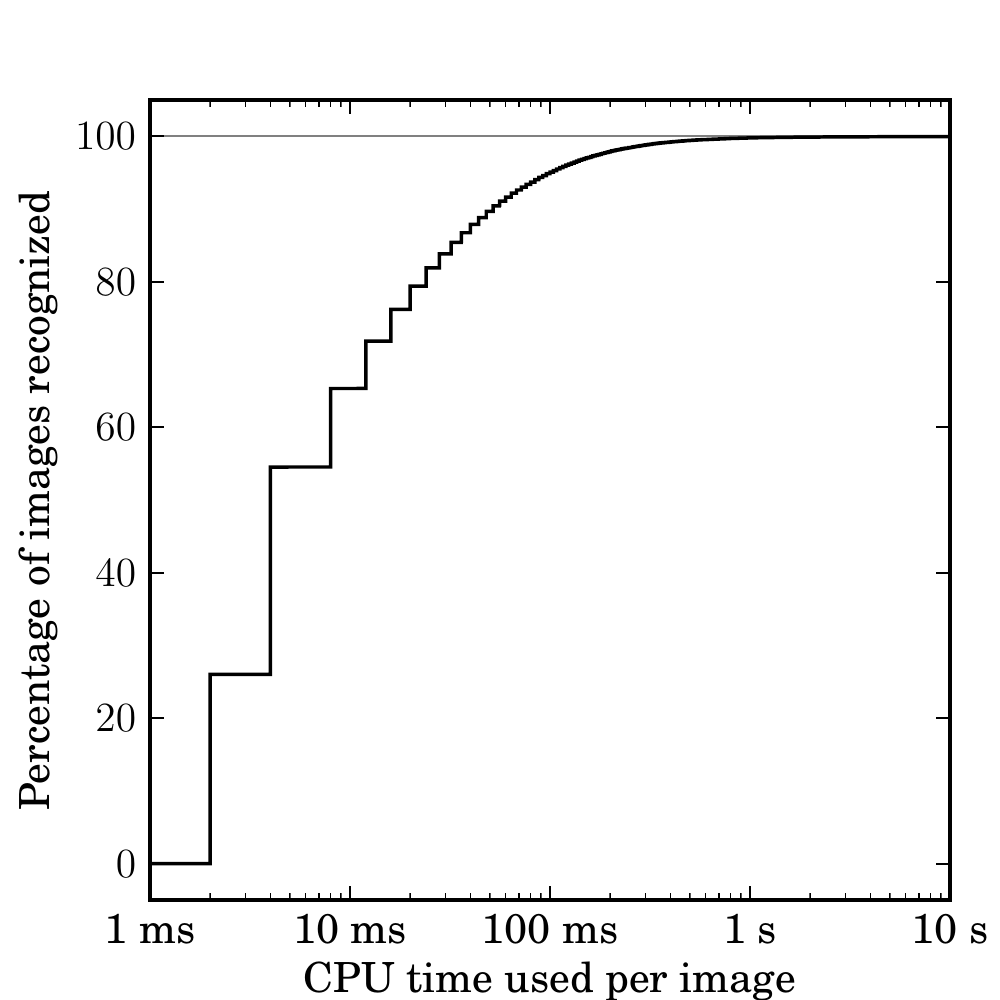}}
\newcommand{\sdssernimagefig}{\includegraphics[width=1.000000\figunit]{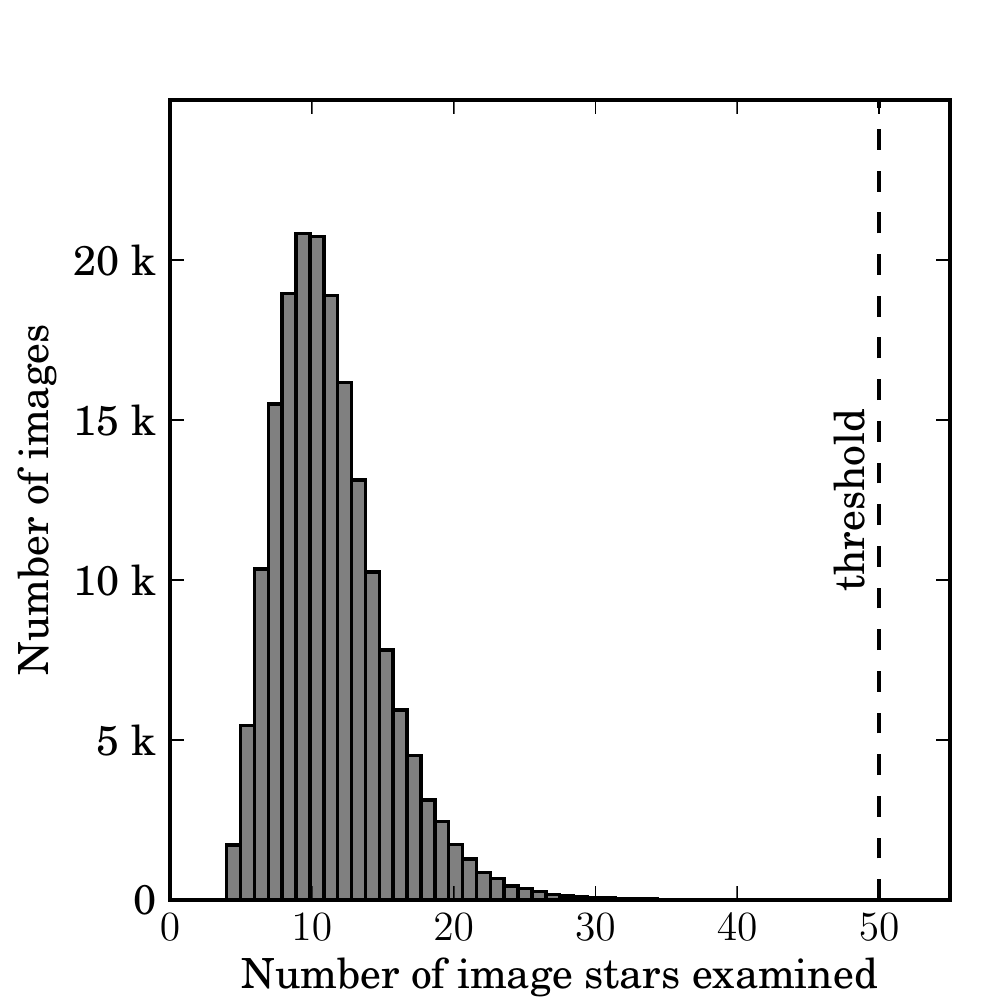}}
\newcommand{\sdssernmatchfig}{\includegraphics[width=1.000000\figunit]{sdss-er-nmatch}}
\newcommand{\sdssercodeerrfig}{\includegraphics[width=1.000000\figunit]{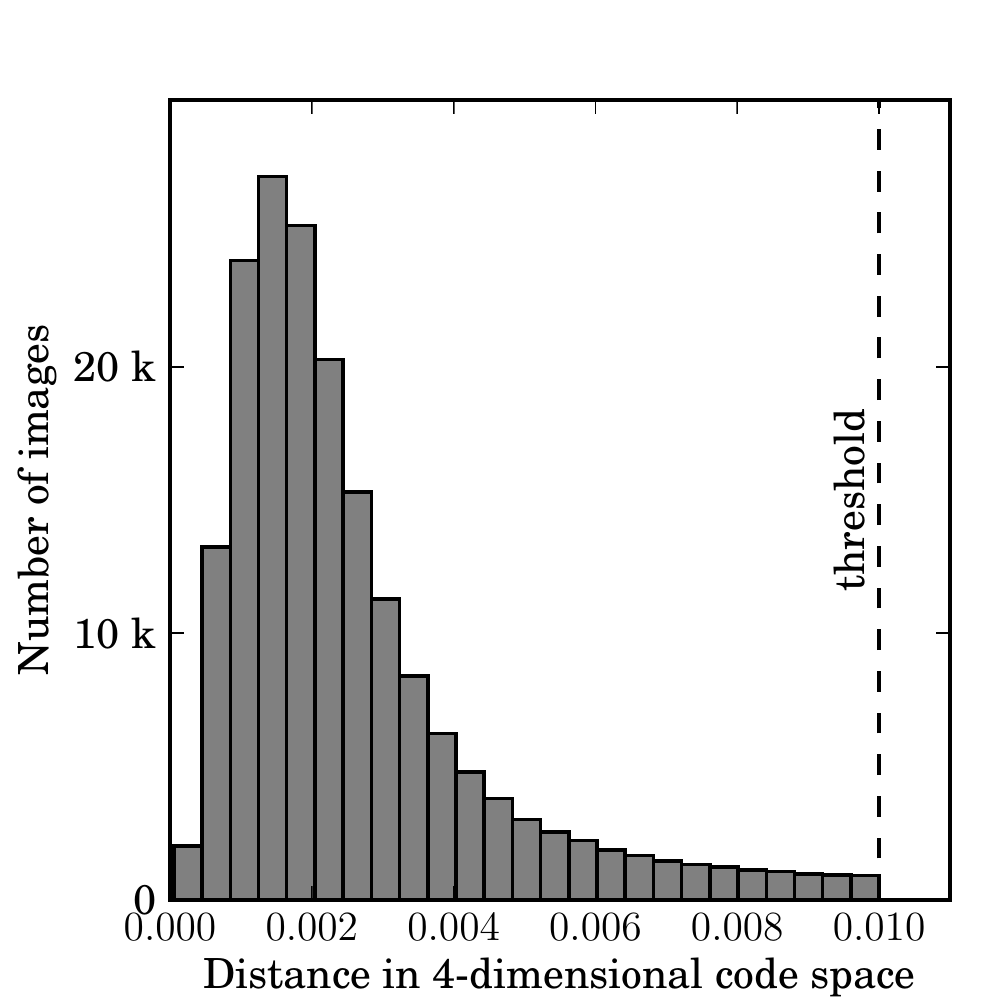}}
\newcommand{\sdsserbayesfig}{\includegraphics[width=1.000000\figunit]{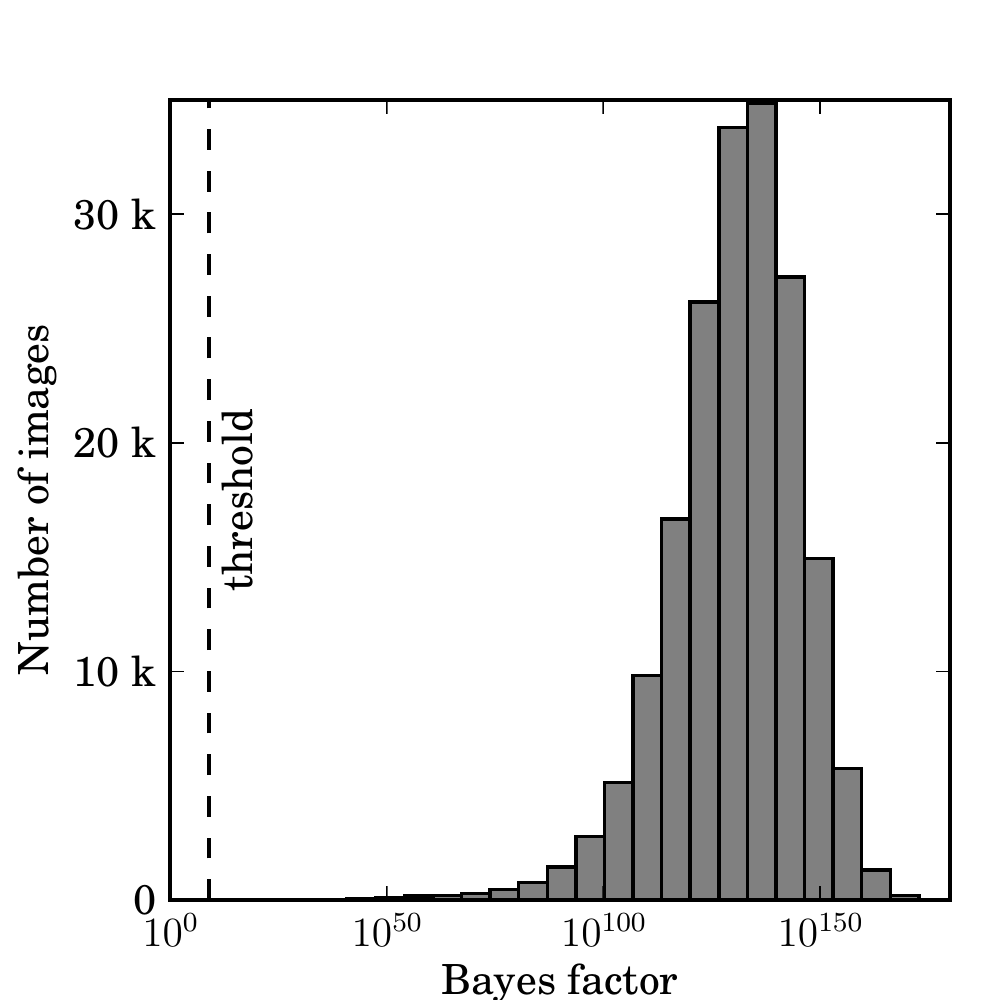}}
\newcommand{\sdsserntoverifyfig}{\includegraphics[width=1.000000\figunit]{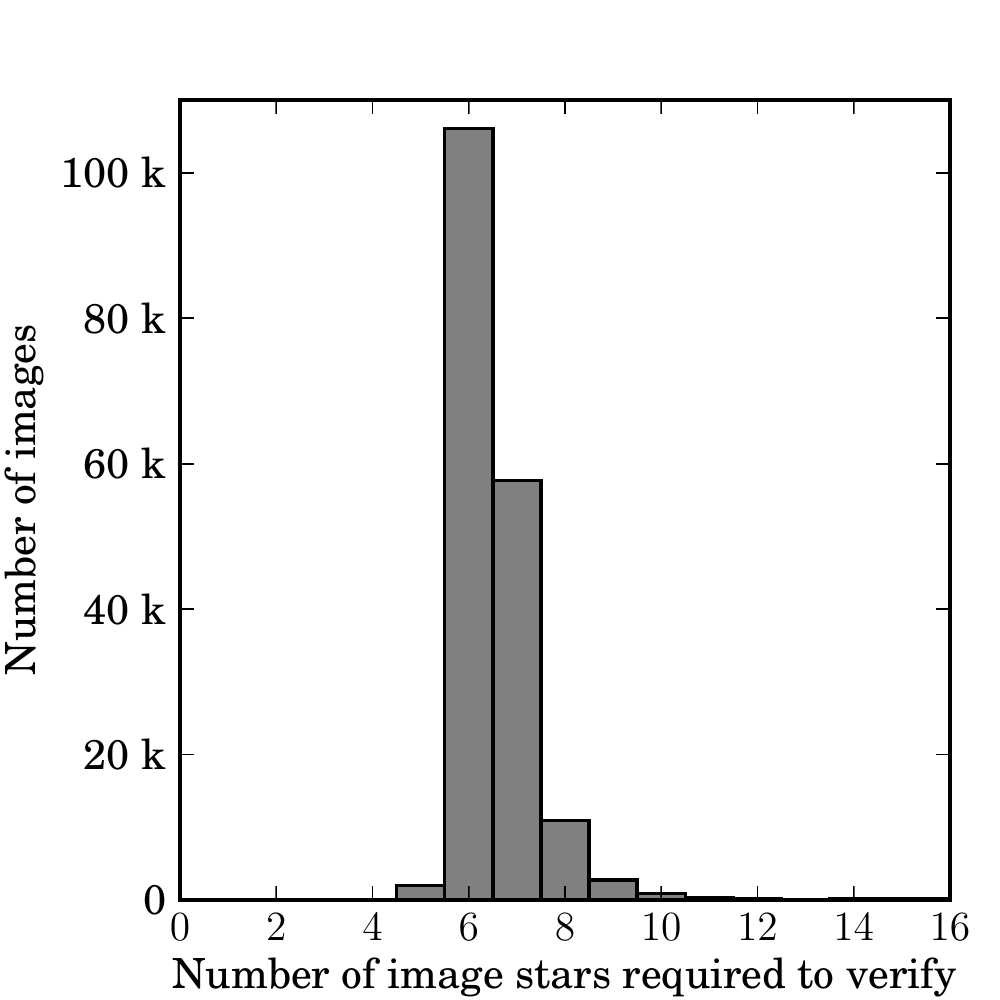}}
\newcommand{\sdssbandtable}{\begin{tabular}{|c|D{.}{.}{3.2}|D{.}{.}{3.2}|D{.}{.}{3.2}|D{.}{.}{3.2}|D{.}{.}{3.2}|}
\cline{2-6}
\multicolumn{1}{c|}{} &\multicolumn{5}{c|}{\textbf{Percentage of images recognized}} \\
\hline
\multicolumn{1}{|c|}{\textbf{CPU time}} & \multicolumn{1}{c|}{$u$} & \multicolumn{1}{c|}{$g$} & \multicolumn{1}{c|}{$r$} & \multicolumn{1}{c|}{$i$} & \multicolumn{1}{c|}{$z$} \\
\hline
\makebox[\pointonesec][r]{$0.1$ s} & 87.80  & 93.88  & 94.87  & 93.59  & 94.36 \\
\makebox[\pointonesec][r]{$1$ s} & 98.58  & 99.73  & 99.78  & 99.73  & 99.75 \\
\makebox[\pointonesec][r]{$10$ s} & 99.82  & 99.96  & 99.97  & 99.96  & 99.96 \\
Final & 99.84  & 99.96  & 99.97  & 99.96  & 99.96 \\
\hline
\end{tabular}
}
\newcommand{\sdssbandsobjsfig}{\includegraphics[width=1.000000\figunit]{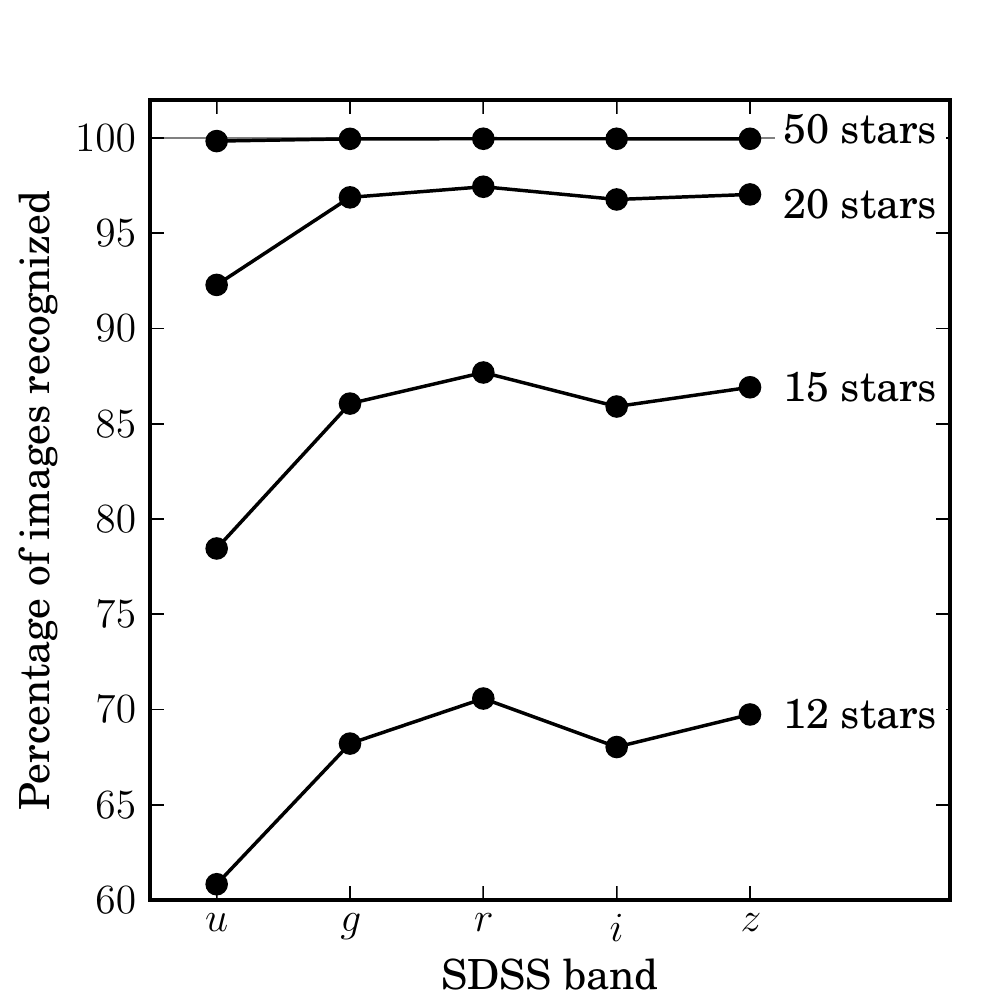}}
\newcommand{\sdssbandstimefig}{\includegraphics[width=1.000000\figunit]{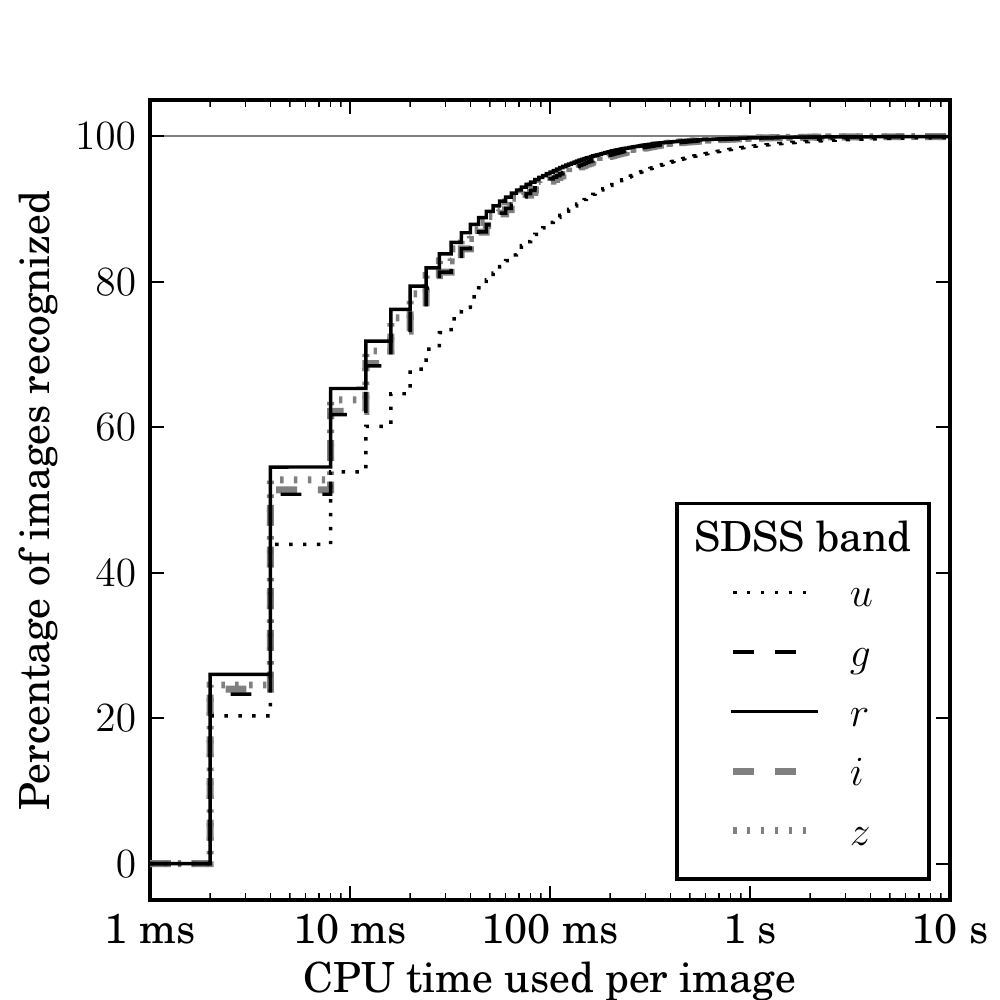}}
\newcommand{\sdssqualtable}{\begin{tabular}{|c|D{.}{.}{3.2}|D{.}{.}{3.2}|D{.}{.}{3.2}|D{.}{.}{3.2}|}
\cline{2-5}
\multicolumn{1}{c|}{} &\multicolumn{4}{c|}{\textbf{Percentage of images recognized}} \\
\hline
\multicolumn{1}{|c|}{\textbf{CPU time}} & \multicolumn{1}{c|}{\makebox[\acceptable][c]{Excellent}} & \multicolumn{1}{c|}{\makebox[\acceptable][c]{Good}} & \multicolumn{1}{c|}{\makebox[\acceptable][c]{Acceptable}} & \multicolumn{1}{c|}{\makebox[\acceptable][c]{Bad}} \\
\hline
\makebox[\pointonesec][r]{$0.1$ s} & 94.87  & 94.85  & 94.57  & 84.11 \\
\makebox[\pointonesec][r]{$1$ s} & 99.78  & 99.74  & 99.64  & 96.58 \\
\makebox[\pointonesec][r]{$10$ s} & 99.97  & 99.94  & 99.94  & 99.11 \\
Final & 99.97  & 99.94  & 99.95  & 99.18 \\
\hline
\end{tabular}
}
\newcommand{\sdssqualtimefig}{\includegraphics[width=1.000000\figunit]{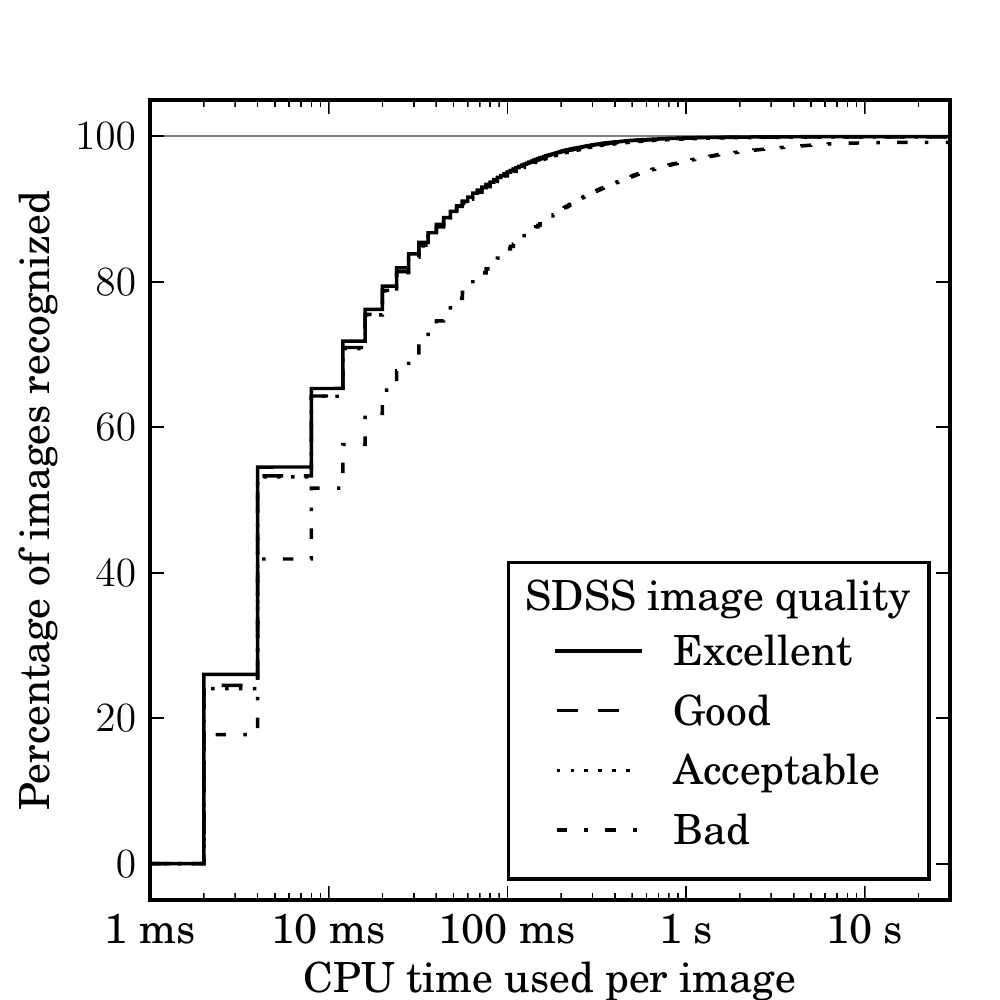}}
\newcommand{\sdssqualobjsfig}{\includegraphics[width=1.000000\figunit]{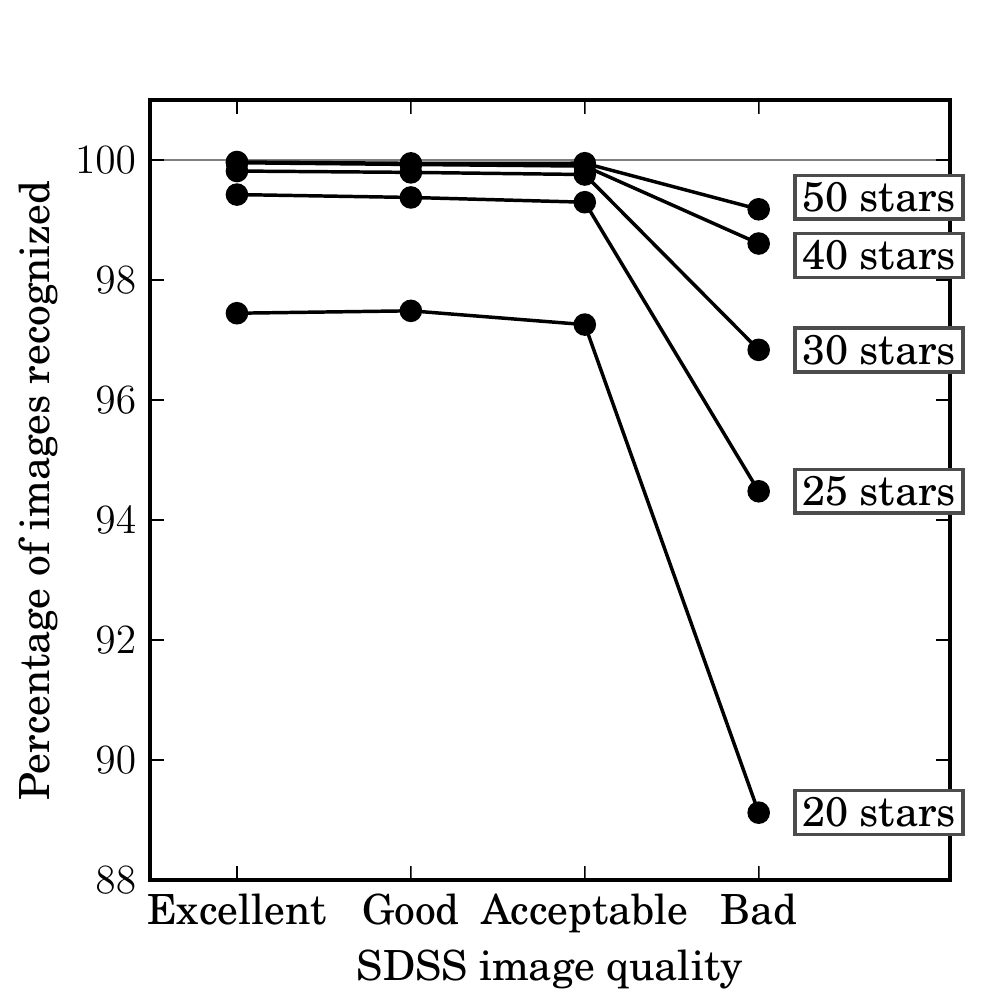}}
\newcommand{\sdssimsizetimefig}{\includegraphics[width=1.000000\figunit]{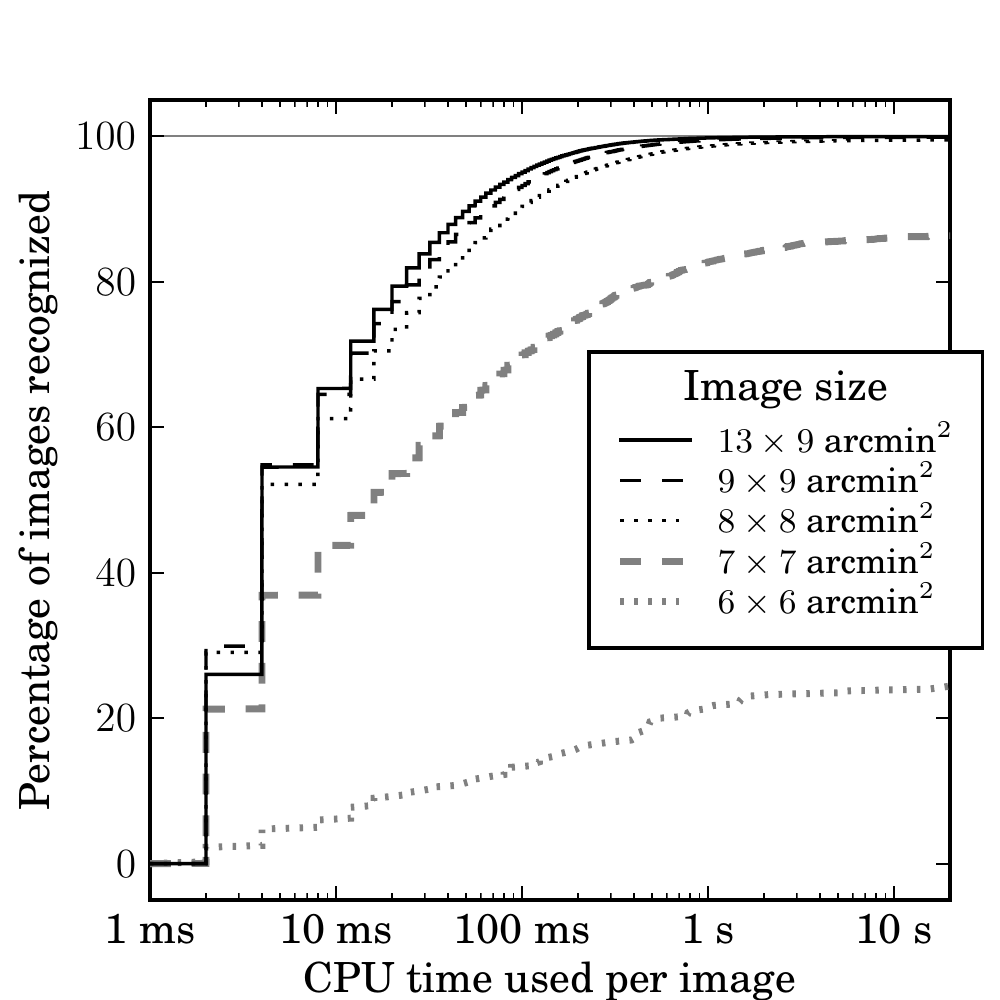}}
\newcommand{\sdssimsizetable}{\begin{tabular}{|D{x}{\times}{2.1}|D{.}{.}{3.2}|}
\hline
\multicolumn{1}{|c|}{\textbf{Image size (arcmin${}^2$)}} &
\multicolumn{1}{c|}{\textbf{Percentage of images recognized}} \\
\hline
13x9 & 99.97\\
9x9 & 99.88\\
8x8 & 99.52\\
7x7 & 86.53\\
6x6 & 24.75\\
\hline
\end{tabular}
}
\newcommand{\sdssimsizeobjsfig}{\includegraphics[width=1.000000\figunit]{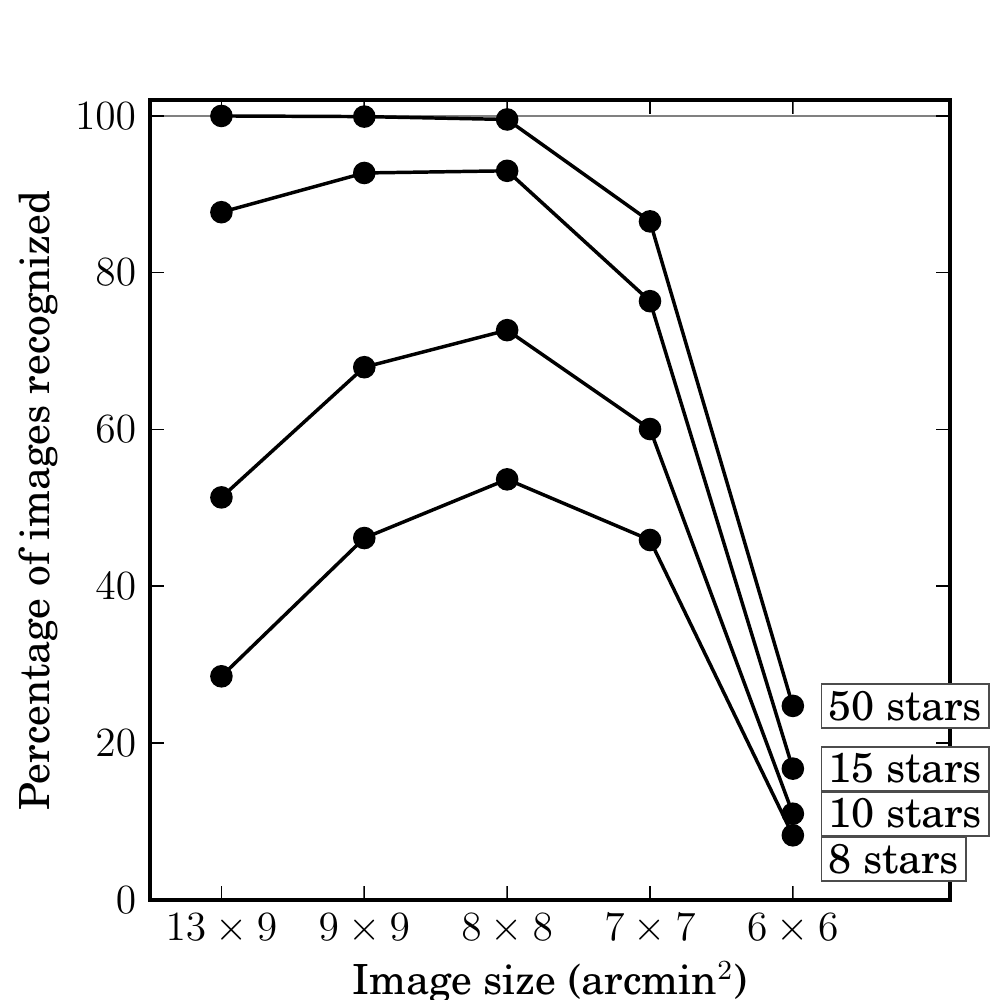}}
\newcommand{\sdsssizehintsreltimefig}{\includegraphics[width=1.000000\figunit]{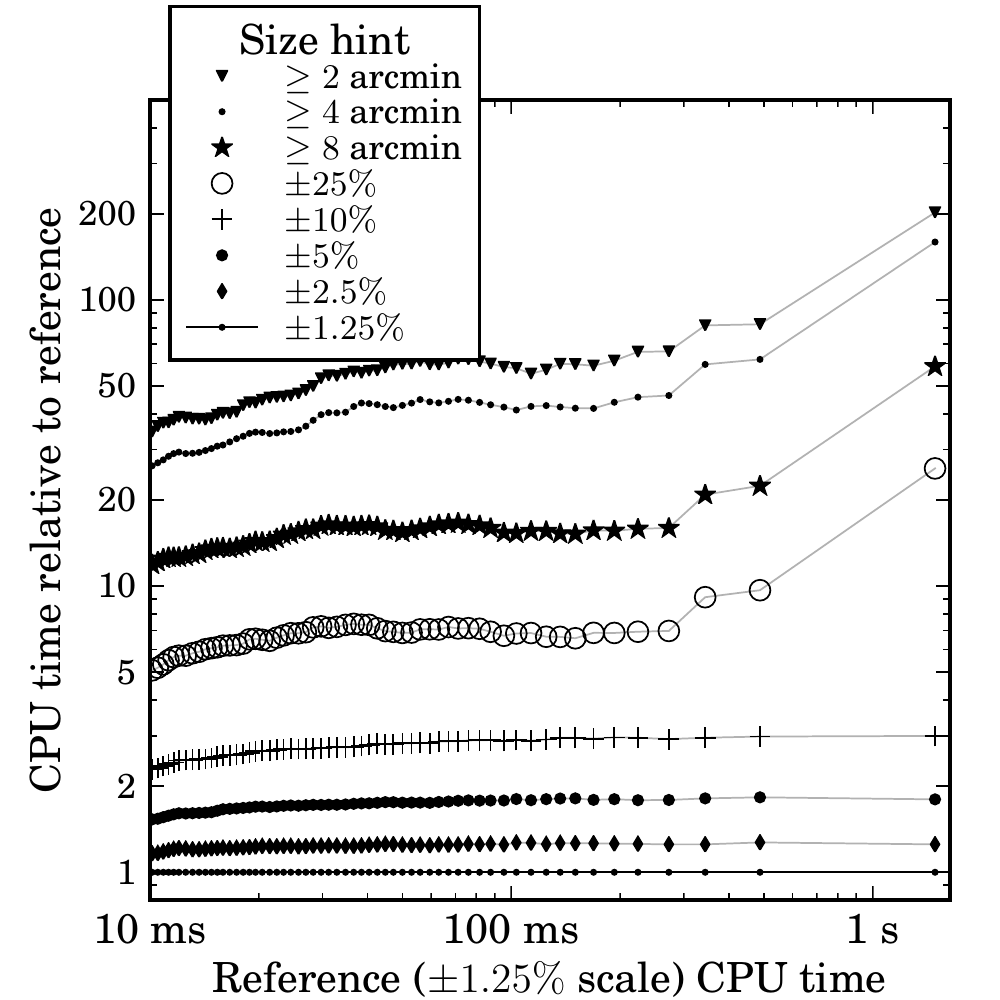}}
\newcommand{\sdsssizehintstimefig}{\includegraphics[width=1.000000\figunit]{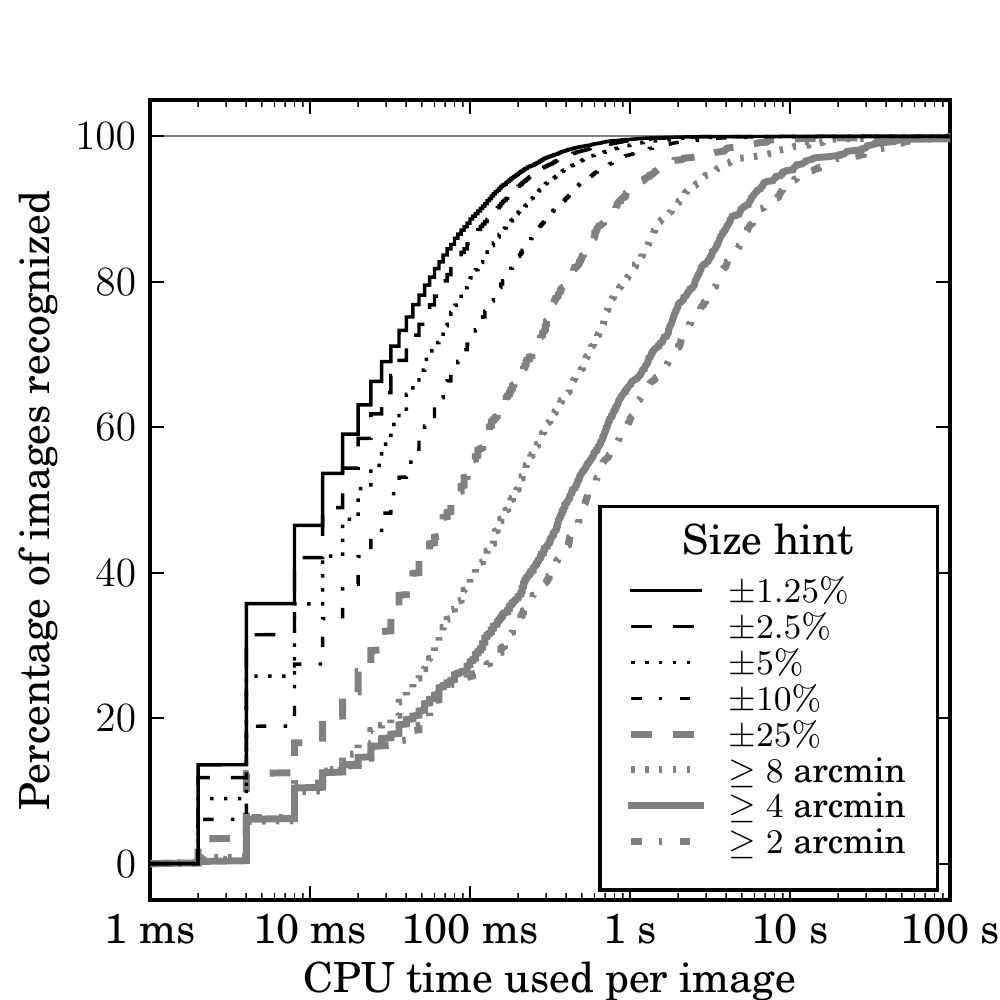}}
\newcommand{\sdsssizehintsindexfig}{\includegraphics[width=1.000000\figunit]{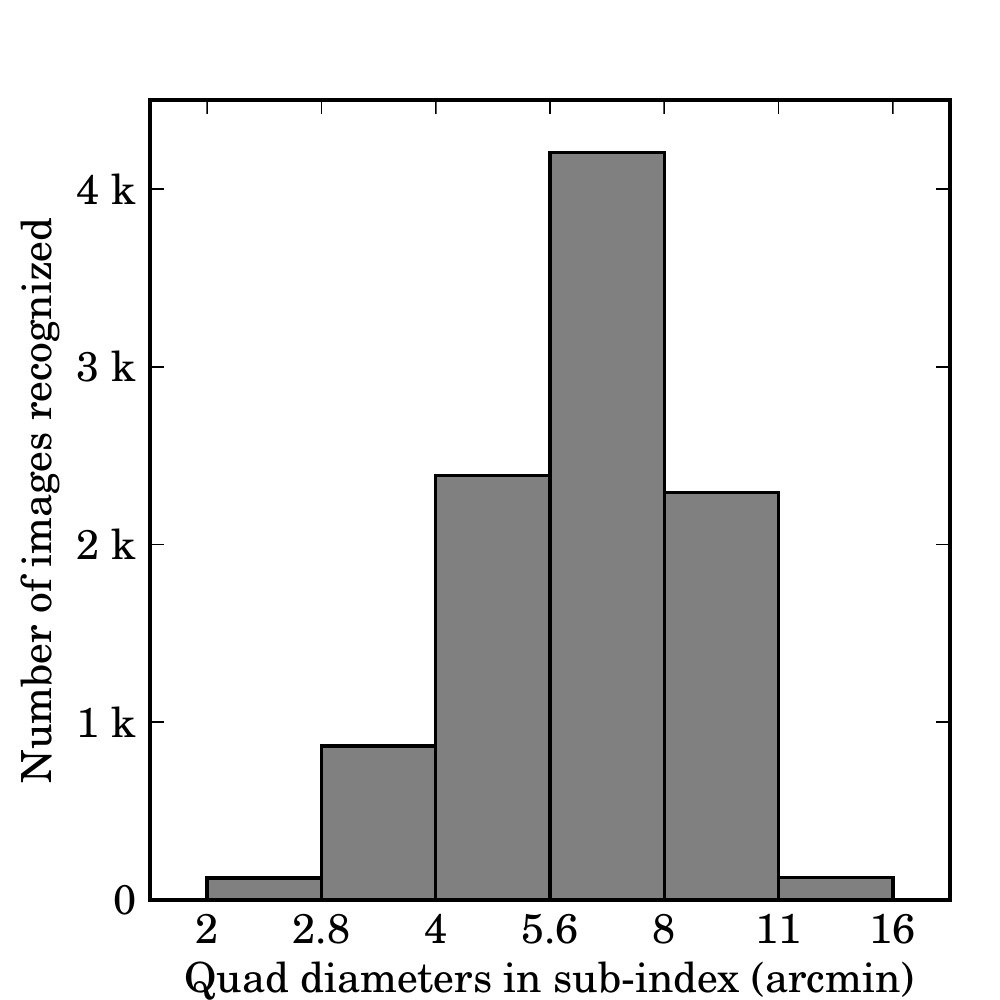}}
\newcommand{\sdssdensityreltimefig}{\includegraphics[width=1.000000\figunit]{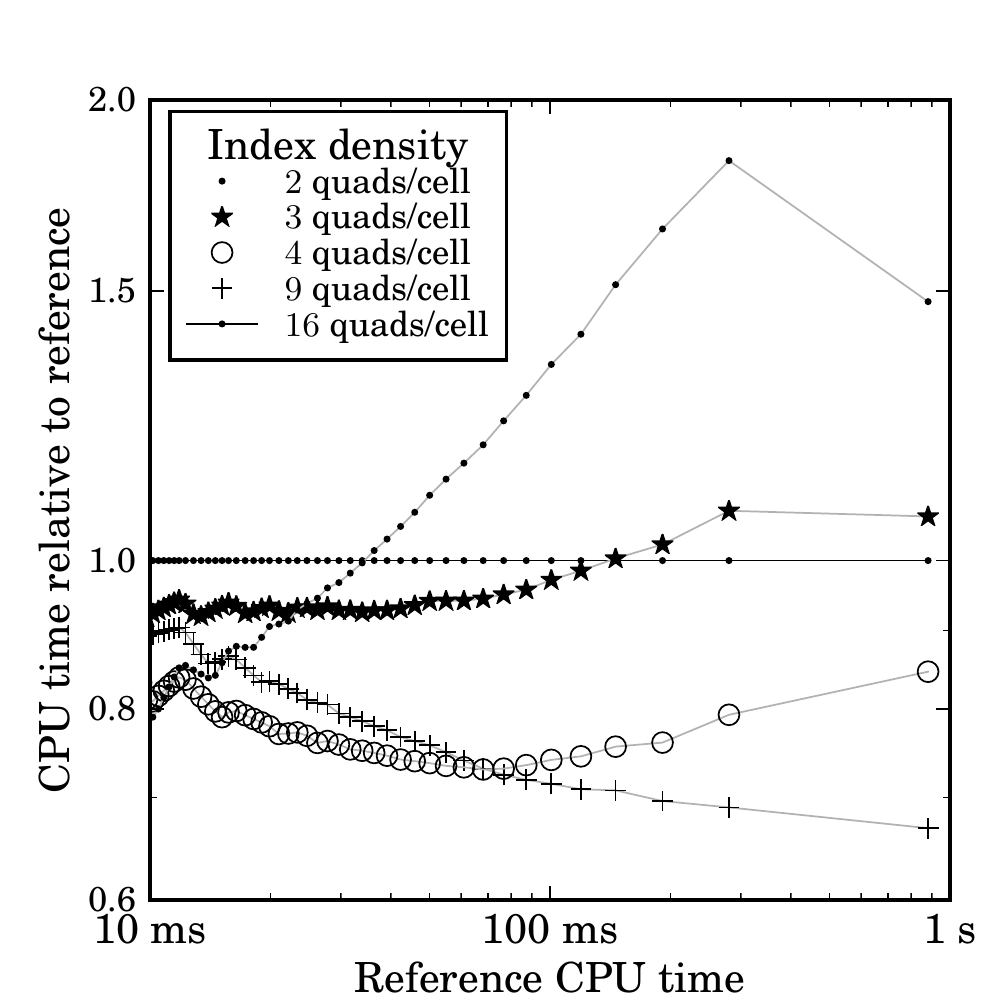}}
\newcommand{\sdssdensitytimefig}{\includegraphics[width=1.000000\figunit]{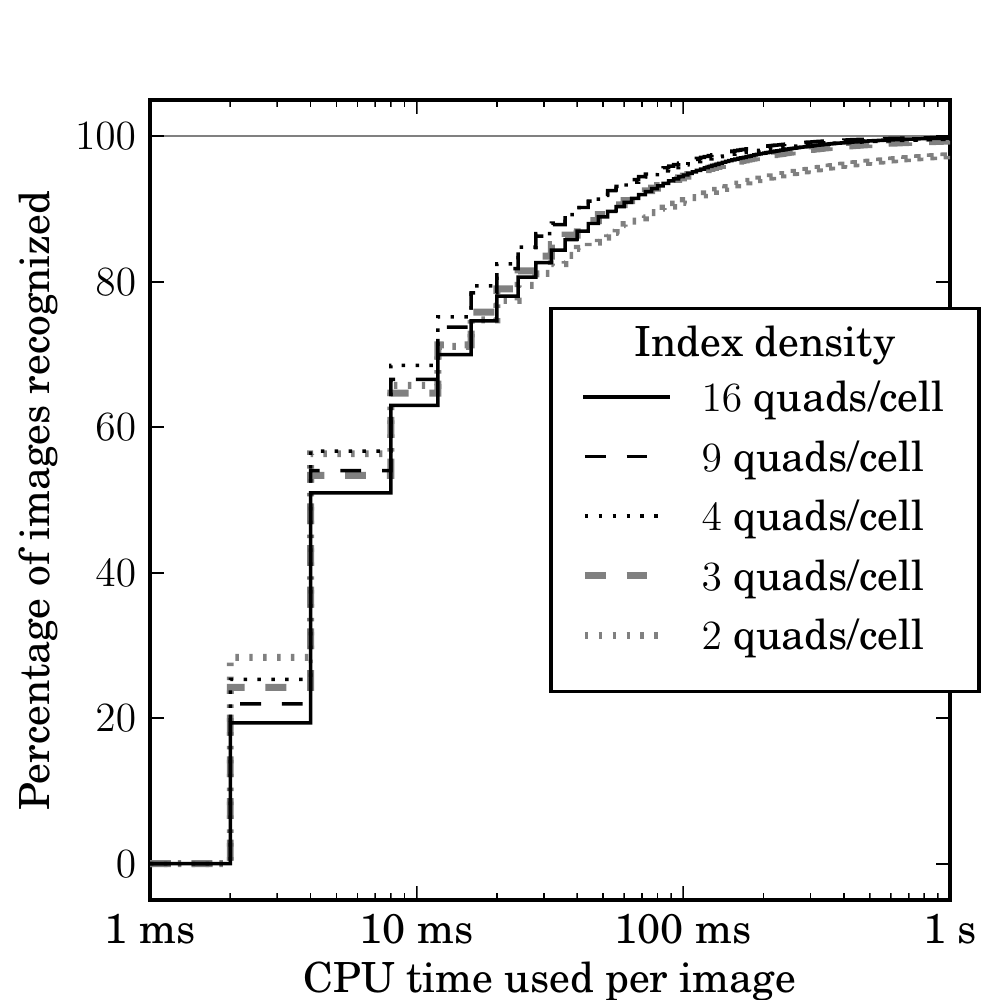}}
\newcommand{\sdssdensitytable}{\begin{tabular}{|c|D{.}{.}{3.2}|D{.}{.}{3.2}|D{.}{.}{3.2}|D{.}{.}{3.2}|D{.}{.}{3.2}|}
\cline{2-6}
\multicolumn{1}{c|}{} &\multicolumn{5}{c|}{\textbf{Percentage of images recognized}} \\
\hline
\multicolumn{1}{|c|}{\textbf{CPU time}} & \multicolumn{1}{c|}{$16$ quads/cell} & \multicolumn{1}{c|}{$9$ quads/cell} & \multicolumn{1}{c|}{$4$ quads/cell} & \multicolumn{1}{c|}{$3$ quads/cell} & \multicolumn{1}{c|}{$2$ quads/cell} \\
\hline
\makebox[3em][r]{$0.1$ s} & 94.44  & 96.32  & 95.89  & 94.30  & 90.94 \\
\makebox[3em][r]{$1$ s} & 99.76  & 99.84  & 99.61  & 99.36  & 97.35 \\
\makebox[3em][r]{$10$ s} & 99.96  & 99.95  & 99.79  & 99.65  & 97.92 \\
\hline
\end{tabular}
}
\newcommand{\sdsstriquinttable}{
\begin{tabular}{|c|D{.}{.}{3.2}|D{.}{.}{3.2}|D{.}{.}{3.2}|}
\cline{2-4}
\multicolumn{1}{c|}{} &\multicolumn{3}{c|}{\textbf{Percentage of images recognized}} \\
\hline
\multicolumn{1}{|c|}{\textbf{CPU time}} & \multicolumn{1}{c|}{\makebox[0.35\cw][c]{Triangles}} & \multicolumn{1}{c|}{\makebox[0.35\cw][c]{Quads}} & \multicolumn{1}{c|}{\makebox[0.35\cw][c]{Quints}} \\
\hline
\makebox[\pointonesec][r]{$0.1$ s} & 0.20  & 57.45  & 23.35 \\
\makebox[\pointonesec][r]{$1$ s} & 28.07  & 92.20  & 36.67 \\
\makebox[\pointonesec][r]{$10$ s} & 78.58  & 99.28  & 72.75 \\
Final & 99.97  & 99.33  & 96.25 \\
\hline
\end{tabular}
}
\newcommand{\sdsstriquintntrynmatchfig}{\includegraphics[width=1.000000\figunit]{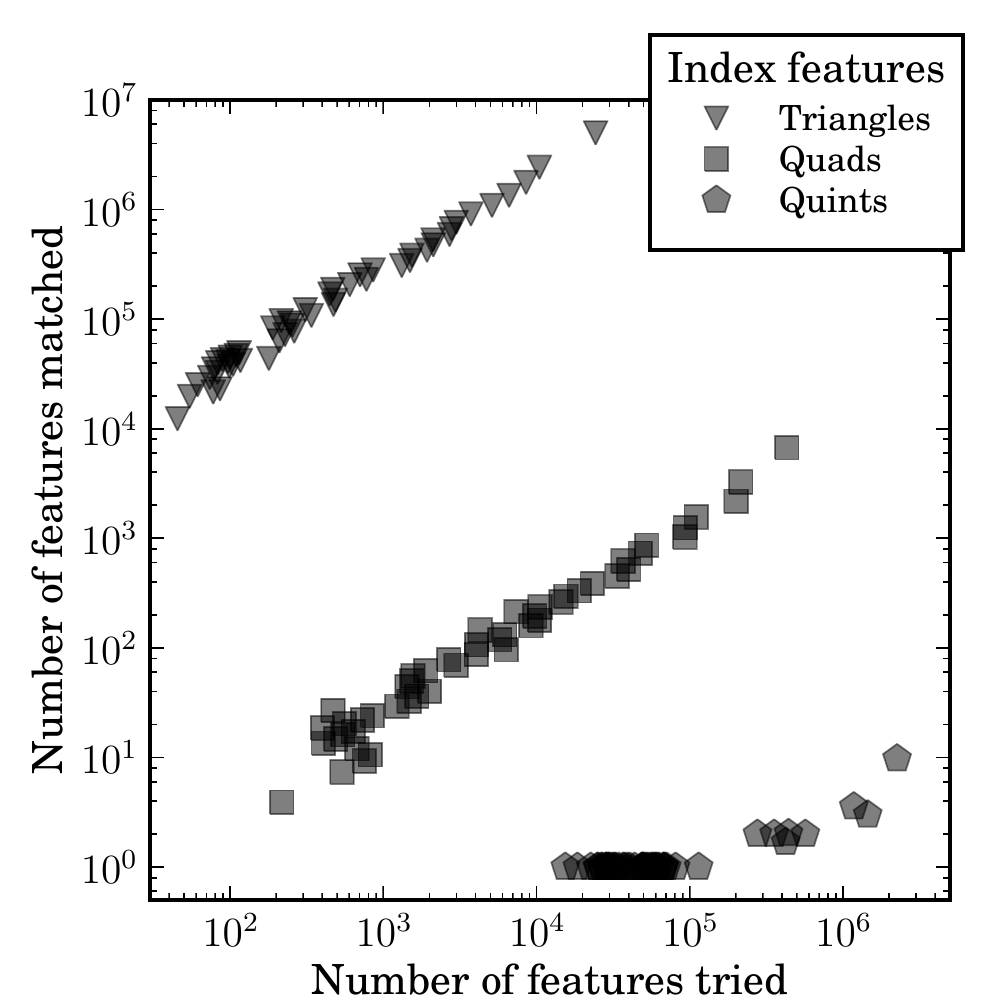}}
\newcommand{\sdsstriquinttimefig}{\includegraphics[width=1.000000\figunit]{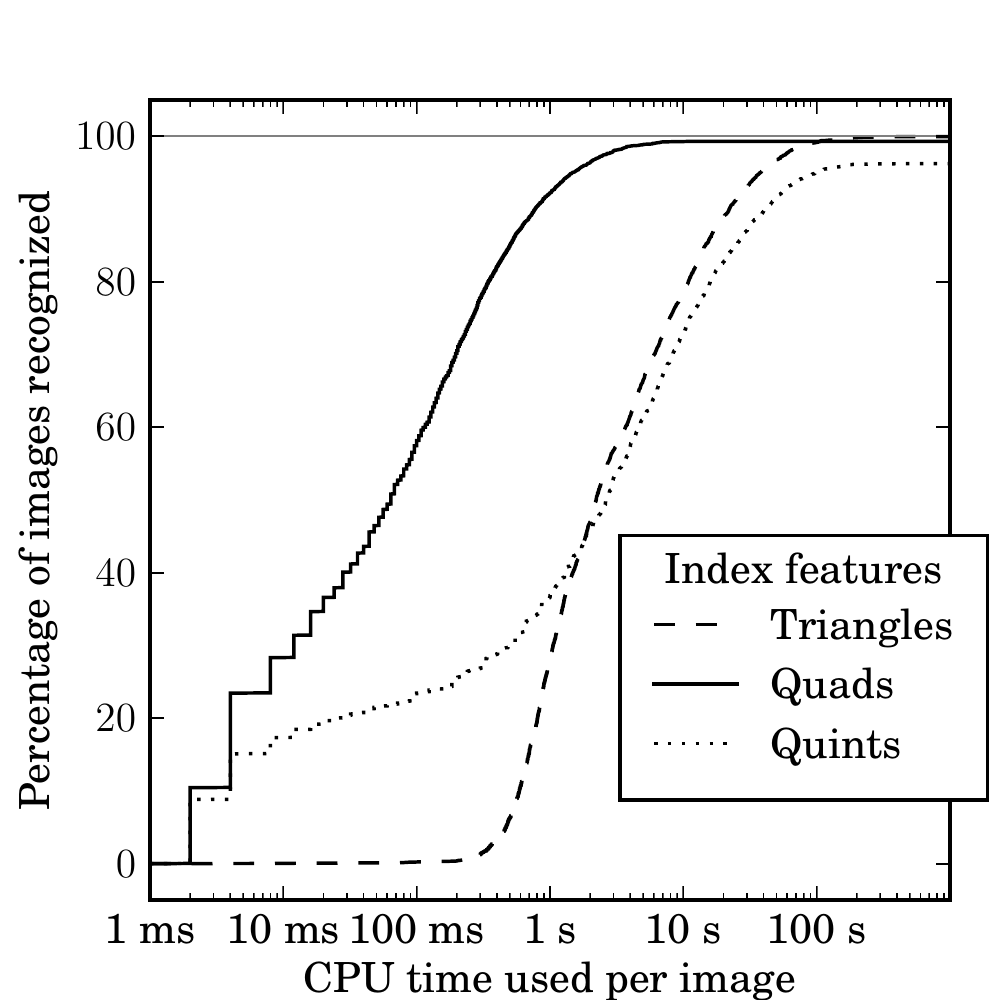}}
\newcommand{\galextable}{\begin{tabular}{|c|c|}
\hline
\textbf{CPU time} & \textbf{Percentage of images recognized} \\
\hline
1 s& 74.46 \\
10 s& 93.56 \\
100 s& 98.95 \\
Final& 99.74 \\
\hline
\end{tabular}
}
\newcommand{\galexcputimefig}{\includegraphics[width=1.000000\figunit]{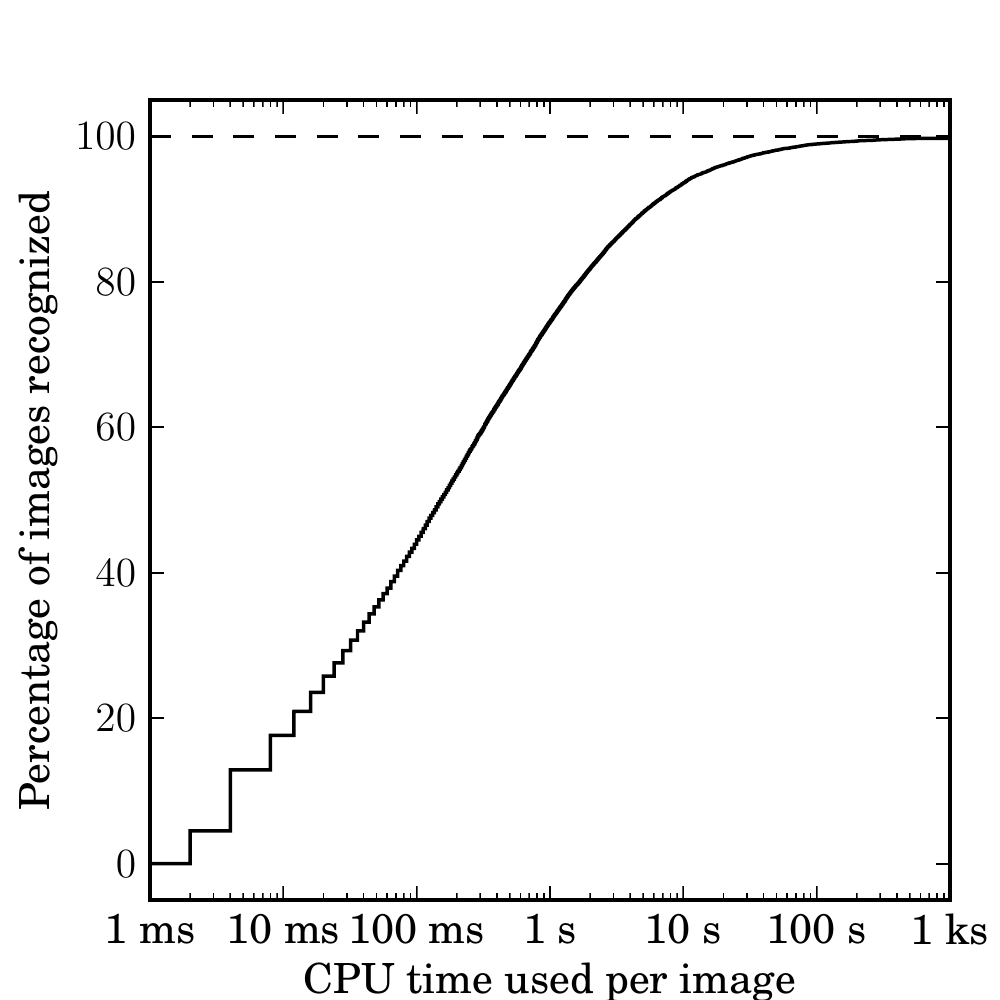}}
\newcommand{\galexindexidfig}{\includegraphics[width=1.000000\figunit]{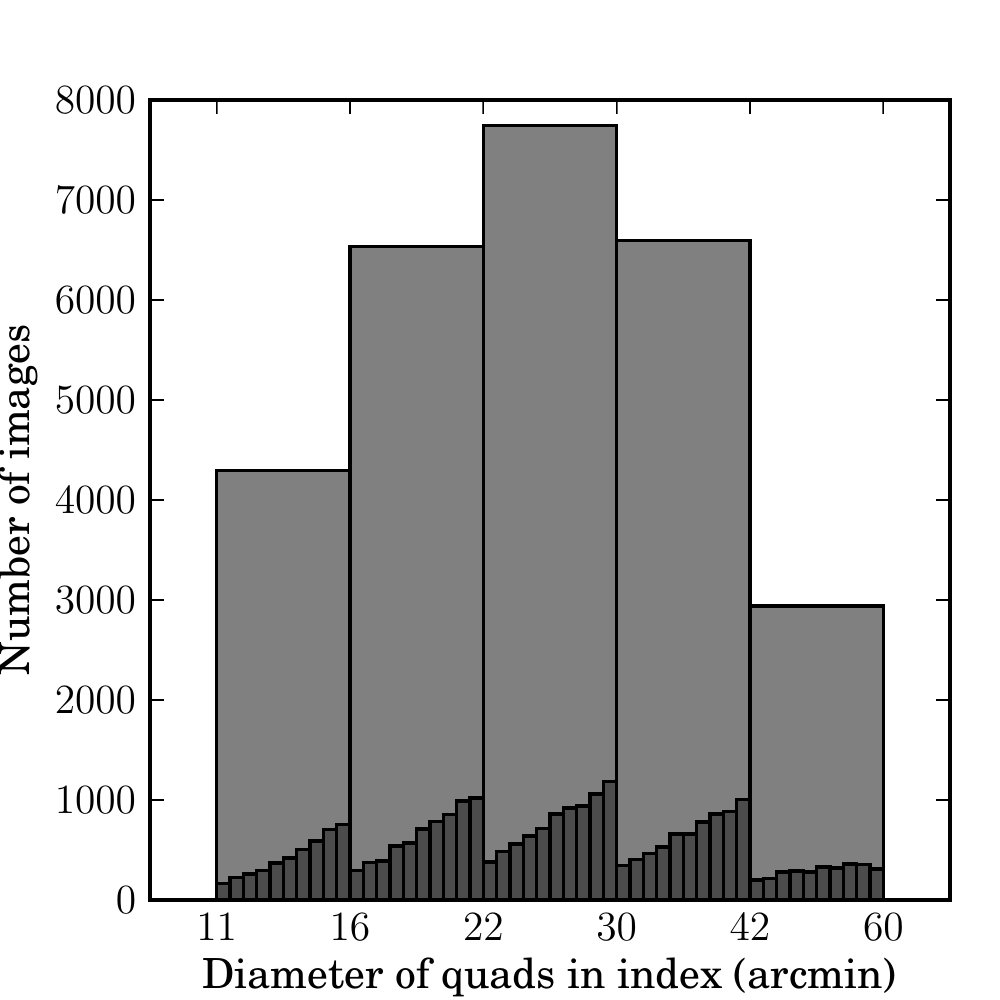}}
\newcommand{\galexquadfig}{\includegraphics[width=1.000000\figunit]{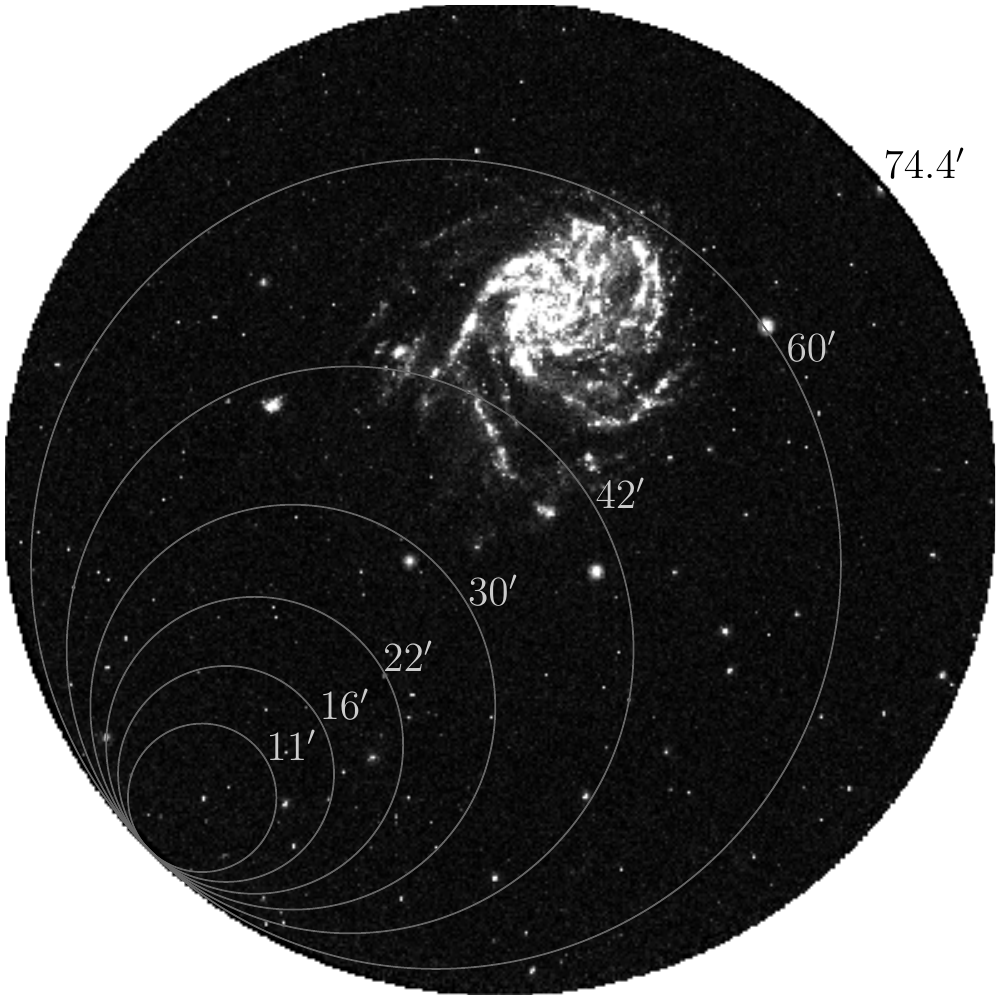}}
\newcommand{\sdssquadfig}{\includegraphics[width=1.000000\figunit]{sdss-quad}}
\newcommand{\aegisacsquadfig}{\includegraphics[width=1.000000\figunit]{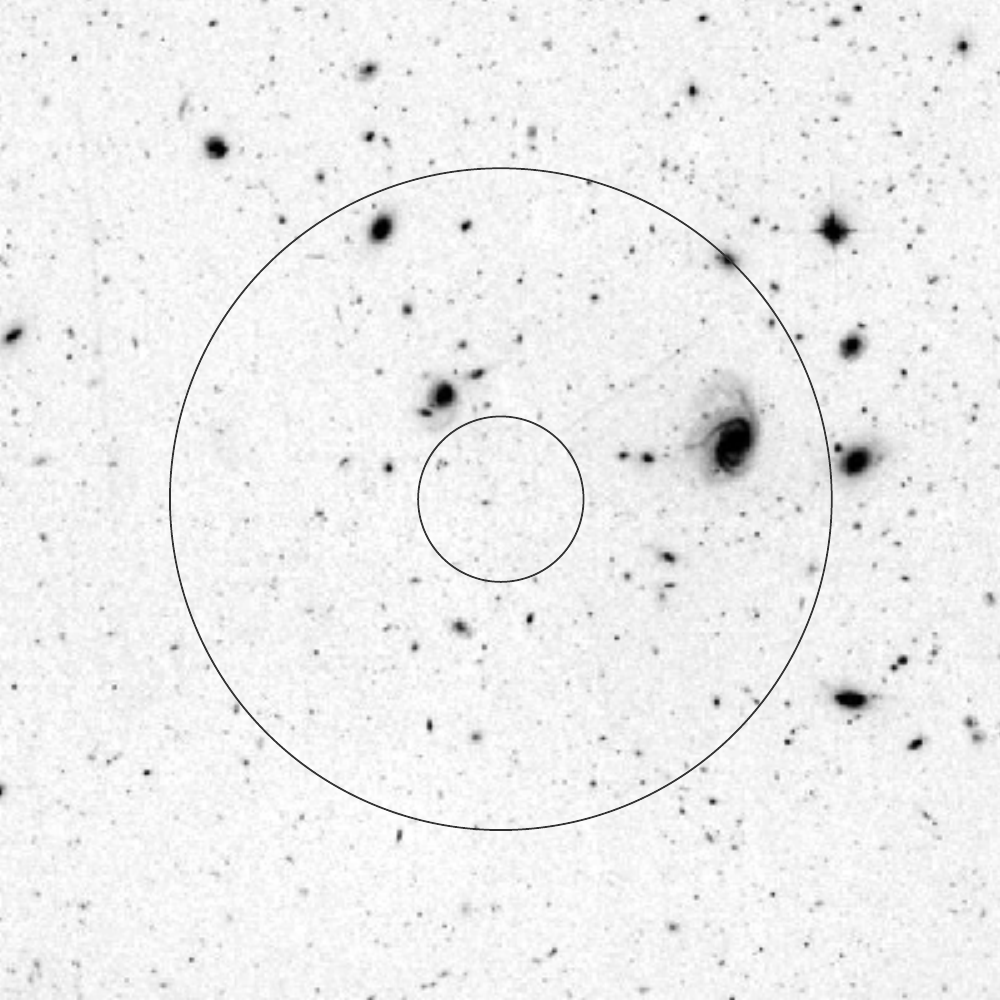}}
\newcommand{\aegisacsquadsizesfig}{\includegraphics[width=1.000000\figunit]{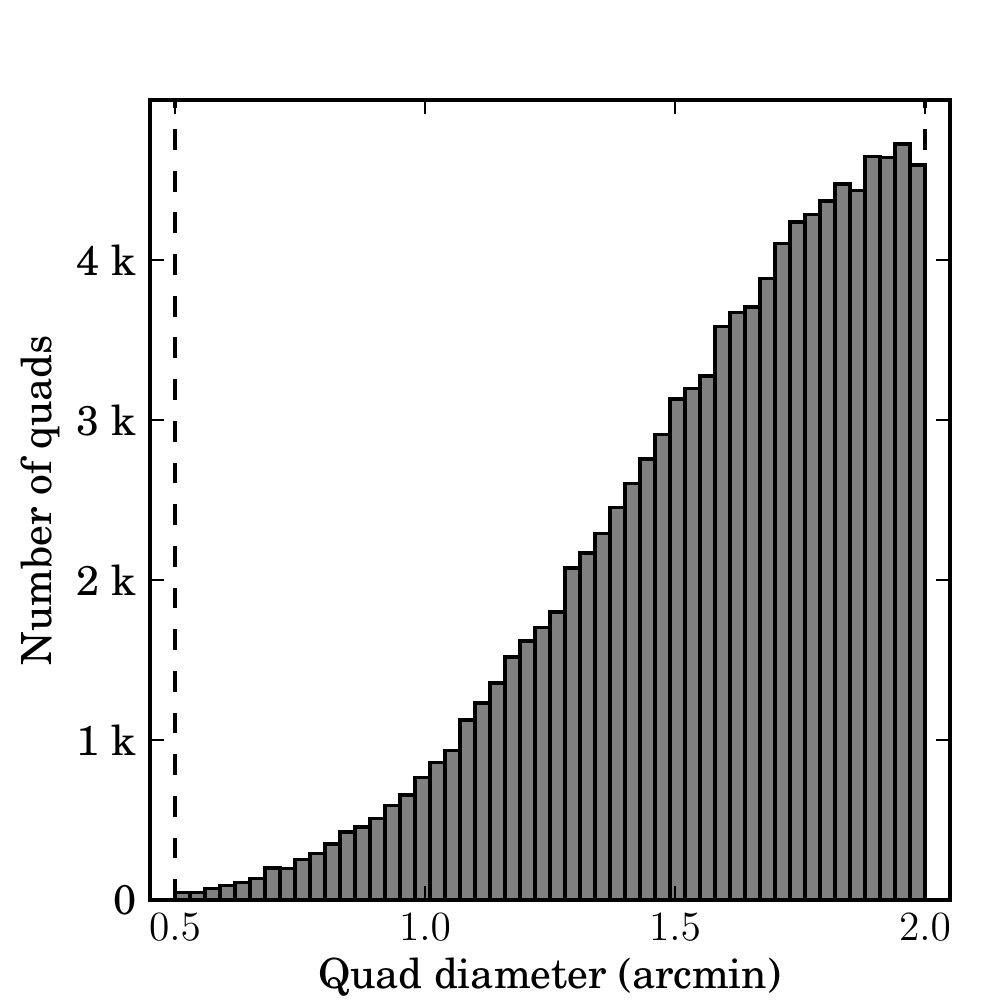}}

\newcommand{\usnob}{USNO-B\xspace}
\newcommand{\twomass}{2MASS\xspace}
\newcommand{\USNOB}{USNO-B catalog\xspace}

\newcommand{\band}[1]{\ensuremath{#1}}
\newcommand{\uband}{\band{u}}
\newcommand{\gband}{\band{g}}
\newcommand{\rband}{\band{r}}
\newcommand{\iband}{\band{i}}
\newcommand{\zband}{\band{z}}

\section{Introduction}

Although there are hundreds of ground- and space-based telescopes
currently operating, and petabytes of stored astronomical images (some
fraction of which are available in public archives), most astronomical
research is conducted using data from a single telescope.  Why do we
as astronomers limit ourselves to using only small subsets of the
enormous bulk of available data?  Three main reasons can be
identified: we don't want to share our data; it's hard to share our
data; and it's hard to use data that others have shared.  The latter
two problems can be addressed by technological solutions, and as
sharing data becomes easier, astronomers will likely become more
willing to do so.

Historically, astronomical data---and, indeed, important scientific
results---have often been closely guarded.  In more recent times,
early access to data has been seen as one of the rewards for joining
and contributing to telescope-building collaborations, and proprietary
data periods typically accompany grants of observing time on
observatories such as the Hubble Space Telescope.  However, this seems
to be changing, albeit slowly.  One of the first large astronomical
data sets to be released publicly in a usable form was the Hubble Deep
Field \citep{hubbledeepfield}.  The Sloan Digital Sky Survey (SDSS;
\citealt{sdsstechnical}) is committed to yearly public data releases, 
and the members of upcoming projects such as the Large Synoptic Survey
Telescope (LSST; \citealt{lsst}) have recognized that the primary
advantage of contributing to the collaboration is not proprietary
access to the data, but rather a deep understanding and familiarity
with the telescope and data, and have (in principle) decided to make
the data available immediately.

Putting aside the issue of \emph{willingness} to share data, there are
issues of our \emph{ability} to share data effectively.  Making use of
large, heterogeneous, distributed image collections requires fast,
robust, automated tools for calibration, vetting, organization, search
and retrieval.

The Virtual Observatory (VO; \citealt{vo}) establishes a framework and
protocols for the organization, search, and retrieval of astronomical
images.  The VO is structured as a distributed system in which many
``publishers'' provide image collections and interfaces that allow
these images to be searched.  This distributed framework allows the VO
to scale, but it also means that any property we might want the images
published through the VO to have must be specified by the standards,
and we must trust all publishers to implement the standards correctly.
Of particular importance are calibration \metadata
\citep{astrometryandvo}.  The draft Simple Image Access Protocol
\citep{siap} states that ``an image should be a calibrated object
frame'' and specifies some loose requirements for astrometric
\metadata.  However, there is neither a requirement, nor a specified
method, for communicating more detailed information about the
calibration processes that have been applied to the image.  A VO user
who wants to know exactly how the raw CCD frame was reduced to the
pixel values in the image must use processes outside the VO
framework---most likely by reading papers and contacting the image
publisher---and this will take much longer than finding and retrieving
the image.

The VO cannot be expected to impose a minimum standard of ``quality''
on images published through VO protocols, for several reasons.  First,
doing so would require making a tradeoff between the quality and
quantity of images that are publishable.  Since different users of the
VO have different needs, there is no objective way to make this
tradeoff.  For example, one researcher might only want images taken
during photometric conditions, while one studying a transient event
might want all available imaging, regardless of quality.  Second,
there is no objective measure of the quality of an image: different
aspects of an image are important to different users.  Finally, even
the most well-intentioned and skilled publisher will occasionally make
mistakes, and this effect will become more pronounced as surveys
become more automated and data rates increase.  Thus, users of the VO
cannot in general rely on the quality or correctness of image data or
calibration \metadata, and individually hand-checking each image does
not scale to the future in which the VO provides access to any
significant fraction of the world's astronomical images.  For the
goals of the VO movement to be achieved, the tools that allow users to
vet, verify, and recalibrate images must be developed, and ideally
these tools will be integrated into the VO system.

In this \doctype we present a system, \an, that automatically
produces astrometric \metadata for astronomical images.  That is,
given an image, our system produces the pointing, scale, and
orientation of the image---the astrometric calibration \metadata or
World Coordinate System (WCS).  The system requires no first guess,
and works with the information in the image pixels alone.  The success
rate is above $99.9~\percent$ for contemporary near-ultraviolet and
visual imaging survey data, with no false positives.

Our system enables an immense amount of ``lost'' astronomical imagery
to be used for scientific purposes.  This includes photographic plate
archives, an immense and growing number of images taken by amateur
astronomers, as well as data from individual professional astronomers
and ground-based observatories whose \metadata are non-existent,
lost, or simply wrong.  While many modern telescopes do produce
correct, standards-compliant \metadata, many others have control
systems that drift relative to the sky, yielding only approximate
astrometric \metadata.  Still others produce no \metadata, or produce
it in some ideosyncratic, non-standards-compliant form.  Even
sophisticated and highly-automated surveys such as SDSS occasionally
have failures in the systems that produce astrometric calibration
information, resulting in perfectly good but ``lost'' images.  A
system that makes these data available for study will effectively
recover a significant amount of lost observing time, fill in gaps in
the astronomical record, and make observers more productive by
eliminating the tedious and often unenlightening task of fixing the
astrometry.  Furthermore, a robust, fully-automated system allows the
data to be trusted, because the calibration \metadata, and their
associated error estimates, have been derived from the images
themselves, not from some unknown, undocumented, unverified or
untrustworthy external source.

Our system can be seen as a specialized kind of image-based search.
Given an image, we can identify and label the objects that appear in
the image, with a very high success rate and no false positives.  We
have achieved the ultimate goal of computer vision, within the domain
of astronomical images.  Our system is based solely on the contents of
the image, in sharp contrast to most contemporary image search systems
(such as Google Image Search), which rely on contextual
information---the text surrounding the image on a web page---rather
than the information in the image itself.

In the literature, the task of recognizing astronomical images is
known as ``blind astrometric calibration'' or the ``lost in space''
problem, since an early application was for estimating the attitude of
a spacecraft using a camera mounted on the spacecraft.  By identifying
the stars that are visible, the pose of the camera can be determined
\citep{liebe1993}.  In such systems, triangles of stars are typically
used as geometric features (for example, \citealt{junkins1977}).  Triangles are
effective in this regime because the images typically span tens of
degrees and contain only dozens of very bright stars: the search space
is small.  Furthermore, these systems are not fully ``blind'' in that
they are designed for particular cameras, the specifications of which
are available to the system designers.

Triangle-based approaches have also been used to fine-tune the astrometry
problem when a good initial estimate is available \citep{pal2006}.
Because the search domain is limited to a small area around the
initial estimate, the triangle-based approach is effective.
Completely blind systems have been attempted previously (for example,
\citealt{harvey2004} and references therein) but none we know of have been
able to achieve the scalability and fidelity of our approach.

Both the triangle matching approach and ours (described below) are
based on ``geometric hashing'' (for example, \citealt{lamdan1990},
\citealt{huttenlocher1990}).  Our system uses the same two-step
approach used by these systems, in which a set of hypotheses are
generated from sparse matches, then a second stage does detailed
verification of the hypotheses.

There are several automated calibration systems that refine the
astrometric calibration of an image to produce a high-precision
alignment to a reference catalog given a good first guess (for
example, \citealt{valdes1995, wcstools4, bertin2005}).  These systems are
reliable and robust, but they require a reasonable first guess about
the image pointing, orientation, and scale.  Our system can be used to
\emph{create} that good first guess.

\section{Methods}

Our approach involves four main components.  First, when given a query
image, we detect astronomical sources (``stars'', hereafter) by
running a number of image-processing steps.  This typically yields a
few hundred or more stars localized to sub-pixel accuracy.  Next, the
system examines subsets of these stars, producing for each subset a
geometric hash code that describes their relative positions.  We
typically use subsets of four stars, which we call ``quads.''  Having
computed the hash code for the query quad, the system then searches in
a large pre-computed index for almost identical hash codes.  Each
matching hash code that is found corresponds to a hypothesized
alignment between the quad in the query image and the quad in the
index, which can be expressed as a hypothesized location, scale, and
orientation of the image on the sky.  The final component is a
verification criterion, phrased as a Bayesian decision problem, which
can very accurately decide if the hypothesized alignment is correct.
The system continues generating and testing hypotheses until we find
one that is accepted by the verification process; we then output that
hypothesis as our chosen alignment.  In some cases, we never find a
hypothesis in which we are sufficiently confident (or we give up
searching before we find one), but our thresholds are set
conservatively enough that we almost never produce a false positive
match.  Each component of our approach is outlined below.

%--- where should this paragraph go?

The primary technical contributions of our system include the use of
the ``geometric hashing'' \citep{lamdan1990, wolfson1997} approach to
solve the huge search problem of generating candidate calibrations; an
index-building strategy that takes into account the distribution of
images we wish to calibrate; the verification procedure which
determines whether a proposed astrometric calibration is correct; and
good software engineering which has allowed us to produce a practical,
efficient system.  All our code is publicly available under the GPL
license, and we are also offering a web service.

\subsection{Star detection}

The system automatically detects compact objects in each input image
and centroids them to yield the pixel-space locations of stars.  This
problem is an old one in astronomy; the challenge \emph{here} is to
perform it robustly, with no human intervention, on the large variety
of input images the system faces.  Luckily, because the rest of the
system is so robust, it can handle detection lists that are missing a
few stars or have some contaminants.

The first task is to identify (detect) localized sources of light in
the image.  Most astronomical images exhibit sky variations,
sensitivity variations, and scattering; we need to find peaks on top
of such variations.  First we subtract off a median-smoothed version
of the image to ``flatten'' it.  Next, to find statistically
significant peaks, we need to know the approximate noise level.  We
find this by choosing a few thousand random pairs of pixels separated
by five rows and columns, calculating the difference in the fluxes for
each pair, and calculating the variance of those differences, which is
approximately twice the variance $\sigma^2$ in each pixel.  At this
point, we identify pixels which have values in the flattened image
that are $>8\,\sigma$, and connect detected pixels into individual
detected objects.

The second task is to find the peak or peaks in each detected object.
This determination begins by identifying pixels that contain larger
values than all of their neighbors.  However, keeping all such pixels
would retain peaks that are just due to uncorrelated noise in the
image and individual peaks within a single ``object.''  To clean the
peak list, we look for peaks that are joined to smaller peaks by
saddle points within $3\,\sigma$ of the larger peak (or $1~\percent$
of the larger peak's value, whichever is greater), and trim the
smaller peaks out of our list.

%%%% FIXME -- what does the code actually do?

Finally, given the list of all peaks, the third task is to centroid
the star position at sub-pixel accuracy. Following previous work
\citep{sdssimaging}, we take a $3\times 3$ grid around each star's peak
pixel, effectively fitting a Gaussian model to the nine values, and
using the peak of the Gaussian. Occasionally this procedure produces a
Gaussian peak outside the $3\times 3$ grid, in which case we default
to the peak pixel, although such cases are virtually always caused by
image artifacts.  This procedure produces a set of $x$ and $y$
positions in pixel coordinates corresponding to the position of
objects in the image.

% Mention more sophisticated options -- modelling the PSF and fitting
% that rather than the rather arbitrary Gaussian -- (which isn't what
% it actually does anyway!  It fits a general paraboloid!)

Compared to other star detection systems such as
SExtractor \citep{sextractor}, our approach is simpler and generally
more robust given a wide variety of images and no human intervention.
For example, while we do a simple median-filtering to remove the
background signal from the image, SExtractor uses sigma-clipping and
mode estimation on a grid of subimages, which are then median-filtered
and spline-interpolated.

\subsection{Hashing of asterisms to generate hypotheses}

Hypotheses about the location of an astronomical image live in the
continuous four-dimensional space of position on the celestial sphere
(pointing of the camera's optical axis), orientation (rotation of the
camera around its axis), and field of view (solid angle subtended by
the camera image).  We want to be able to recognize images that span
less than one-millionth the area of the sky, so the effective number
of hypotheses is large; exhaustive search will be impractical.  We
need a fast search heuristic: a method for proposing hypotheses that
almost always proposes a correct hypothesis early enough that we have
the resources to discover it.

\begin{figure}[htp]
  \begin{center}
    \includegraphics[width=\quadfigwidth]{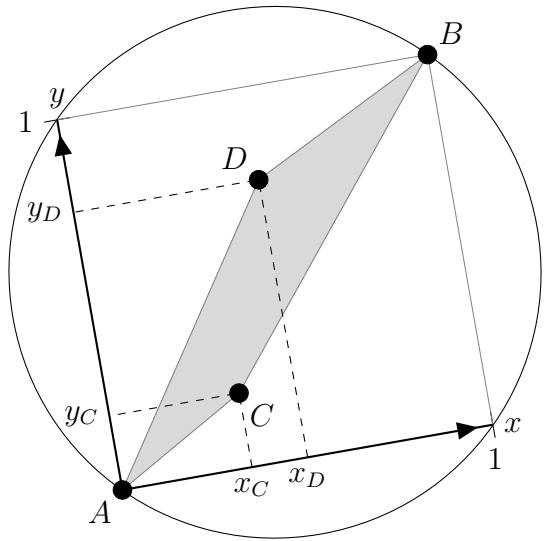}
  \end{center} 
  \caption{The geometric hash code for a ``quad'' of stars, $\starA$,
      $\starB$, $\starC$, and $\starD$.  Stars $\starA$ and $\starB$
      define the origin and \mbox{$(1,1)$,} respectively, of a local
      coordinate system, in which the positions of stars $\starC$ and
      $\starD$ are computed.  The
      coordinates \mbox{$(\xC,\yC,\xD,\yD)$} become our geometric hash
      code that describes the relative positions of the four stars.
	  The hash code is invariant under translation, scaling, and
      rotation of the four stars.  \label{fig:quad}}
\end{figure}

Our fast search heuristic uses a continuous geometric hashing
approach.  Given a set of stars (a ``quad''), we compute a local
description of the shape---a geometric hash code---by mapping the
relative positions of the stars in the quad into a point in a
continuous-valued, 4-dimensional vector space (``code space'').
\Figref{fig:quad} shows this process.  Of the four stars comprising
the quad, the most widely-separated pair are used to define a local
coordinate system, and the positions of the remaining two stars in
this coordinate system serve as the hash code.  We label the most
widely-separated pair of stars ``$\starA$'' and ``$\starB$''.  These
two stars define a local coordinate system.  The remaining two stars
are called ``$\starC$'' and ``$\starD$'', and their positions in this
local coordinate system are $(\xC,\yC)$ and $(\xD,\yD)$.  The
geometric hash code is simply the 4-vector \mbox{$(\xC,\yC,\xD,\yD)$}.
We require stars $\starC$ and $\starD$ to be within the circle that
has stars $\starA$ and $\starB$ on its diameter.  This hash code has
some symmetries: swapping $\starA$ and $\starB$ converts the code to
\mbox{$(1\!-\!\xC,1\!-\!\yC,1\!-\!\xD,1\!-\!\yD)$} while swapping
$\starC$ and $\starD$ converts \mbox{$(\xC,\yC,\xD,\yD)$} into
\mbox{$(\xD,\yD,\xC,\yC)$.}  In practice, we break this symmetry by
demanding that \mbox{$\xC \le \xD$} and that \mbox{$\xC + \xD \le 1$;}
we consider only the permutation (or relabelling) of stars that
satisfies these conditions (within noise tolerance).

This mapping has several properties that make it well suited to our
indexing application. First, the code vector is \emph{invariant} to
translation, rotation and scaling of the star positions so that it can
be computed using only the relative positions of the four stars in any
conformal coordinate system (including pixel coordinates in a query
image).  Second, the mapping is \emph{smooth}: small changes in the
relative positions of any of the stars result in small changes to the
components of the code vector; this makes the codes resilient to small
amounts of positional noise in star positions.  Third, if stars are
uniformly distributed on the sky (at the angular scale of the quads
being indexed), codes will be uniformly distributed in (and thus make
good use of) the 4-dimensional code-space volume.

Noise in the image and distortion caused by the atmosphere and
telescope optics lead to noise in the measured positions of stars in
the image.  In general this noise causes the stars in a quad to move
slightly with respect to each other, which yields small changes in the
hash code (\ie, position in code space) of the quad.  Therefore, we
must always match the image hash code with a \emph{neighborhood} of
hash codes in the index.

The standard geometric hashing ``recipe'' would suggest using
triangles rather than quads.  However, the positional noise level in
typical astronomical images is sufficiently high that triangles are
not distinctive enough to yield reasonable performance.  An important
factor in the performance of geometric hashing system is the
``oversubscription factor'' of code space.  The number of hash codes
that must be contained in an index is determined by the effective
number of objects that are to be recognized by the system: if the goal
is to recognize a million distinct objects, the index must contain at
least a million hash codes.  Each hash code effectively occupies a
volume in code space: since hash codes can vary slightly due to
positional noise in the inputs (star positions), we must always search
for matching codes within a volume of code space.  This volume is
determined by the positional noise levels in the input image and the
reference catalog.  The oversubscription factor of code space, if it
is uniformly populated, is simply the number of codes in the index
multiplied by the fractional volume of code space occupied by each
code.  If triangles are used, the fractional volume of code space
occupied by a single code is large, so the code space becomes heavily
oversubscribed.  Any query will match many codes by coincidence, and
the system will have to reject all of these false matches, which is
computationally expensive.  By using quads instead of triangles, we
nearly \emph{square} the distinctiveness of our features: a quad can
be thought of as two triangles that share a common edge, so a quad
essentially describes the co-occurrence of two triangles.  A much
smaller fraction of code space is occupied by each quad, so we expect
fewer coincidental (false) matches for any given query, and therefore
fewer false hypotheses which must be rejected.

We could use quintuples of stars, which are even more distinctive than
quads.  However, there are two disadvantages to increasing the number
of stars in our asterisms.  The first is that the probability that all
$k$ of the stars in an indexed asterism appear in the image and that
all $k$ of the stars in a query asterism appear in the index both
decrease with increasing $k$.  For images taken at wavelengths far
from the catalog wavelength, or shallow images, this consideration can
become severe.  The second disadvantage is that near-neighbor lookup,
even with a \kdtree, becomes more time-consuming with increasing
dimensionality. The dimensionality of the code space for quintuples is
$6$-dimensional, compared to the $4$-dimensional code space of quads.
We test triangle- and quintuple-based indices in
\secref{sec:triquint} below.

When presented with a list of stars from an image to calibrate, the
system iterates through groups of four stars, treating each group as a
quad and computing its hash code.  Using the computed code, we perform
a neighborhood lookup in the index, retrieving all the indexed codes
that are close to the query code, along with their corresponding
locations on the sky.  Each retrieved code is effectively a
\emph{hypothesis}, which proposes to identify the four reference
catalog stars used to create the code at indexing time with the four
stars used to compute the query code. Each such hypothesis is
evaluated as described below.

The question of which hypotheses to check and when to check them is a
purely heuristic one. One could chose to wait until a hypothesis has
two or more ``votes'' from independent codes before checking it or
check every hypothesis as soon as it is proposed, whichever is faster.
In our experiments, we find that it is faster, and much less
memory-intensive, to simply check every hypothesis rather than
accumulate votes.

\subsection{Indexing the sky}

As with all geometric hashing systems, our system is based around a
pre-computed index of known asterisms.  Building the index begins with
a reference catalog of stars.  We typically use an all-sky (or
near-all-sky) optical survey such as USNO-B1 \citep{usnob,
barroncleaning} as our reference catalog, but we have also used the
infrared 2MASS catalog \citep{twomass} and the ultraviolet catalog from
GALEX \citep{galex}, as well as non-all-sky catalogs such as SDSS.
From the reference catalog we select a large number of quads (using a
process described below).  For each quad, we store its hash code and a
reference to the four stars of which it is composed.  We also store
the positions of those four stars.  Given a query quad, we compute its
hash code and search for nearby codes in the index.  For each nearby
code, we look up the corresponding four stars in the index, and create
the hypothesis that the four stars in the query quad correspond to the
four stars in the index.  By looking up the positions of the query
stars in image coordinates and the index stars in celestial
coordinates, we can express the hypothesis as a pointing, scale, and
rotation of the image on the sky.

In order for our approach to be successful, our index must balance
several properties.  We want to be able to recognize images from any
part of the sky, so we want to choose quads uniformly over the sky.
We want to be able to recognize images with a wide range of angular
sizes, so we want to choose quads of a variety of sizes.  We expect
that brighter stars will be more likely to be found in our query
images, so we want to build quads out of bright stars preferentially.
However, we also expect that some stars, even the brightest stars,
will be missing or mis-detected in the query image (or the reference
catalog), so we want to avoid over-using any particular star to build
quads.

We handle the wide range of angular sizes by building a series of
sub-indices, each of which contains quads whose quads have scales
within a small range (for example, a factor of two).  At some level
this is simply an implementation detail: we could recombine the
sub-indices into a single index, but in what follows it is helpful to
be able to assume that the sub-index will be asked to recognize query
images whose angular sizes are similar to the size of the quads it
contains.  Since we have a set of sub-indices, each of which is tuned
to an overlapping range of scales, we know that at least one will be
tuned to the scale of the query image.

\begin{figure}[htp]
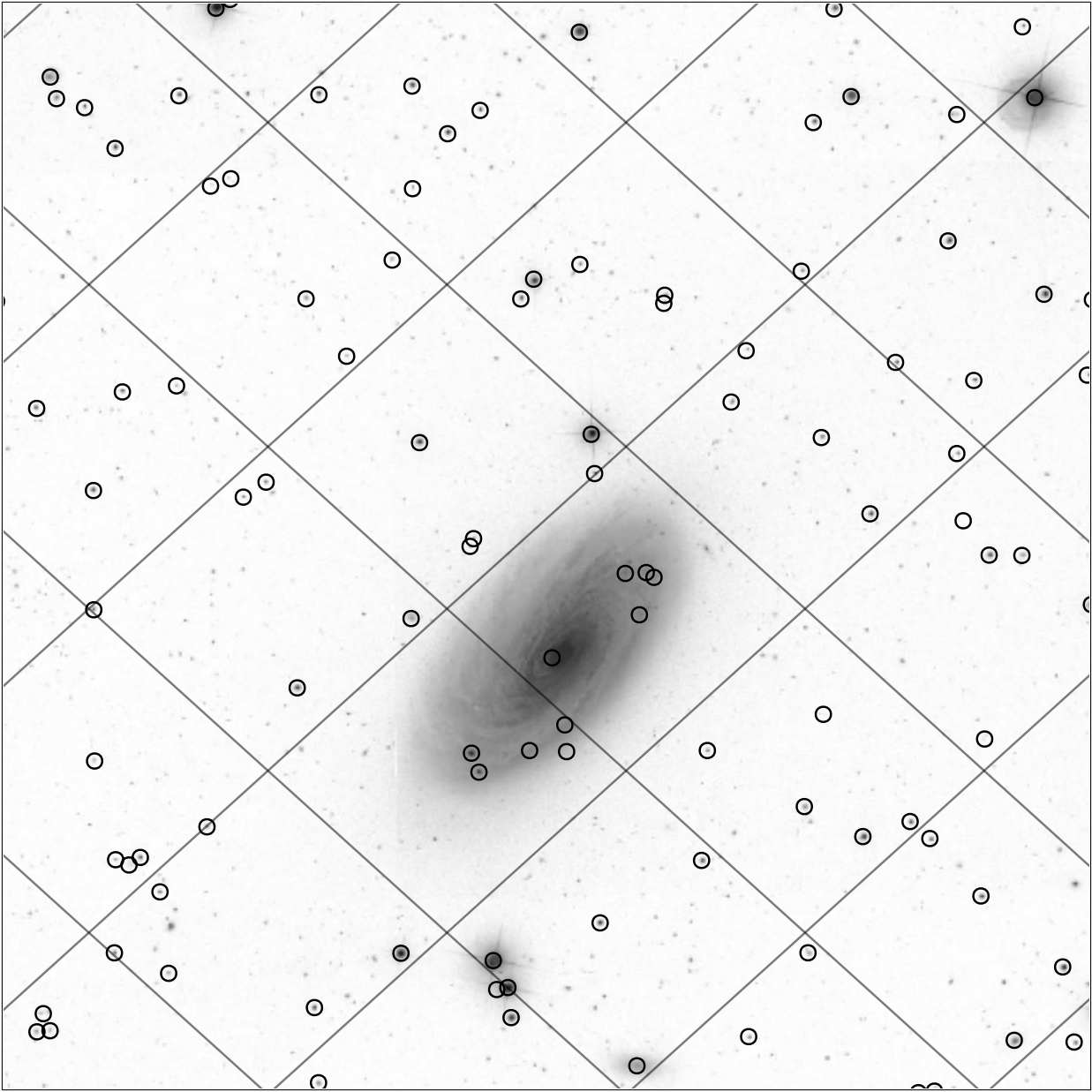

\begin{center}
% This and the following figs are generated by sdss-tests/cut-index-fig.sh
\setlength{\fboxsep}{0.5pt}
\framebox{\includegraphics[width=0.99\textwidth]{\filesuffix{\bigorsmallfig{cutb}{cut-small}}{pdf}}}
\end{center}
\caption{A small region of sky (about $0.3\times0.3~\degrees$ centered on
$(\RA, \Dec) = (188, 14.45)~\degrees$), showing the \healpix grid, and
the brightest $5$ stars that we select from each cell.  The image
shown is from the Sloan Digital Sky Survey.
\label{fig:cut}}
\end{figure}

We begin by selecting a spatially-uniform and bright subset of stars
from our reference catalog.  We do this by placing a grid of
equal-area patches (``HEALPixels''; \citealt{healpix}) over the sky and
selecting a fixed number of stars, ordered by brightness, from each
grid cell.  The grid size is chosen so that grid cells are a small
factor smaller than the query images.  Typically we choose the grid
cells to be about a third of the size of the query images, and select
$10$ stars from each grid cell, so that most query images will contain
about a hundred query stars.  \Figref{fig:cut} illustrates this
process on a small patch of sky.

\begin{figure}[htp]
\begin{center}
\setlength{\fboxsep}{0.5pt}
\framebox{\includegraphics[width=0.99\textwidth]{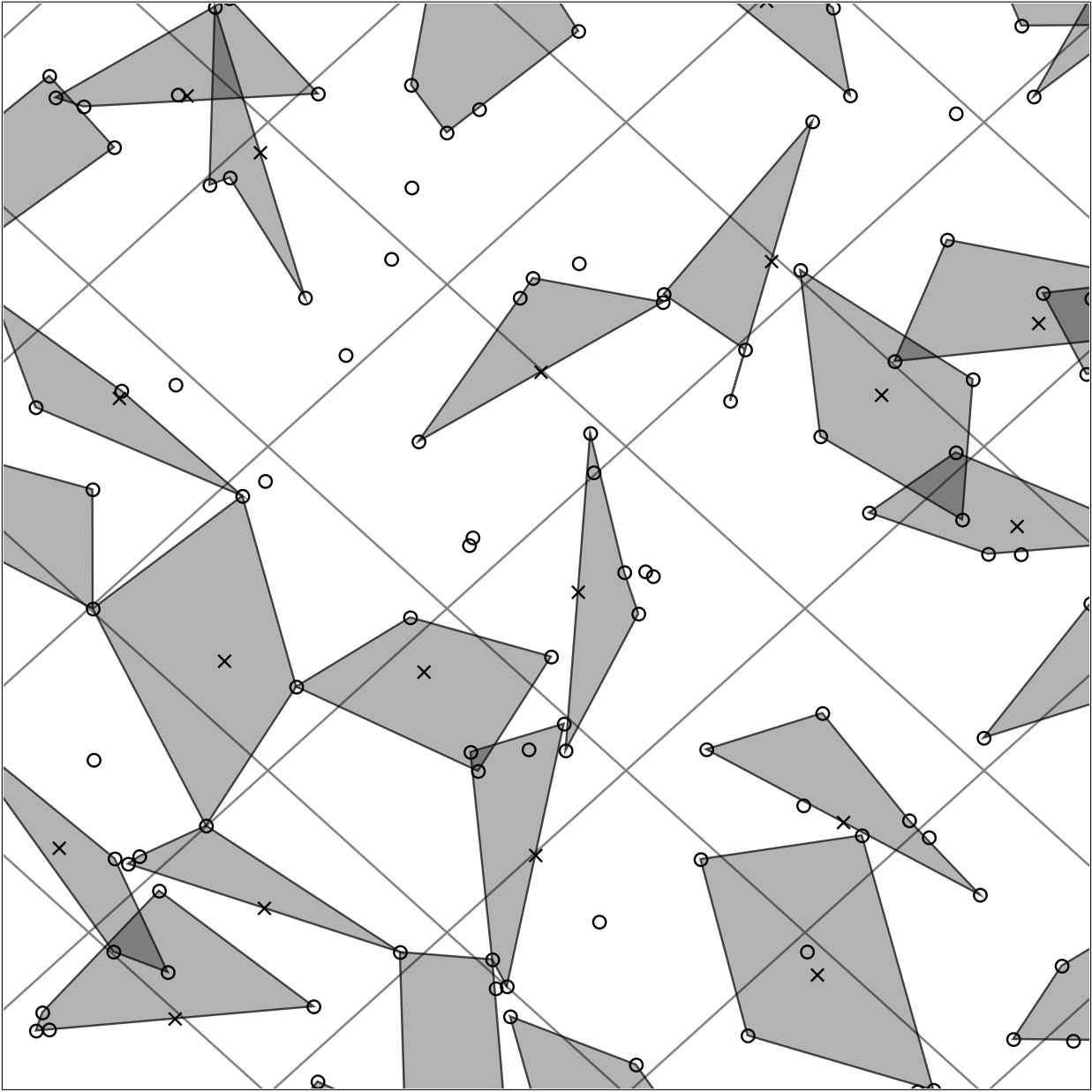}}
\end{center}
\caption{The same region of sky as shown in the previous figure, showing the
\healpix grid, and the quads that are created during the first pass through
the grid cells.  The quads must have a diameter (the distance between
the two most distant stars) within a given range---in this case, $1$
to $\sqrt{2}$ times the side-length of the grid cells.  In each grid
cell, the system attempts to build a quad whose center (\ie, midpoint
of the diameter line)---marked with an $\mathbf{\times}$ in the
figure---is within the grid cell.
\label{fig:quad1}}
\end{figure}

%% FIXME - shrink ``quads2'' and add ``quads4'' side-by-side ?
%% -mark 'X'es too

\begin{figure}[htp]
\begin{center}
\setlength{\fboxsep}{0.5pt}
\framebox{\includegraphics[width=0.99\textwidth]{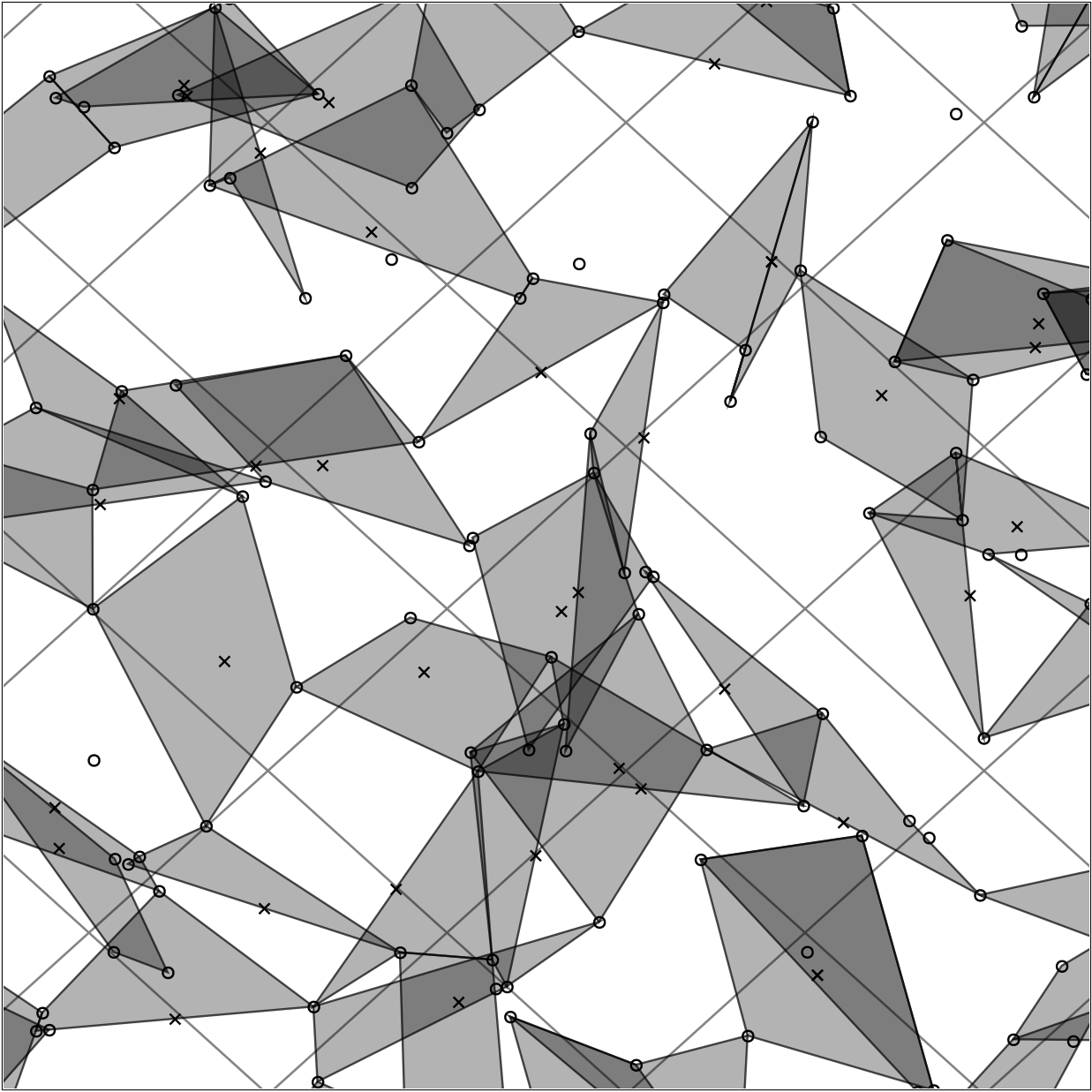}}
\end{center}
\caption{The same region of sky as in the previous figures, showing the
quads that are created after the second round of attempting to build a
quad in each grid cell.
\label{fig:quad2}}
\end{figure}

Next, we visit each grid cell and attempt to find a quad within the
acceptable range of angular sizes and whose center lies within the
grid cell.  We search for quads starting with the brightest stars, but
for each star we track the number of times it has already been used to
build a quad, and we skip stars that have been used too many times
already.  We repeat this process, sweeping through the grid cells and
attempting to build a quad in each one, a number of times.  In some
grid cells we will be unable to find an acceptable quad, so after this
process has finished we make further passes through the grid cells,
removing the restriction on the number of times a star can be used,
since it is better to have a quad comprised of over-used stars than no
quad at all.  Typically we make a total of $16$ passes over the grid
cells, and allow each star to be used in up to $8$ quads.  \Figs
\ref{fig:quad1} and \ref{fig:quad2} show the quads built during the first
two rounds of quad-building in our running example.

In principle, an index is simply a list of quads, where for each quad
we store its geometric hash code, and the identities of the four stars
of which it is composed (from which we can look up their positions on
the sky).  However, we want to be able to search quickly for all hash
codes near a given query hash code.  We therefore organize the hash
codes into a \kdtree data structure, which allows rapid retrieval of
all quads whose hash codes are in the neighborhood of any given query
hash code.  In order to carry out the verification step we also keep
the star positions in a \kdtree, since for each matched quad we need
to find other stars that should appear in the image if the match is
true.  Since none of the available \kdtree implementations were
satisfactory for our purposes, we created a fast, memory-efficient,
pointer-free \kdtree
\thesisonly{implementation.}%
\notthesisonly{implementation \citep{dstnthesis}.}
\thesisonly{Our \kdtree implementation is presented in detail
in \chapref{chap:kdtree}.}

\begin{figure}[htp]
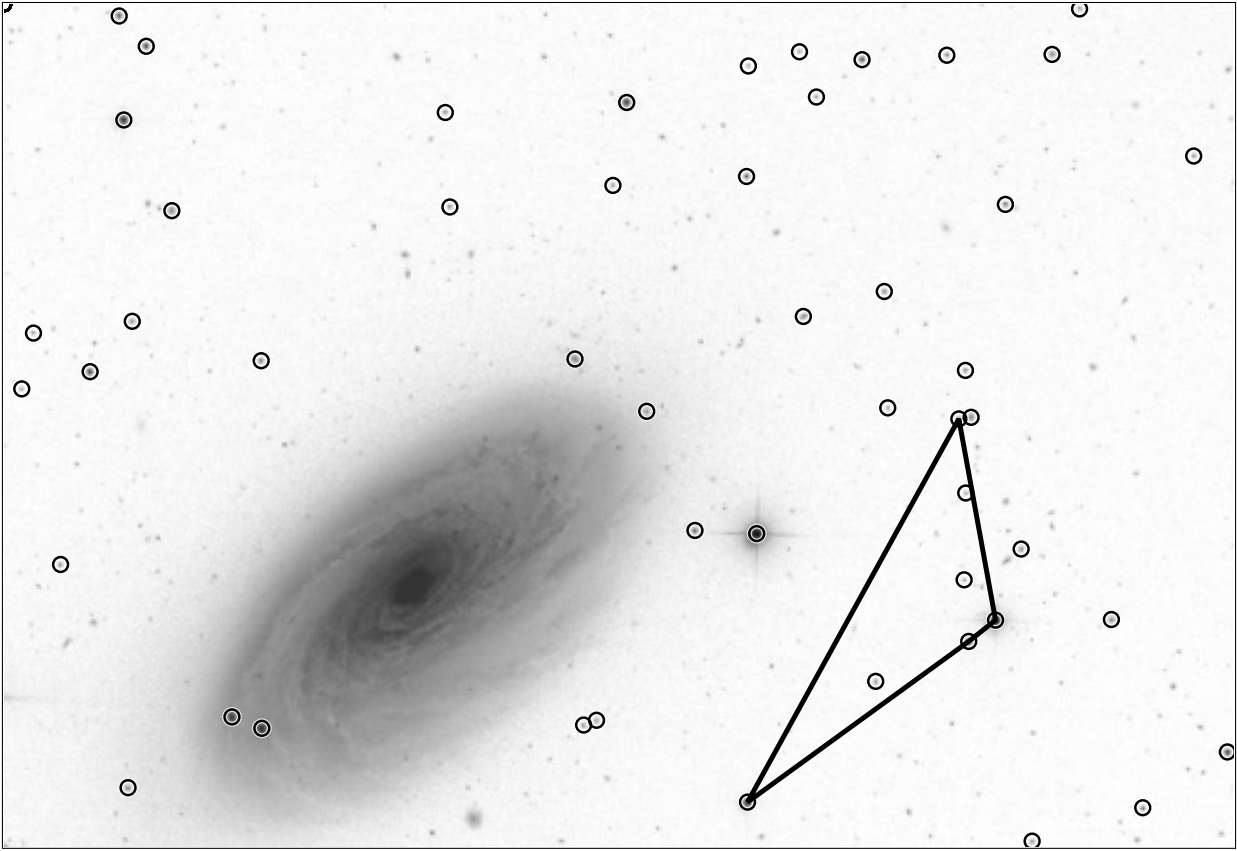

\begin{center}
\setlength{\fboxsep}{0.5pt}
\framebox{\includegraphics[width=0.99\textwidth]{\filesuffix{\bigorsmallfig{m88-quadb}{m88-quad-small}}{pdf}}}
\end{center}
\caption{A sample query image, with the brightest
$100$ sources our system detects (circles), and a quad in the image to
which our system will search for matches in the index.  This quad
looks like a triangle because two of its stars are nearly collinear.
Image credit: Sloan Digital Sky Survey.
\label{fig:imagequad}}
\end{figure}

\begin{figure}[htp]
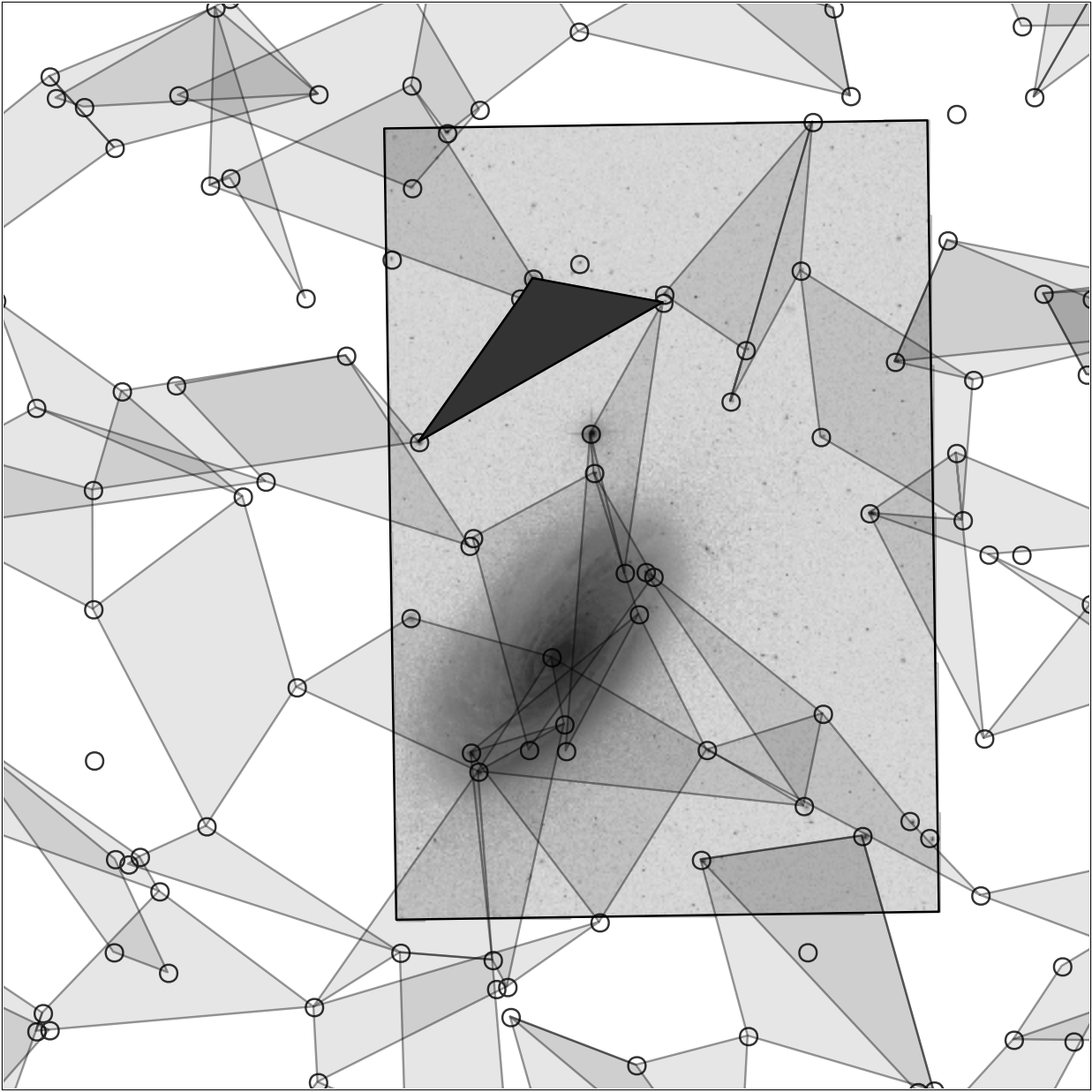

\begin{center}
\setlength{\fboxsep}{0.5pt}
\framebox{\includegraphics[width=0.99\textwidth]{\filesuffix{\bigorsmallfig{matchb}{match-small}}{pdf}}}
\end{center}
\caption{Our example index, showing a quad in the index that matches 
a quad from the query image (shaded solid).  The image is shown
projected in its correct orientation on the sky.
\label{fig:quadmatch}}
\end{figure}

Given a query image, we detect stars as discussed above, and sort the
stars by brightness.  Next, we begin looking at quads of stars in the
image.  For each quad, we compute its geometric hash code and search
for nearby codes in the index.  For each matching code, we retrieve
the positions of the stars that compose the quad in the index, and
compute the hypothesized alignment---the World Coordinate System---of
the match.  We then retrieve other stars in the index that are within
the bounds of the image, and run the verification procedure to
determine whether the match is true or false.  We stop searching when
we find a true match in which we are sufficiently confident, or we
exhaust all the possible quads in the query image, or we run out of
time.  See \figref{fig:imagequad}.

\subsection{Verification of hypotheses}

The indexing system generates a large number of hypothesized
alignments of the image on the sky.  The task of the verification
procedure is to reject the large number of false matches that are
generated, and accept true matches when they are found.  Essentially,
we ask, ``if this proposed alignment were correct, where else in the
image would we expect to find stars?'' and if the alignment has very
good predictive power, we accept it.

We have framed the verification procedure as a Bayesian decision
process.  The system can either accept a hypothesized match---in which
case the hypothesized alignment is returned to the user---or the
system can reject the hypothesis, in which case the indexing system
continues searching for matches and generating more hypotheses.  In
effect, for each hypothesis we are choosing between two models: a
``foreground'' model, in which the alignment is true, and a
``background'' model, in which the alignment is false.  In Bayesian
decision-making, three factors contribute to this decision: the
relative abilities of the models to explain the observations, the
relative proportions of true and false alignments we expect to see,
and the relative costs or \emph{utilities} of the outcomes resulting
from our decision.

The \emph{Bayes factor} is a quantitative assessment of the relative
abilities of the two models---the foreground model $\fg$ and the
background model $\bg$---to produce or explain the observations.  In
this case the observations, or data, $\data$, are the stars observed
in the query image.  The Bayes factor
\begin{equation}
K = \frac{p(\data \given \fg)}{p(\data \given \bg)}
%\label{eq:bayesf}
\end{equation}
is the ratio of the marginal likelihoods.  We must also include in our
decision-making the prior $p(\fg)/p(\bg)$, which is our \emph{a
priori} belief, expressed as a ratio of probabilities, that a proposed
alignment is correct.  Since we typically examine many more false
alignments than true alignments (because we stop after the first true
alignment is found), this ratio will be small.
\notthesisonly{We typically set it, conservatively, to $10^{-6}$.}

The final component of Bayesian decision theory is the \emph{utility}
table, which expresses the subjective value of each outcome.  It is
good to accept correctly a true match or reject correctly a false
match (``true positive'' and ``true negative'' outcomes,
respectively), and it is bad to reject a true match or accept a false
match (``false negative'' and ``false positive'' outcomes,
respectively).  In the \an setting, we feel it is very bad to produce
a false positive: we would much rather fail to produce a result rather
than produce a false result, because we want the system to be be able
to run on large data sets without human intervention, and we want to
be confident in the results.
\notthesisonly{%
Our utility table is shown here.
\begin{center}
  \newcommand{\spacer}{\hspace{0.7em}}
\centering
\begin{tabular}{|c|c|r@{$\ $}l|r@{$\ $}l|}
  \cline{3-6}
    \multicolumn{2}{c|}{} & \multicolumn{4}{c|}{Reality \tstrut} \\
  \cline{3-6}
  \multicolumn{2}{c|}{} & \multicolumn{2}{c|}{True Alignment} & \multicolumn{2}{c|}{False Alignment \tstrut} \\
  \hline
  \multirow{2}{*}{Decision} &
  Accept & \spacer $u(\truepos) =$ & $+1$ & \spacer $u(\falsepos) =$ & $-1999$ \tstrut \\
  \cline{2-6}
  &
  Reject & $u(\falseneg) =$ & $-1$ & $u(\trueneg) =$ & $+1$ \tstrut \\
  \hline
\end{tabular}
\end{center}
}

\notthesisonly{%
Applying Bayesian decision theory, we make our decision to accept or
reject the hypothesized alignment by computing the expected utility
$\expect{u}$ of each decision.  The expected utility of accepting the
alignment is:
\begin{align}
\expect{u \given \mathrm{Accept}, \data}
&= u(\truepos) \, p(\truepos \given \data) + u(\falsepos) \, p(\falsepos \given \data) \\
&= u(\truepos) \, p(\fg \given \data) + u(\falsepos) \, p(\bg \given \data)
\end{align}
while the expected utility of rejecting the alignment is:
\begin{align}
\expect{u \given \mathrm{Reject}, \data}
&= u(\falseneg) \, p(\falseneg \given \data) + u(\trueneg) \, p(\trueneg \given \data) \\
&= u(\falseneg) \, p(\fg \given \data) + u(\trueneg) \, p(\bg \given \data) \quad .
\end{align}
We should accept the hypothesized alignment if:
\begin{align}
\expect{u \given \mathrm{Accept}, \data} 
&>
\expect{u \given \mathrm{Reject}, \data} \\
u(\truepos) \, p(\fg \given \data) + u(\falsepos) \, p(\bg \given \data)
&>
u(\falseneg) \, p(\fg \given \data) + u(\trueneg) \, p(\bg \given \data) \\
\frac{p(\fg \given \data)}{p(\bg \given \data)}
&>
\frac{u(\trueneg) - u(\falsepos)}{u(\truepos) - u(\falseneg)} \\
K &> \frac{p(\bg)}{p(\fg)} \ \frac{u(\trueneg) - u(\falsepos)}{u(\truepos) - u(\falseneg)}
\label{eq:kthresh}
\end{align}
where we have applied Bayes' theorem to get
\begin{equation}
\frac{p(\fg \given \data)}{p(\bg \given \data)} = K \ \frac{p(\fg)}{p(\bg)} \quad .
\end{equation}
}

\notthesisonly{%
With the prior and utilities given above, we find that we should
accept a hypothesis if:
\begin{align}
K &> \frac{p(\bg)}{p(\fg)} \ \frac{u(\trueneg) - u(\falsepos)}{u(\truepos) - u(\falseneg)} \\
K &> 10^6 \ \frac{1 - -1999}{1 - -1} \\
K &> 10^9
\end{align}
}

\notthesisonly{
That is, we accept a proposed alignment if the Bayes factor of the
foreground model to the background model exceeds a threshold that is
set based on our desired operating characteristics.  In our case, the
threshold is large so the foreground model (in which the alignment is
true) must be far better at explaining (\ie, predicting) the observed
positions of stars in the query image than the background model (in
which the alignment is false).}

\thesisonly{
Applying Bayesian decision theory given our desired operating
characteristics, we find that we should accept a proposed alignment
only if the Bayes factor is exceedingly large.  That is, the
foreground model in which the alignment is true must be far better at
explaining the observed positions of stars in the query image than the
background model that the alignment is false.  We typically set the
Bayes factor threshold to $10^9$ or $10^{12}$, but as will be shown in
the experiments below, we could set it even higher.  This threshold is
computed from our (subjective) desired operating characteristics, so
is not derivable from first principles.
}

In the foreground model, $\fg$, the four stars in the query image and
the four stars in the index are aligned.  We therefore expect that
other stars in the query image will be close to other stars in the
index.  However, we also know that some fraction of the stars in the
query image will have no counterpart in the index, due to occlusions
or artifacts in the images, errors in star detection or localization,
differences in the spectral bandpass, or because the query image
``star'' is actually a planet, satellite, comet, or some other
non-star, non-galaxy object.  True stars can be lost, and false stars
can be added.  Our foreground model is therefore a mixture of a
uniform probability that a star will be found anywhere in the
image---a query star that has no counterpart in the index---plus a
blob of probability around each star in the index, where the size of
the blob is determined by the combined positional variances of the
index and query stars.

Under the background model, $\bg$, the proposed alignment is false, so
the query image is from some unknown part of the sky; the index is not
useful for predicting the positions of stars in the image.  Our simple
model therefore places uniform probability of finding stars anywhere
in the test image.  \notthesisonly{We have experimented with a more
sophisticated background models that adapts to the observed
distribution of image stars, but we do not discuss that work here.}

The verification procedure evaluates stars in the query image, in
order of brightness, under the foreground and background models.  The
product of the foreground-to-background ratios is the Bayes factor.
We continue adding query stars until the Bayes factor exceeds our
threshold for accepting the match, or we run out of query stars.

\thesisonly{There are some subtleties in the verification process
which are explored in depth in \chapref{chap:verify}.}

\section{Results}

\subsection{Blind astrometric calibration of the Sloan Digital Sky Survey}

We explored the potential for automatically organizing and annotating
a large real-world data set by taking a sample of images generated by
the Sloan Digital Sky Survey and considering them as an unstructured
set of independent queries to our system.  For each SDSS image, we
discarded all \metadata, including all positional and rotational
information and the date on which the exposure was taken.  We allowed
ourselves to look only at the two-dimensional positions of detected
``stars'' (most of which were in fact stars but some of which were
galaxies or detection errors) in the image.  Normally, our system
would take \emph{images} as input, running a series of image
processing steps to detect stars and localize their positions.  The
SDSS data reduction pipeline already includes such a process, so for
these experiments we used these detected star positions rather than
processing all the raw images ourselves.  Further experiments have
shown that we would likely have achieved similar, if not better,
results by using our own image processing software.

\begin{figure}[htp]
\begin{center}
%\sdssquadfig
\includegraphics[width=\figunit]{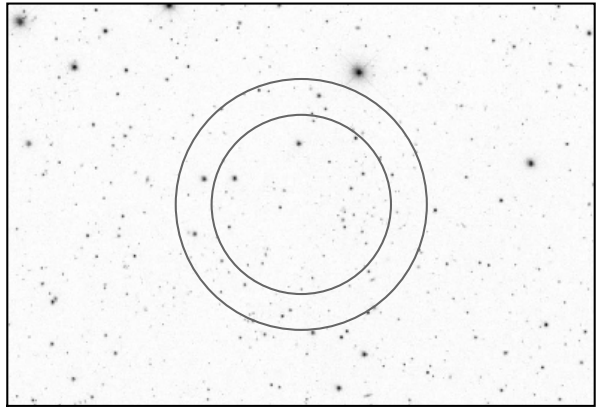}
\end{center}
%\caption{A typical SDSS image, with the \iband-, \rband-, and
%\gband-band exposures mapped to the red, green, and blue channels of the image.
%The range of quad diameters that we use in the experiments below is
%shown by the circles.
\caption{A typical image from the Sloan Digital Sky Survey (SDSS).
The range of quad diameters that we use in the experiments below is
shown by the circles.  Image credit: Sloan Digital Sky Survey.
\label{fig:sdss}}
\end{figure}

Each SDSS image has $2048\times1489$ pixels and covers
$9\times13~\arcmin^2$, slightly less than one-millionth the area of
the sky.  Each image measures one of five bandpasses, called \uband,
\gband, \rband, \iband, and \zband, spanning the near-infrared through
optical to near-ultraviolet range.  Each band receives a $54$-second
exposure on a $2.5$-meter telescope.  A typical image is shown in
\figref{fig:sdss}.

The SDSS image-processing pipeline assigns to each image a quality
rating: ``excellent'', ``good'', ``acceptable'', or ``bad''.  We
retrieved the source positions (\ie, the list of objects detected by
the SDSS image-processing pipeline) in every image within the main
survey (Legacy and SEGUE footprints), using the public Catalog Archive
Server (CAS) interface to Data Release 7 \citep{sdssdr7}.  We retrieved
only sources categorized as ``primary'' or ``secondary'' detections of
stars and galaxies, and required that each image contained at least
$300$ objects.  The images that are excluded by these cuts contain
either very bright stars or peculiarities that cause the SDSS
image-processing pipeline to balk.  The number of images in Data
Release 7 and our cut is given in the table below.

%% FIXME --- histogram figure showing the 300 cut is reasonable.

%% 427,343 - distinct run,field,camcol in Field (fieldrfc)
%% 426,440 - distinct run,field,camcol in PhotoObjAll (allrfc)
%%    (903) - mostly bright stars; non-photometric
%% 421,381 - goodfields
%%  (5,059) - mostly mode=3, some mode=5, one mode=2
%%            --> quality	count
%%            --> 0	        3668   (bad)
%%            --> 1	        319    (acc)
%%            --> 2	        481    (good)
%%            --> 3	        591    (exc)
%%  see http://trac.astrometry.net/wiki/SdssCasNotes
%
% MyTable_3
%% 3	183359
%% 2	101490
%% 1	48802
%% 0	93692

\nonumberparagraphs
\begin{center}
\begin{tabular}{|l|D{,}{,}{3.3}|D{,}{,}{3.3}|}
\hline
\tableheaderx{Quality} & \tableheader{Total number of images} &
\tableheader{Number of images in our cut} \\
\hline
excellent & 183,359 & 182,221 \\
good & 101,490 & 100,763 \\
acceptable & 48,802 & 48,337 \\
bad & 93,692 & 89,219 \\
\hline
total & 427,343 & 420,540 \\
\hline
\end{tabular}
\end{center}
\numberparagraphs

\subsubsection{Performance on excellent images}

%%%%   The stats quoted in this section are using index 702.
%%%%      runs sdss-{1,11,12}
%%%%   I also ran 1302 and got almost exactly the same results.
%%%%      runs sdss-{16,17,18}

\begin{figure}[htp]
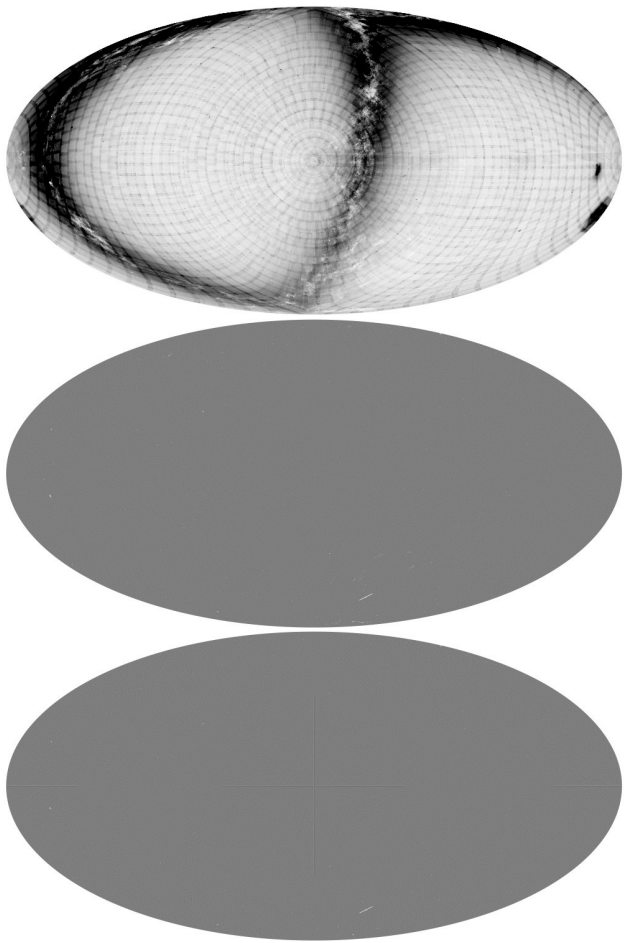

\begin{center}
\bigorsmallfig{\filesuffix{density-fig}{pdf}}{%
    \includegraphics[width=0.5\textwidth]{\filesuffix{usnob-density-smaller}{jpg}} \\
    \includegraphics[width=0.5\textwidth]{\filesuffix{cut-402-density-smaller}{jpg}} \\
    \includegraphics[width=0.5\textwidth]{\filesuffix{index-1302-density-smaller}{jpg}}%
    }
\end{center}
\vspace{-20pt}
\caption{\captionpart{Top:} Density of sources in the \USNOB 
(in Hammer-Aitoff projection).
Dark colors indicate high density.  The north
    celestial pole is in the center of the image.  
	The dark strip through the center and
    around the edges is the Milky Way; lower-density dust lanes can be
    seen.
	% The two dense blobs at the right are the Large and Small
    % Magellanic Clouds.
	%
	The \USNOB was created by scanning
    photographic plates, and the places where the plates overlap are
    clearly visible as concentric rings and spokes of overdensities.
    \captionpart{Middle:} Density of sources in our spatially uniform
    cut of the \USNOB.  Most of the sky is
    very uniformly covered.  A few small bright (low-density) areas
    are visible, including a line near the bottom.  These are areas
    where the \USNOB is underdense due to defects.
    \captionpart{Bottom:} Density of quads in the index used in most
    of the experiments presented here.  Again, most of the sky is uniformly
	covered with quads.
\label{fig:density}}
\end{figure}

In order to show the best performance achievable with our system, we
built an index that is well-matched to SDSS \rband-band images.
Starting with the red bands from a cleaned version of the \USNOB, we
built an index containing stars drawn from a \healpix grid with cell
sizes about $4\times4~\arcmin^2$, and $10$ stars per cell.  We then
built quads with diameters of $4$ to $5.6~\arcmin$.  For each grid
cell, we searched for a quad whose center was within the grid cell,
starting with the brightest stars but allowing each star to be used at
most $8$ times.  We repeated this process $16$ times.  The index
contains a total of about $100$ million stars and $150$ million quads.
\Figref{fig:density} shows the spatial distribution of the stars and
quads in the index.

% index 702, cut 402: nside=880 -> ~4 arcmin.  4-5.6' quads.
% (for x in index-702-*.fits; do modhead $x NSTARS; done) | gawk 'BEGIN{N=0} {N+=$3} END{print N}'

\begin{figure}[htp]
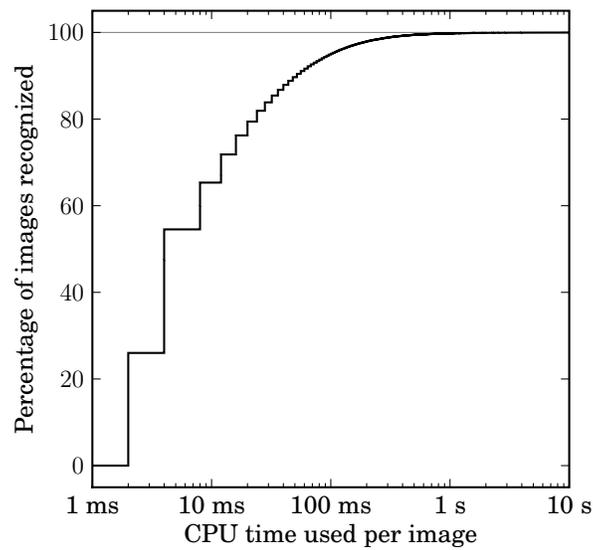

\begin{center}
    \sdssercputimefig
\end{center}
    \caption{Results from the excellent \rband-band SDSS images and
    \usnob-based index.  The percentage of images that are recognized
    correctly (\ie, astrometrically calibrated) with respect to the
    CPU time spent per image.  Many images are recognized rapidly, but
    there is a heavy tail of ``hard'' images.  Spending more and more
    CPU results in sharply diminishing returns.  After $1$ second,
    over $99.7~\percent$ of images are recognized, and after $10$
    seconds, over $99.97~\percent$ are recognized.  The steps are due
    to the finite resolution of the CPU timer we are using.
	\label{fig:sdsser1}}
\end{figure}

\begin{figure}[htp]
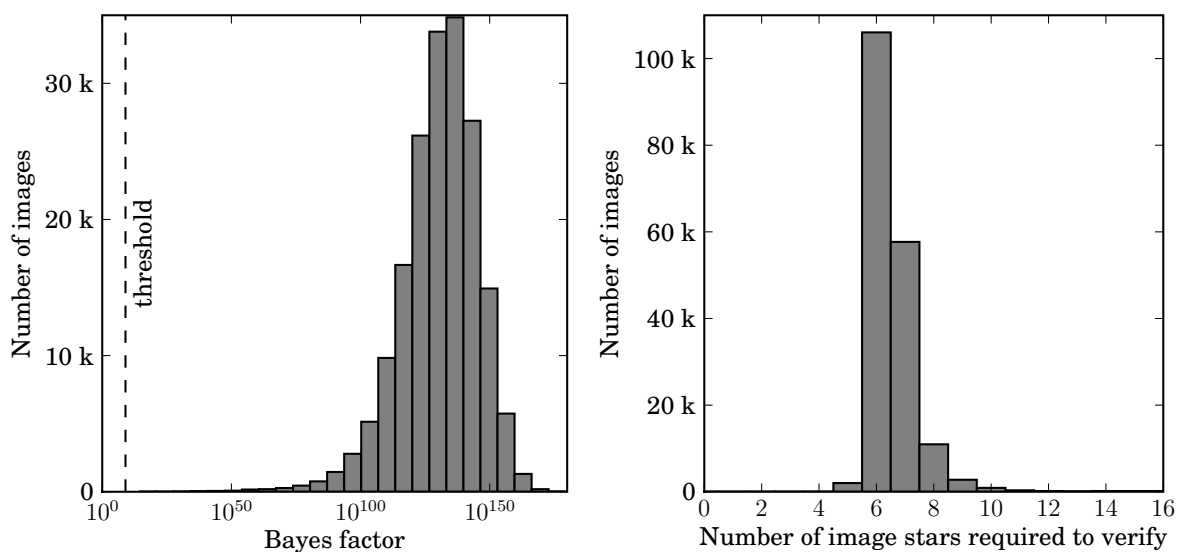

\begin{center}
\begin{tabular}{@{}c@{}c@{}}
\sdsserbayesfig & \sdsserntoverifyfig
\end{tabular}
\end{center}
\caption{Results from the excellent \rband-band SDSS images, continued.
	 \captionpart{Left:} The Bayes factor of the foreground model
    versus the background model for the hypotheses that we accept.
    The dashed line shows the threshold implied by our desired
    operating characteristics.  These excellent-quality images yield
    incredibly high Bayes factors---when we find a correct match is it
    unequivocal.  \captionpart{Right:} The number of stars in the
    query image that had to be examined before the Bayes-factor
    threshold was reached.  \label{fig:sdsser2}}
\end{figure}

\begin{figure}[htp]
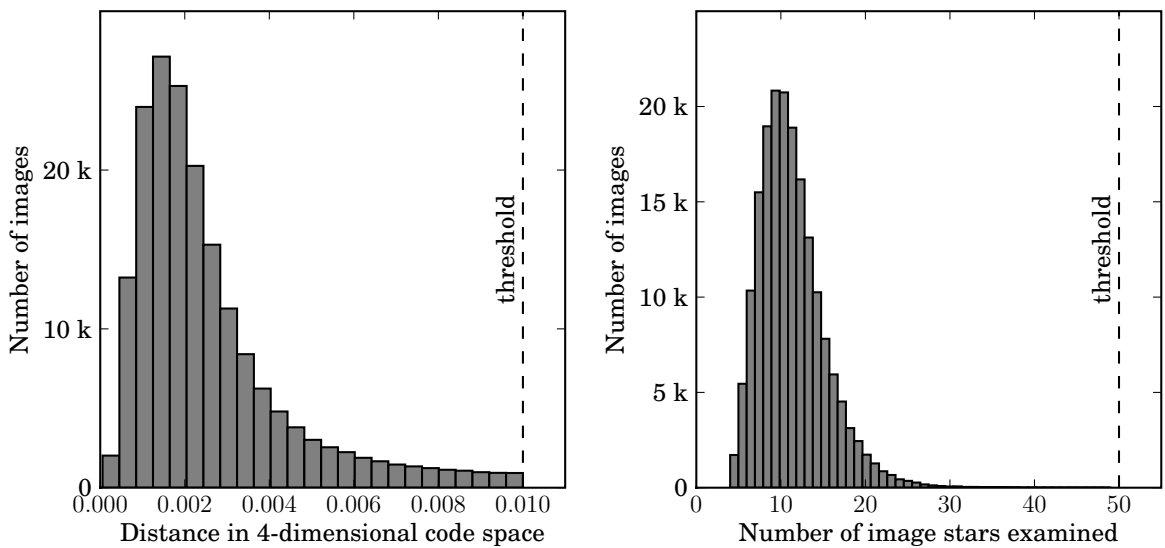

\begin{center}
\begin{tabular}{@{}c@{}c@{}}
    \sdssercodeerrfig & \sdssernimagefig \\
\end{tabular}
\end{center}
    \caption{Results from the excellent \rband-band SDSS images,
    continued.  \captionpart{Left:} The distance in four-dimensional
    geometric hash code space between the query quad and the first
    correctly-matched index quad.  In this experiment we searched for
    matches within distance $0.01$: well into the tail of the
    distribution.  \captionpart{Right:} The number of stars in the
    query image that the system built quads from before finding the
    first correct match.  In a few cases, the brightest $4$ stars
    formed a valid quad which was matched correctly to the index: we
    got the correct answer after our first guess!
\label{fig:sdsser3}}
\end{figure}

\begin{figure}[htp]
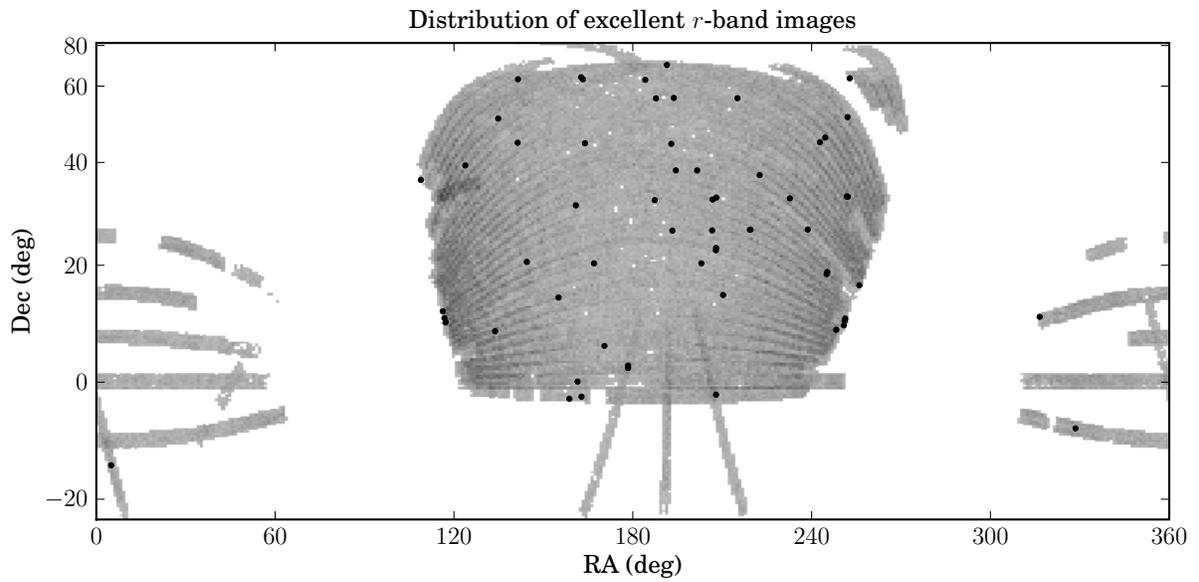

\begin{center}
\sdsserradecfig
\end{center}
    \caption{Distribution of the excellent-quality \rband-band SDSS
    images on the sky.  The gray footprint represents the
    correctly-recognized images, while the black dots show the images
    that were not recognized using the \usnob-based index.  There is a
    slight overdensity of failures near the beginnings and ends of
    runs, but otherwise no apparent spatial structure.
\label{fig:sdsser4}}
\end{figure}

% Could plot: number of quads tried wrt number of stars examined.

We randomized the order of the excellent-quality \rband-band images,
discarded all astrometric \metadata---leaving only the pixel positions
of the brightest $300$ stars---and asked our system to recognize each
one.  We allowed the system to create quads from only the brightest
$50$ stars.  All $300$ stars were used during the hypothesis-checking
step, but since the Bayes factors tend to be overwhelmingly large, we
would have found similar results if we had kept only the $50$
brightest stars.  We also told our system the angular size of the
images to within about $1~\percent$, though we emphasize that this was
merely a means of reducing the computational burden of this
experiment: we would have achieved exactly the same results (after
more compute time) had we provided no information about the scale
whatsoever; we show this in \secref{sec:sizehints} below.

\nonumberparagraphs
\begin{center}
\sdssertable
\end{center}
\numberparagraphs

\begin{figure}[htp]
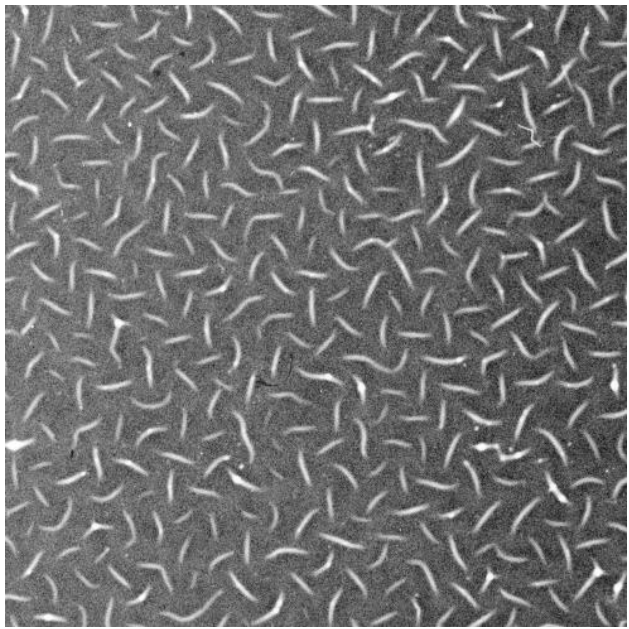

\begin{center}
% python get_usnob.py -n "Worms" --fits 243 36
% an-fitstopnm -i usnob-243-36-*_se0275.000.fits -r -v | pnminvert | pnmtojpeg > worms.jpg
\includegraphics[width=0.5\textwidth]{\filesuffix{\bigorsmallfig{worms}{worms-smaller}}{jpg}}
\end{center}
\caption{``Worms'' in \usnob.  We found these unusual artifacts
by looking at one of the places where our system failed to recognize
SDSS images.  The image is of the photographic plate POSS-IE 275,
centered on $(\RA,\Dec) = (243, 36)~\degrees$ and $15\times15~\arcmin$
in size.  Image credit: Copyright Palomar Observatory, National
Geographic Society, and California Institute of Technology; courtesy
of USNO Image and Catalogue Archive.
\label{fig:worms}}
\end{figure}

The results, shown in the table above and in \figs \ref{fig:sdsser1},
\ref{fig:sdsser2}, \ref{fig:sdsser3}, and \ref{fig:sdsser4}, are
that we can successfully recognize over $99.97~\percent$ of the
images.  We then examined the \usnob reference catalog at the true
locations of the images that were unrecognized.  Some of these
locations contained unusual artifacts.  For example,
\figref{fig:worms} shows a region where the 
\usnob catalog contains ``worm'' features.  The cause of these
artifacts is unknown (David Monet, personal communication), but they
affect several square degrees of one of the photographic plates that
were scanned to create the \usnob catalog.

In order to determine the extent to which our failure to recognize
images is due to problems with the \usnob reference catalog, we built
an index from the Two-Micron All-Sky Survey (2MASS) catalog, using the
same process as for the \usnob index, using the 2MASS
\band{J}-band rather than the \usnob red bands.  We then asked our
system to recognize each of the SDSS images that were unrecognized
using the \usnob-based index.  Of the $61$ images, $51$ were
recognized correctly, leaving only $10$ images unrecognized.
Examining these images, we found that some contained bright, saturated
stars which had been flagged as unreliable by the SDSS image-reduction
pipeline.  We retrieved the original data frames and asked our system
to recognize them.  All $10$ were recognized correctly: our source
extraction procedure was able to localize the bright stars correctly,
and with these the indexing system found a correct hypothesis.  With
these three processing steps, we achieve an overall performance of
$100~\percent$ correct recognition of all $182,221$ excellent images,
with no false positives.  This took a total of about $80$ minutes of
CPU time.  The index is about $5$ gigabytes in size, and once it is
loaded into memory, multiple CPU cores can use it in parallel, so the
wall-clock time can be a fraction of the total CPU time.  During this
experiment, a total of over $180$ million quads were tried, resulting
in about $77$ million matches to quads in the index.  Many of these
matches were found to result in image scales that were outside the
range we provided to the system, so the verification procedure was run
only $6$ million times.

% 702-slim: 4.7 GB.

% sdss-20

For completeness, we also checked the images that were rated as
excellent but failed our selection cut.  We retrieved the original
images and used our source extraction routine and both the \usnob- and
2MASS-based indexes.  Our system was able to recognize correctly all
$1138$ such images.

\subsubsection{Performance on images of varying bandpass}

% sdss-2

In order to investigate the performance of our system when the
bandpass of the query image is different than that of the index, we
asked the system to recognize images taken through the SDSS
filters \uband, \gband, \rband, \iband, and \zband.  We used only the
images rated ``excellent''.

\begin{figure}[htp]
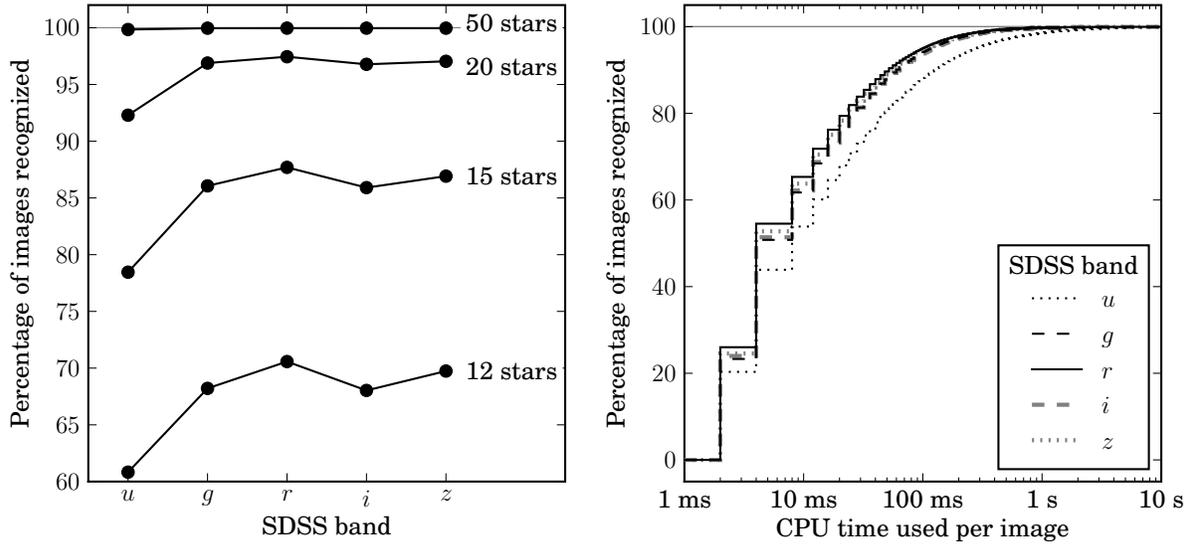

\begin{center}
\begin{tabular}{@{}c@{}c@{}}
    \sdssbandsobjsfig & \sdssbandstimefig
\end{tabular}
\end{center}
\caption{Performance on images taken through different SDSS bandpass filters.
\captionpart{Left:} The percentage of images recognized after building quads
from a given number of the brightest stars in each image.  The
\rband-band is the best match to the \usnob-based index we are using.  Generally
the recognition rate drops with the distance between the bandpass of
the index and the bandpass of the image.  The \iband-band performance
in this instance is lower than expected.
\captionpart{Right:} The percentage of images recognized as the amount of
CPU time spent per image increases.  The \rband-band images are most
quickly recognized, \gband-, \iband-, and \zband-band images take
slightly more effort, and \uband-band images take considerably more
CPU time.  The asymptotic recognition rates are nearly identical
except for \uband-band, which is slightly lower.
\label{fig:sdssbands}}
\end{figure}

\nonumberparagraphs
\begin{center}
\sdssbandtable
\end{center}
\numberparagraphs

The results, shown in the table above and in \figref{fig:sdssbands},
demonstrate that as the difference between the query image bandpass
and the index bandpass increases, the amount of CPU time required to
recognize the same fraction of images increases.  This performance
drop is more pronounced on the blue (\uband) side than the red
(\zband) side.  After looking at the brightest $50$ stars, the system
is able to recognize essentially the same fraction of images.  As
shown by the first experiment in this section, this asymptotic
recognition rate is largely due to defects in the reference catalog
from which the index is built.

\subsubsection{Performance on images of varying quality}

% sdss-3

In order to characterize the performance of the system as image
quality degrades, we asked the system to recognize \rband-band images
that were classified as ``excellent'', ``good'', ``acceptable'', or
``bad'' by the SDSS image-reduction pipeline.

\nonumberparagraphs
\begin{center}
\sdssqualtable
\end{center}
\numberparagraphs

\begin{figure}[htp]
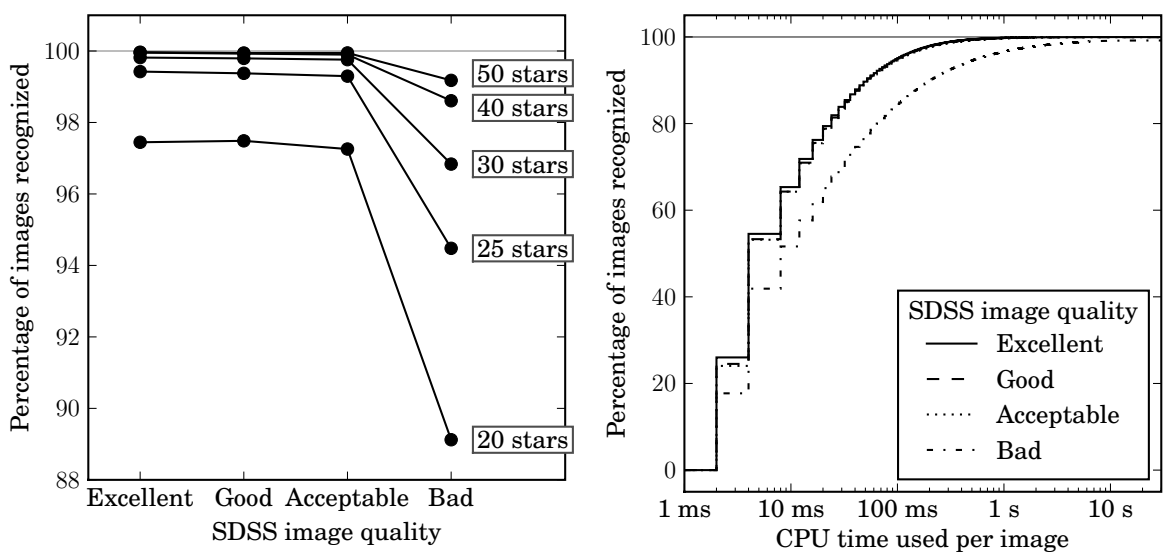

\begin{center}
\begin{tabular}{@{}c@{}c@{}}
    \sdssqualobjsfig & \sdssqualtimefig
\end{tabular}
\end{center}
\caption{Performance of the system given images of varying quality.
\captionpart{Left:} The percentage of images recognized after looking at a
given number of stars in each image, for excellent-, good-,
acceptable-, and bad-quality images from SDSS.  There is a small drop
in performance for good and acceptable images, and a more significant
drop for bad ones; all but the bad reach approximately the same
asymptotic recognition rate.  \captionpart{Right:} CPU time per image
for each quality rating.  All but the bad images show nearly
indistinguishable performance.
\label{fig:sdssqual}}
\end{figure}

The results, shown in the table above and in \figref{fig:sdssqual},
show almost no difference in performance between excellent, good, and
acceptable images.  Bad images show a significant drop in performance,
though we are still able to recognize over $99~\percent$ of them.

\subsubsection{Performance on images of varying angular size}

%%% FIXME - show image sizes with quad sizes superimposed.

% 13x9: sdss-1
% 9x9: sdss-4
% 8x8: sdss-5
% 7x7: sdss-8
% 6x6: sdss-9

\newcommand{\thirteenbynine}{\ensuremath{13\times9}}
\newcommand{\ninebynine}{\ensuremath{9\times9}}
\newcommand{\eightbyeight}{\ensuremath{8\times8}}
\newcommand{\sevenbyseven}{\ensuremath{7\times7}}
\newcommand{\sixbysix}{\ensuremath{6\times6}}
\newcommand{\arcminsquare}{\ensuremath{\textrm{arcmin}^2}}

We investigated the performance of our system with respect to the
angular size of the images by cropping out all but the central region
of the excellent-quality \rband-band images and running our system on
the sub-images.  Recall that the original image size is
$\thirteenbynine~\arcminsquare$, and that the index we are using
contains quads with diameters between $4$ and $5.6~\arcminsquare$.  We
cut the SDSS images down to sizes $\ninebynine$, $\eightbyeight$,
$\sevenbyseven$, and $\sixbysix~\arcminsquare$.  The $\sevenbyseven$
and $\sixbysix$ images required much more CPU time, so we ran only
small random subsamples of the images of these sizes.

\begin{figure}[htp]
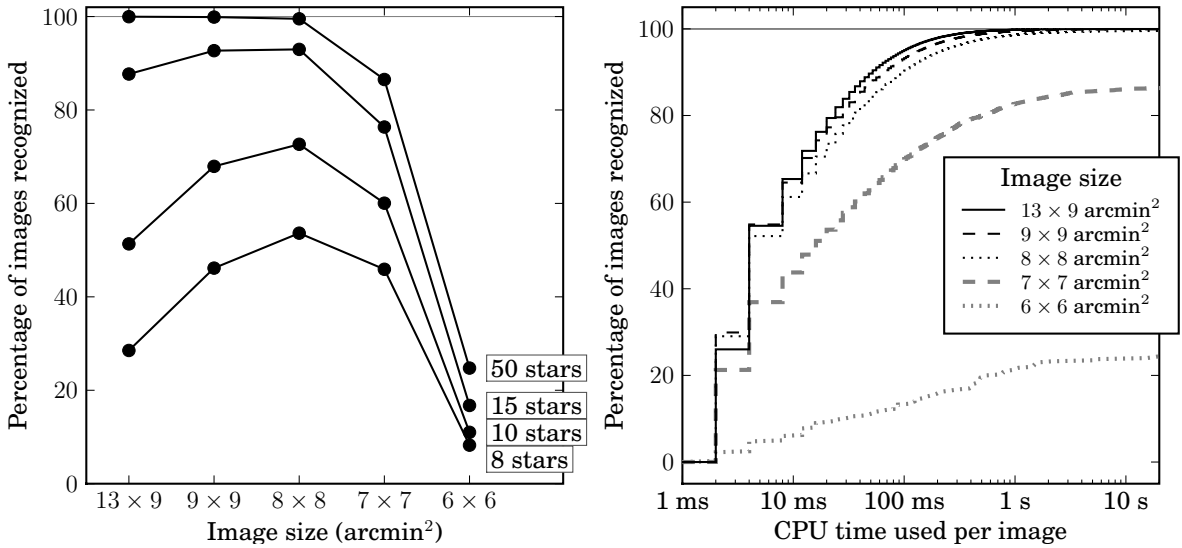

\begin{center}
\begin{tabular}{@{}c@{}c@{}}
    \sdssimsizeobjsfig & \sdssimsizetimefig
\end{tabular}
\end{center}
\caption{Performance of the system given images of varying angular sizes.
\captionpart{Left:} The percentage of images recognized after looking at a
given number of stars in each image, for images of the given sizes.
Perhaps surprisingly, more of the $\eightbyeight$ images are
recognized correctly during the first few milliseconds, but as more
time elapses, the larger-scale images are more likely to be
recognized.
\captionpart{Right:}  CPU time per image required to recognize images of
each angular size.  After $10~\milliseconds$, the larger images are
recognized more quickly.  The $\sevenbyseven$ and $\sixbysix$ images
appear to be reaching asymptotic recognition rates far below
$100~\percent$.
\label{fig:sdssimsize}}
\end{figure}

\nonumberparagraphs
\begin{center}
\sdssimsizetable
\end{center}
\numberparagraphs

The results, presented in the table above and in
\figref{fig:sdssimsize}, show that performance degrades slowly for
images down to $\eightbyeight~\arcminsquare$, and then degrades
sharply.  This is not surprising given the size of quads in the index
used in this experiment: in the smaller images, only stars near the
edges of the image can possibly be $4$ to $5.6~\arcmin$ away from
another star, so the set of stars that can form the `backbone' of a
quad is small.  Observe that images below $2.8\times2.8~\arcminsquare$
cannot possibly be recognized by this index, since no pair of stars
can be $4~\arcmin$ or more away from each other.

This does not imply that small images cannot be recognized by our
system: given an index containing smaller quads, we may still be able
to recognize them.  The point is simply that for any given index there
is some threshold of angular size below which the image recognition
rate will drop, and another threshold below which the recognition rate
will be exactly zero.

\subsubsection{Performance with varying image scale hints}
\label{sec:sizehints}

% sdss-15

In all the experiments above, we told our system the angular scale (in
arcseconds per pixel) to within $\pm1.25~\percent$ of the true value,
and we used an index containing only quads within a small range of
diameters.  In this experiment, we show that these hints merely make
the recognition process faster without affecting the general results.

We created a set of sub-indices, each covering a range of $\sqrt{2}$
in quad diameters.  The smallest-scale sub-index contains quads of $2$
to $2.8~\arcmin$, and the largest contains quads of about $20$ to $30$
degrees in diameter.  Each sub-index is built using the same
methodology as outlined above for the $4$ to $5.6~\arcmin$ index, with
the scale adjusted appropriately.  The smallest-scale sub-index
contains only $6$ stars per \healpix grid cell rather than $10$ as in
the other sub-indices, because the \usnob catalog does not contain
enough stars: a large fraction of the smallest cells contain fewer
than $10$ stars.  In the smallest-scale sub-index we then do $9$
rounds of quad-building, reusing each star at most $5$ times, as
opposed to $16$ rounds reusing each star at most $8$ times as in the
rest of the sub-indices.

\begin{figure}[htp]
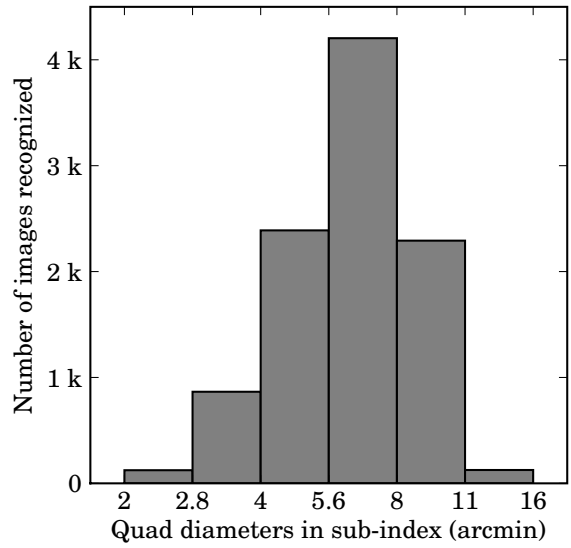

\begin{center}
%\begin{tabular}{@{}c@{}c@{}}
    \sdsssizehintsindexfig
%\end{tabular}
\end{center}
\caption{The sub-index that is first to recognize SDSS images, when all
the sub-indices are run in lock-step.  Although we used sub-indices
containing quads up to $30$ degrees in diameter, clearly only the ones
that contain quads that can possibly be found in an SDSS image (which
are about $16~\arcmin$ across the diagonal) can generate correct
hypotheses that will recognize the image.  Perhaps surprisingly, each
of these sub-indices is first to recognize some subset of the images,
though a strong tuning effect is clear.
\label{fig:sdsssizehintsindex}}
\end{figure}

Each time our system examines a quad in the image, it searches each
sub-index in turn for matching quads, and evaluates each hypothesized
alignment generated by this process.  The system proceeds in
lock-step, testing each quad in the image against each sub-index in
turn.  The first quad match that generates an acceptably good
alignment is taken as the result and the process stops.  Note that
several of the sub-indices may be able to recognize any given image.
Indeed, \figref{fig:sdsssizehintsindex} shows that every sub-index
that contains quads that can possibly be found in SDSS images is first
to recognize some of the images in this experiment.  Different
strategies for ordering the computation---for example, spending equal
amounts of CPU time in each sub-index rather than proceeding in
lock-step---might result in better overall performance.

\begin{figure}[htp]
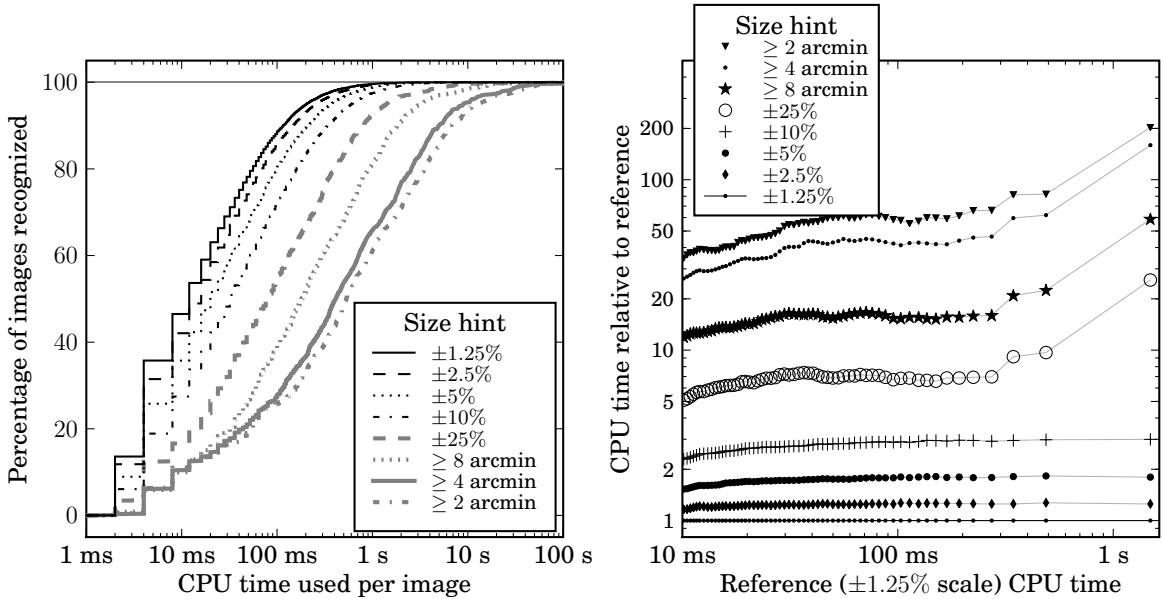

\begin{center}
\begin{tabular}{@{}c@{}c@{}}
    \sdsssizehintstimefig & \sdsssizehintsreltimefig
\end{tabular}
\end{center}
\caption{Performance of the system given varying limits on the image size.
\captionpart{Left:} CPU time per image required to recognize images, given
various limits on the angular size of the image.  Our system achieves
the same asymptotic recognition rate in each case: giving the system
less information about the true scale of the image simply means that
it must evaluate and reject more false hypotheses before finding a
true one.  The ``$\ge8~\arcmin$'' hint indicates that we told the
system that the image width is between $8~\arcmin$ and $180~\degrees$.
The upper limit has much less impact on performance than the lower
limit, since the sub-indices that cover large angular scales contain
many fewer quads and are therefore much faster to search, and generate
fewer coincidental matches.  \captionpart{Right:} CPU time per image
relative to the $\pm1.25~\percent$ case.  We divided the CPU times for
each case into percentiles; the mean time within each percentile is
plotted.  Generally, giving the system less information about the size
of the images results in an approximately constant-factor increase in
the CPU time required.  Although there appears to be a sharp upward
trend for the ``loosest'' four size ranges, this may be an effect of
small sample size: since these cases take so long to run, we tested
only $1000$ images, while for the rest of the cases we tested $10,000$
images.  The CPU time distribution is heavy-tailed, so the expected
variance is large.
\label{fig:sdsssizehints}}
\end{figure}

\subsubsection{Performance with varying index quad density}

% sdss-1  (index 702)
% sdss-10 (index 1102, half as dense)
% sdss-13 (index 1202, quarter as dense)

The fiducial index we have been using in these experiments contains
about $16$ quads per \healpix grid cell.  Since each SDSS image has
an area of about $7$ cells, we expect each image to contain about
$100$ quad centers.  The number of complete quads (\ie, quads for
which all four stars are contained in the image) in the image will be
smaller, but we still expect each image to contain many quads.  This
gives us many chances of finding a correct match to a quad in the
image.  This redundancy comes at a cost: the total number of quads in
the index determines the rate of false matches---since the code-space
volume is fixed, packing more quads into the space results in a larger
number of matches to any given query point---which directly affects
the speed of each query.  By building indices with fewer quads, we can
reduce the redundancy but increase the speed of each query.  This does
not necessarily increase the overall speed, however: an index
containing fewer quads may require more quads from the image to be
checked before a correct match is found.  In this experiment, we vary
the quad density and measure the overall performance.

\nonumberparagraphs
\begin{center}
\sdssdensitytable
\end{center}
\numberparagraphs

\begin{figure}[htp]
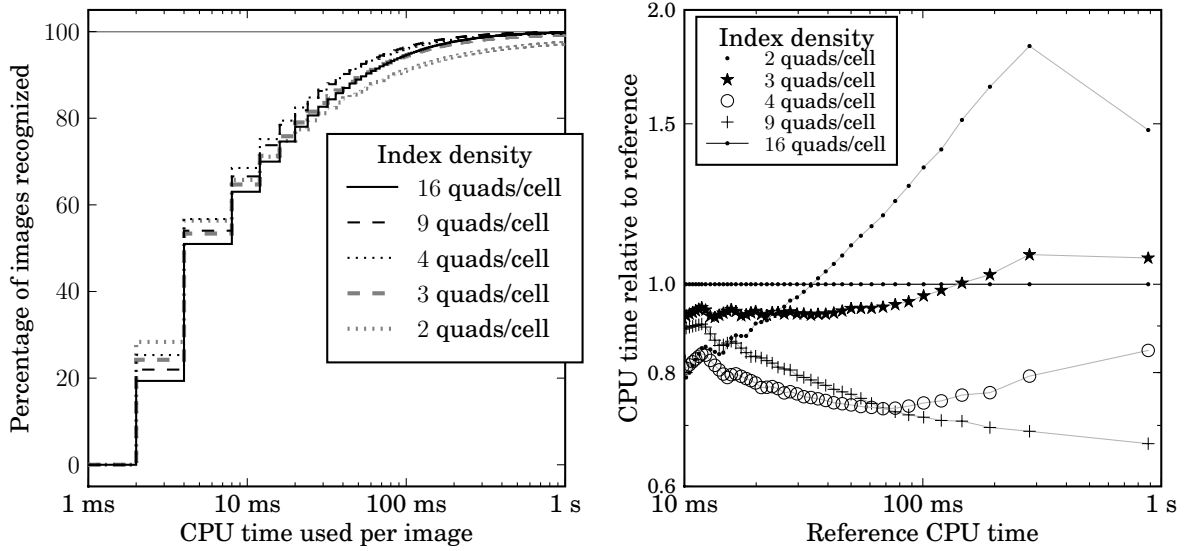

\begin{center}
\begin{tabular}{@{}c@{}c@{}}
    \sdssdensitytimefig & \sdssdensityreltimefig
\end{tabular}
\end{center}
\caption{Performance of the system using indices containing varying
densities of quads.  (The total number of quads in an index is
proportional to the density of quads.)
\captionpart{Left:} CPU time per image required to recognize images.
\captionpart{Right:} Relative CPU time to recognize images.  We split
the set of images into percentiles and have plotted the mean time
within each percentile, relative to the 16-quad-per-cell reference
index.  The indices containing fewer quads are faster to search per
quad query, but may require more quads to be tried before a correct
match is found.  The smaller indices are also able to recognize fewer
images, because some images will simply not contain a quad that
appears in the index.  For the high-quality SDSS images we are using,
the smallest of the indices here results in a $2~\percent$ drop in
recognition rate (from nearly $100~\percent$ to about $98~\percent$),
but for poorer-quality images the drop could be larger.
\label{fig:sdssdensity}}
\end{figure}

As shown in the table above and \figref{fig:sdssdensity}, reducing the
density from $16$ to $9$ quads per \healpix grid cell has almost no
effect on the recognition rate but takes only two-thirds as much CPU
time.  Reducing the density further begins to have a significant
effect on the recognition rate, and actually takes \emph{more} CPU
time overall.

\subsubsection{Performance on indices built from triangles and quints}
\label{sec:triquint}

% sdss-32, 33, 34

% nquads
% ntries-brute
% time

We tested the performance of our quad-based index against a
triangle-based index and a quintuple-based (``quint'') index.  The
index-building processes were exactly as in our quad-based indices.

In the experiments above we searched for all matches within a distance
of $0.01$ in the quad feature space.  Since the triangle- and
quint-based indices have feature spaces of different dimensionalities
($2$ for triangles, $6$ for quints), we first ran our system with the
matching tolerance set to $0.02$ in order to measure the distribution
of distances of correct matches in the three feature spaces.  We then
set the matching tolerance to include $95~\percent$ of each
distribution.  For the triangle-based index, we found this matching
tolerance to be $0.0064$, for the quad-based index it was $0.0095$,
and for the quint-based index, $0.011$.  These experiments were
performed on a random subset of $4000$ images, because they are quite
time-consuming.

\nonumberparagraphs
\begin{center}
\sdsstriquinttable
\end{center}
\numberparagraphs

\begin{figure}[htp]
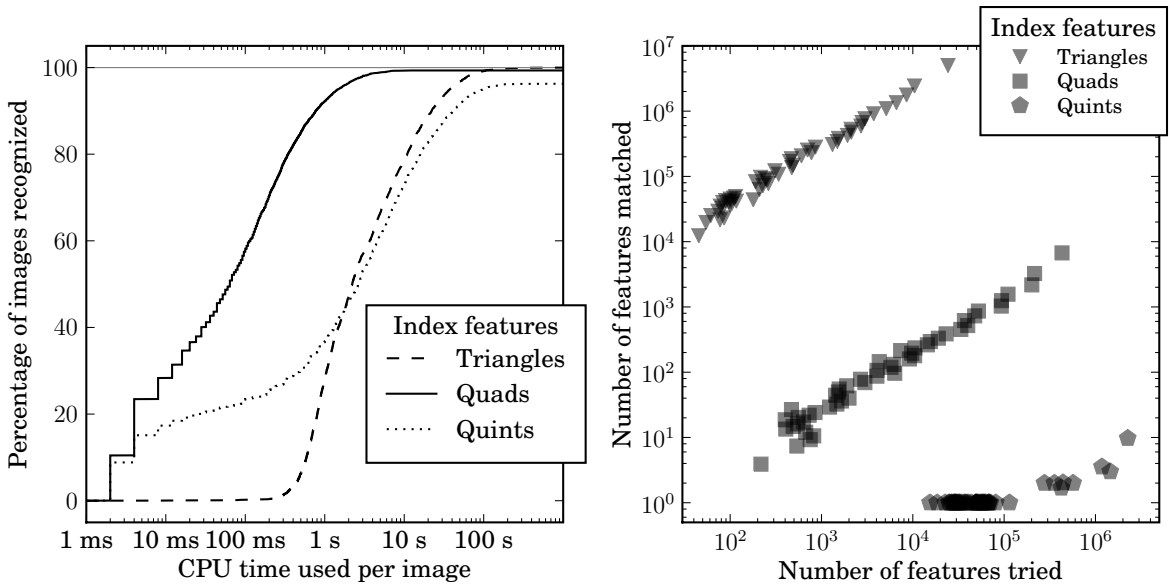

\begin{center}
\begin{tabular}{@{}c@{}c@{}}
\sdsstriquinttimefig & \sdsstriquintntrynmatchfig
\end{tabular}
\end{center}
\caption{Performance of the system using indices containing triangles,
quads, and quintuples of stars.  \captionpart{Left:} CPU time per
image required to recognize images.  The quad-based index vastly
outperforms the triangle- and quint-based indices.
\captionpart{Right:} The number of features tried (\ie, the number
of triangles, quads, or quints of stars from the image that were
tested), and the number of matches to features in the index that
resulted.  Each plotted point summarizes $2~\percent$ of the
correctly-recognized images, sorted by the number of features tried.
As expected, the triangle-based index produces many more matches for
any given query, because the same number of features are packed into a
lower-dimensional feature space.  Fewer features have to be tried
before the first correct matches are found, because only three
corresponding stars have to be found.  Quints, on the other hand, are
very distinctive: for over $80~\percent$ of the images that were
correctly recognized, the first matching quint ever found was a
correct match.  However, many quints from the image have to be tested
before this match is found.
\label{fig:sdsstriquint}}
\end{figure}

The results are given in the table above and in
\figref{fig:sdsstriquint}.  After looking at the brightest
$50$ stars in each image, the triangle-based index is able to
recognize the largest number of images, but both the triangle- and
quint-based indices take significantly more time than the quad-based
index.  It seems that quad features strike the right balance between
being distinctive enough that any given query does not generate too
many coincidental (false) matches---as the triangle-based index
does---but containing few enough stars that it does not take long to
find a feature that is in both the image and the index---as the
quint-based index does.

We expect that the relative performance of triangle-, quad-, and
quint-based indices depends strongly on the angular size of the images
to be recognized.  An index designed to recognize images of large angular
size requires fewer features, so the code space is less densely filled
and fewer false matches are generated.  For this reason, we expect
that above some angular size, a triangle-based index will recognize
images more quickly than a quad-based index.

%Perhaps surprisingly, the small decrease in the matching
%distance in feature space from $0.01$ to $0.0095$ for the quad-based
%index resulted in a $>0.6~\percent$ drop in recognition rate.

\subsection{Blind astrometric calibration of Galaxy Evolution Explorer data}

% GAL-4

\begin{figure}[htp]
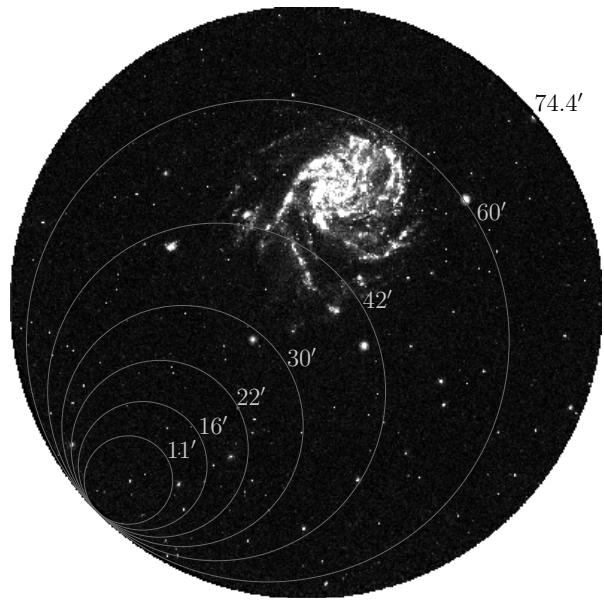

\begin{center}
\galexquadfig
\end{center}
\caption{A sample GALEX field (AIS-101-sg32-nuv),
with circles showing the sizes of the quads in the indices we use in
this experiment.  The images produced by the GALEX pipeline are
$3840\times3840$ pixels, and the near-UV field of view is a circle of
diameter $1.24~\degrees$, or about $74.4~\arcmin$.  Image credit:
Courtesy NASA/JPL-Caltech.
\label{fig:galexquad}}

\end{figure}

To show the performance of our system on significantly larger images,
at a bandpass quite far from that of our index, we experimented with
data from the Galaxy Evolution Explorer (GALEX).  GALEX is a space
telescope that observes the sky through near- and far-ultraviolet
bandpass filters.  Although it is fundamentally a photon-counting
device, the GALEX processing pipeline renders images (by collapsing
the time dimension and histogramming the photon positions), and these
are the data products used by most researchers.  The images are
circular, with a diameter of about $1.2$ degrees.  See
\figref{fig:galexquad}.  In this experiment, rather than using the
images themselves, we use the catalogs (lists of sources found in each
image) that are released along with the images.  These catalogs are
produced by running a standard source extraction program (SExtractor)
on the images.  We retrieved the near-UV catalogs for all $28,182$
images in Galex Release 4/5 that have near-UV exposure.

% indices 705-709
% (for x in index-70[56789].fits; do modhead $x NSTARS; done) | gawk 'BEGIN{N=0} {N+=$3} END{print N}'
% 29,358,529
% (for x in index-70[56789].fits; do modhead $x NQUADS; done) | gawk 'BEGIN{N=0} {N+=$3} END{print N}'
% 36,146,408

We built a series of indices of various scales, each spanning roughly
$\sqrt{2}$ in scale.  The smallest contained quads with diameters from
$11$ to $16~\arcmin$, the next contained quads between $16$ and
$22~\arcmin$, followed by $22$ to $30$, $30$ to $42$, and $42$ to
$60~\arcmin$.  Each index was built according to the ``standard
recipe'' given above, using stars from the red bands of \usnob as
before.  In total these indices contain about $30$ million stars and
$36$ million quads.

\nonumberparagraphs
\begin{center}
\galextable
\end{center}
\numberparagraphs

\begin{figure}[htp]
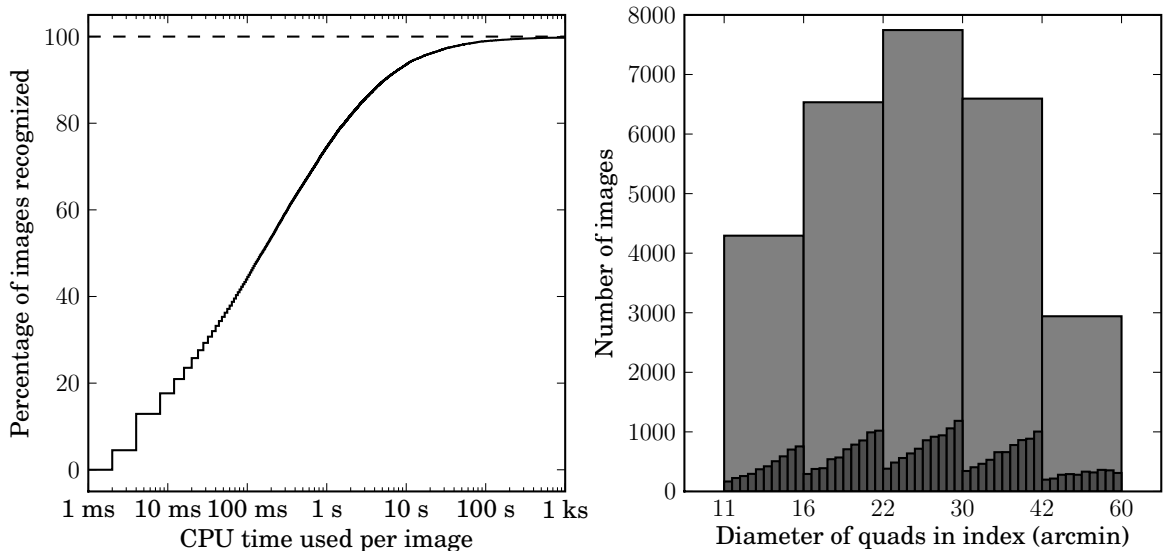

\begin{center}
\begin{tabular}{@{}c@{}c@{}}
    \galexcputimefig & \galexindexidfig
\end{tabular}
\end{center}
\caption{Results on GALEX near-UV images.  \captionpart{Left:} CPU
time used per image.  The shape of the curve is very similar to that
in the previous SDSS experiments, though the ``knee'' of diminishing
returns occurs after more CPU time (possibly because we gave the
system much less information about the correct scale of the images).
\captionpart{Right:} The number of images recognized by each of the
indices (identified by the range of sizes of quads they contain).  For
each quad in the image, we search for matches in each of the indices,
stopping after the first match that is confirmed by the verification
test.  The histogram therefore shows only which index recognized the
image \emph{first}, rather than which indices might have recognized it
given more time.  The inset histograms show the distribution of quad
sizes within each index (on a linear scale).  Generally the larger
quads are more successful, except in the largest index, where the size
of the largest quads approaches that of the whole image.
\label{fig:galex}}
\end{figure}

In this experiment, we told our system that the images were between
$1$ and $2$ degrees wide, and we allowed it to build quads from the
first $100$ sources in each image.  The results are shown in the table
above and \figref{fig:galex}.  The recognition rate is quite similar
to that of the excellent-quality SDSS \rband-band fields, suggesting
that even though the near-UV bandpass of these images is quite far
from the bandpass of the index, which we would expect to make the
system less successful at recognizing these images, their larger
angular size seems to compensate.  The system is significantly slower
at recognizing GALEX images, but this is partly because we gave it a
fairly wide range of angular scales, and because we used several
indices rather than the single one whose scale is best tuned to these
images.

% Plot GALEX NUV along with SDSS ugriz.

\subsection{Blind astrometric calibration of Hubble Space Telescope data}

In order to demonstrate that there is nothing intrinsic in our method
that limits us to images of a particular scale, we retrieved a set of
Hubble Space Telescope (HST) images and built a custom index to
recognize them.  We chose the All-wavelength Extended Groth strip
International Survey (AEGIS; \citealt{aegis}) footprint because it has
many HST exposures and is within the SDSS footprint.

\begin{figure}[htp]
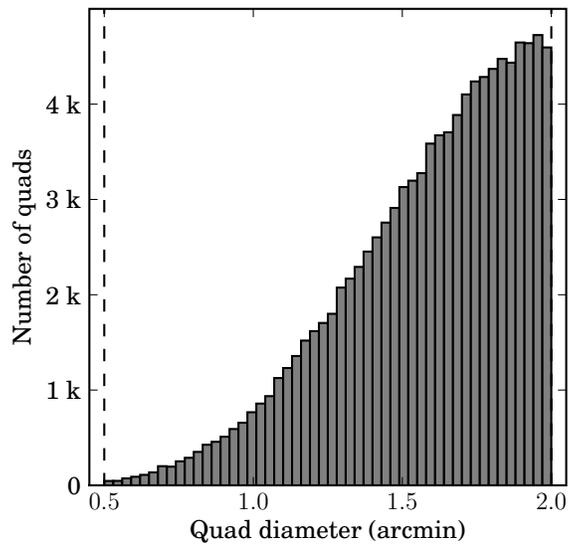

\begin{center}
\aegisacsquadsizesfig
\end{center}
\caption{The diameters of quads in our custom index, built from SDSS
stars, for recognizing Hubble Space Telescope Advanced Camera for
Surveys (ACS) images in the AEGIS footprint.  Although we allowed
quads with diameters between $0.5$ and $2~\arcmin$, the size
distribution of the quads that were created is heavily biased toward
the large end of the range, because the density of SDSS stars---about
$4$ per square arcminute---is not high enough to build many tiny
quads.
\label{fig:aegisquadsizes}}
\end{figure}

To build the custom index for this experiment, we retrieved stars and
galaxies from SDSS within a $2\times2~\deg$ square centered on
$\RA=215~\deg$, $\Dec=52.7~\deg$ with measured \rband-band brightness
between $15$ and $22.2~\mag$, yielding about $57,000$ sources.  We
created the index as usual, using a \healpix grid with cells of size
$0.5~\arcmin$, and building quads with diameters between $0.5$ and
$2~\arcmin$.  This yielded just over $100,000$ quads.  Since the
density of stars and galaxies in our index is only about $4$ sources
per square arcminute, the system tended to produce quads with sizes
strongly skewed toward the larger end of the allowed range of scales.
See \figref{fig:aegisquadsizes}.

\begin{figure}[htp]
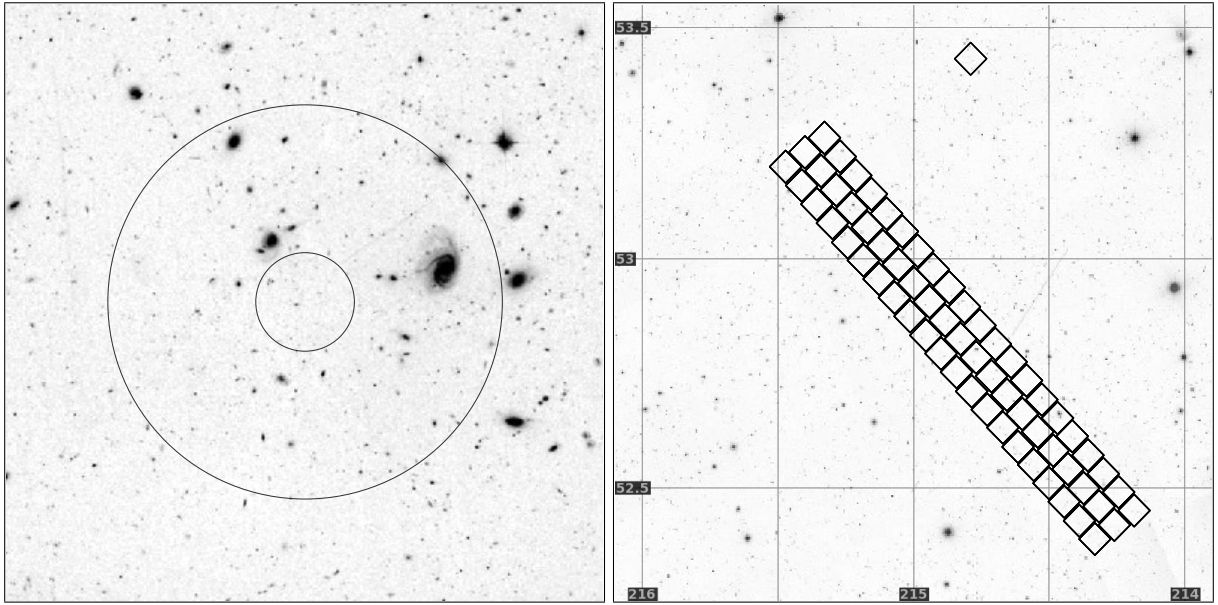

\begin{center}
\setlength{\fboxsep}{0.5pt}
\begin{tabular}{@{}c@{\hspace{3pt}}c@{}}
\framebox{\aegisacsquadfig} &
% The code that generates this figure is documented in
% sdss-tests/README-AEGIS
\bigorsmallfig{%
    \includegraphics[width=\figunit]{\filesuffix{aegis-pdf}{pdf}}%
    }{%
    \framebox{\includegraphics[width=\figunit]{\filesuffix{aegis-small}{pdf}}}%
    }
\end{tabular}
\end{center}
\caption{\captionpart{Left:} A typical cutout of a Hubble Space Telescope
Advanced Camera for Surveys (ACS) image as used in our experiment.
The overlaid circles show the range of diameters of the quads in the
index we use to recognize these images.  ACS images are about
$3.4~\arcmin$ square and $4096\times4096$ pixels, but the cutouts we
use in our experiment are about $3~\arcmin$ and have been downsampled
to $600\times600$ pixels.  The quad diameters are from $0.5$ to
$2~\arcmin$.  Image credit: Courtesy NASA/JPL-Caltech.
\captionpart{Right:} A $\sim2$ square degree region
(part of the $\sim4$ square degree region of SDSS from which we built
our index), overlaid with the footprints of the $191$ ACS images that
were recognized by our system.  There are only $64$ unique footprint
regions because some of the $191$ images are observations of the same
region through different bandpass filters.  The grid lines show $\RA$
and $\Dec$.
\label{fig:aegisacs}}
\end{figure}

We queried the Hubble Legacy Archive \citep{hla} for images taken by
the Hubble Advanced Camera for Surveys (ACS; \citealt{acs}) within the
AEGIS footprint.  Since individual ACS exposures contain many cosmic
rays, we requested only ``level 2'' images, which are created from
multiple exposures and have cosmic rays removed.  A total of $191$
such images were found.  We retrieved $600\times600$-pixel JPEG
previews---see \figref{fig:aegisacs} for an example---and gave these
images to our system as inputs.

Our system successfully recognized $100~\percent$ of the $191$ input
images, taking an average of $0.3$ seconds of CPU time per image (not
including the time required to perform source extraction).  The
footprints of the input images are shown in \figref{fig:aegisacs}.
Although there are $191$ images, there are only $64$ unique
footprints, because images were taken through several different
bandpass filters for many of the footprints.  Although the index
contained quads with diameters from $0.5$ to $2~\arcmin$, the smallest
quad that was used to recognize a field was about $0.9~\arcmin$ in
diameter.

\subsection{Blind astrometric calibration of other imagery}

%%% FIXME -- examples of backyard digicam, astrophoto, Harvard plate.

%%%--------------------------------------------------------------

% DASCH: 5379 scans.  Using Astrometry.net as of June 26, 2008.

\begin{figure}[htp]
\begin{center}
\setlength{\fboxsep}{0.5pt}
% Created with:
%   wget http://hea-www.harvard.edu/DASCH/gallery/ir12723.jpg
%   jpegtopnm ir12723.jpg | pnmcut 0 500 1000 700 | pnminvert | pnmnorm -wvalue 130 | pnmtojpeg > dasch-ir12723-bw.jpg
\framebox{\includegraphics[width=0.7\textwidth]{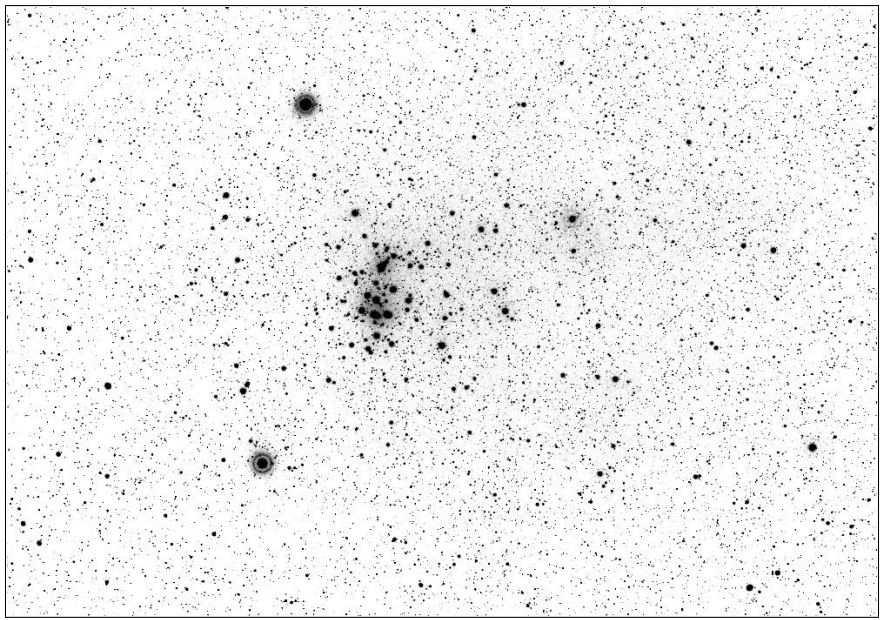}}
\end{center}
\caption{A sample DASCH scan of part of one of the photographic
glass plates of the Harvard Observatory archives.  The initial
astrometric calibration of these plates is being computed by \an.
Praesepe, also known as the Beehive cluster or Messier 44, appears in
this image.  Image credit: DASCH team; Harvard College
Observatory.\label{fig:dasch}}
\end{figure}

The Harvard Observatory archives contain over $500,000$ photographic
glass plates exposed between 1880 and 1985.  See \figref{fig:dasch}
for an example.  The Digital Access to a Sky-Century at Harvard (DASCH)
project is in the process of scanning these plates at high
resolution \citep{harvardplates,dasch}.  Since the original astrometric
calibration for these plates consists of hand-written entries in log
books, a blind astrometric calibration system was required to add
calibration information to the digitized images.  DASCH has been using
the \an system for the past year to create an initial astrometric
calibration which is then refined by WCSTools \citep{wcstools4}.

The DeepSky project \citep{deepsky} is reprocessing the data taken as
part of the Palomar-QUEST sky survey and Nearby Supernova
Factory \citep{palomarquest, nearbysnfactory}.  Since many of the images
have incorrect astrometric \metadata, they are using \an to do the
astrometric calibration.  Over $14$ million images have been
successfully processed thus far (Peter Nugent, personal
communication).

We have also had excellent success using the same system for blindly
calibrating a wide class of other astronomical images, including
amateur telescope shots, photographs from consumer digital SLR
cameras, some of which span tens of $\degrees$.  We have also
calibrated videos posted to YouTube.  These images have very different
exposure properties, capture light in the optical, infrared and
ultraviolet bands, and often have significant distortions away from
the pure tangent-plane projection of an ideal camera.  Part of the
remarkable robustness of our algorithm, which allows it to calibrate
all such images using the same parameter settings, comes from the fact
that the hash function is scale invariant so that even if the center
of an image and the edges have a significantly different pixel scale
(solid angle per pixel), quads in both locations will match properly
into the index (although our verification criterion may conservatively
decide that a true solution with substantial distortion is not
correct). Furthermore, no individual quad or star, either in the query
or the index is essential to success. If we miss some evidence in one
part of the image or the sky we have many more chances to find it
elsewhere.

\subsection{False positives}

\begin{figure}[htp]
\begin{center}
\setlength{\fboxsep}{0.5pt}
\setlength{\fboxrule}{0.25pt}
\framebox{\includegraphics[width=0.52\textwidth]{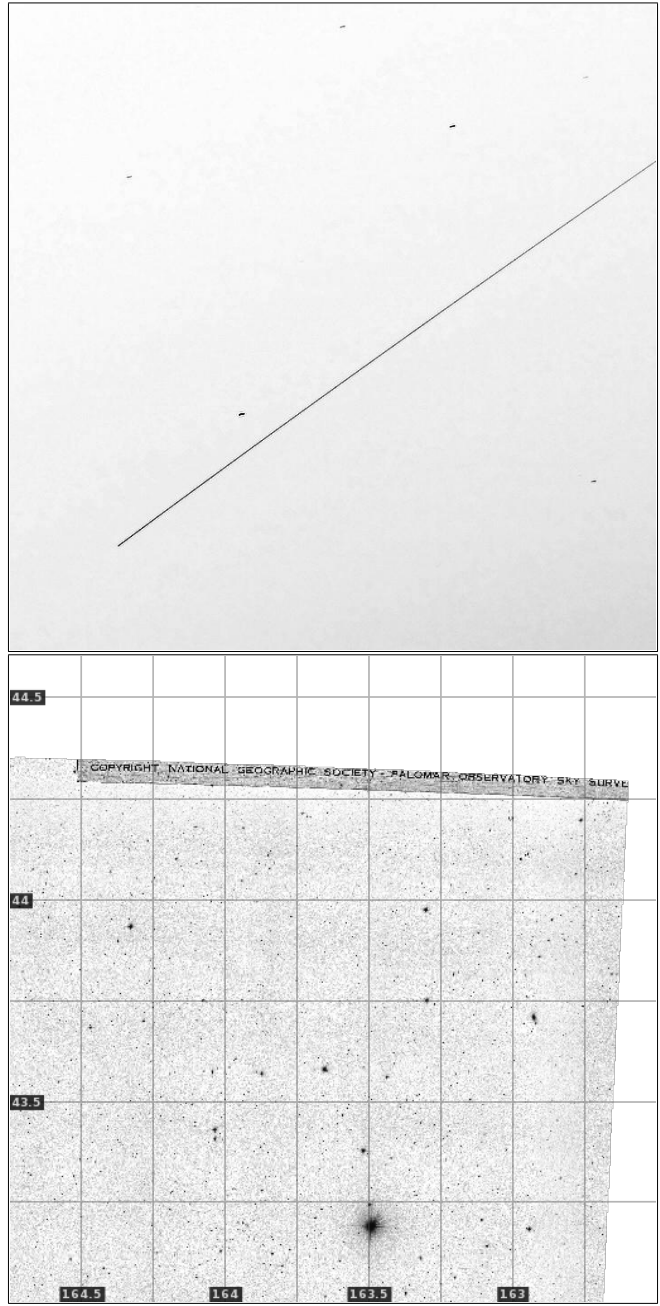}}%
\\
\framebox{\includegraphics[width=0.52\textwidth]{\filesuffix{\bigorsmallfig{usnob-iss}{usnob-iss-small}}{pdf}}}%
\end{center}
\caption{A false-positive match.  \captionpart{Top:} The input image
contains a linear feature: the International Space Station streaked
across the image.  Image credit: copyright Massimo Matassi.
\captionpart{Bottom:} The USNO-B scanned photographic plate has
writing on the corner.  Image credit: copyright Palomar Observatory,
National Geographic Society, and California Institute of Technology;
courtesy of USNO Image and Catalogue Archive.
\label{fig:falseposA}}
\end{figure}

\begin{figure}[htp]
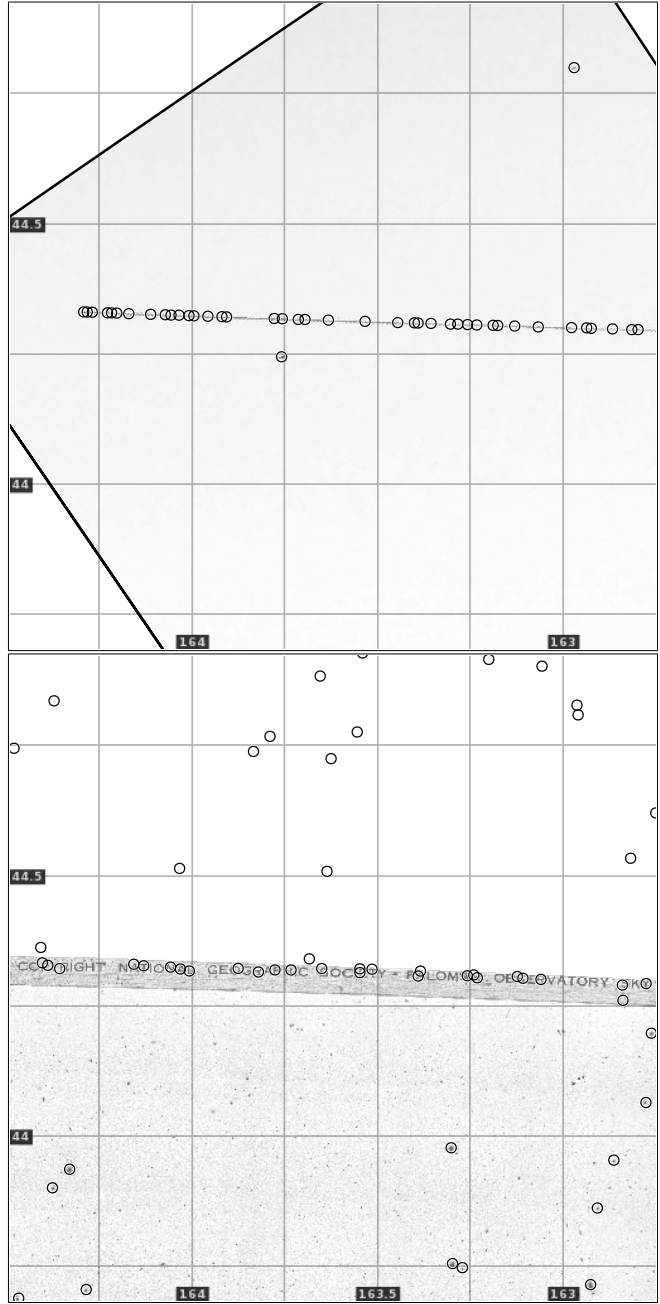

\begin{center}
\setlength{\fboxsep}{0.5pt}
\setlength{\fboxrule}{0.25pt}
\framebox{\includegraphics[width=0.52\textwidth]{\filesuffix{\bigorsmallfig{iss-img-xy}{iss-img-xy-small}}{pdf}}}%
\\
\framebox{\includegraphics[width=0.52\textwidth]{\filesuffix{\bigorsmallfig{iss-usnob-xy}{iss-usnob-xy-small}}{pdf}}}
\end{center}
\caption{A false-positive match (continued).  \captionpart{Top:} The
image, rotated to the alignment that our system found.  The circles
show the sources that were detected.  The linear feature becomes a
line of false sources.  \captionpart{Bottom:} The corresponding region
of the USNO-B plate.  The USNO-B source detection algorithm identifies
many false sources in the region of text.  The lines of sources in the
two images are aligned in this (false) match.
\label{fig:falseposB}}
\end{figure}

Although we set our operating thresholds to be very conservative in
order to avoid false positive matches, images that do not conform to
the assumptions in our model can yield false positive matches at much
higher rates than predicted by our analysis.  In particular, we have
found that images containing linear features that result in lines of
detected sources are often matched to linear flaws in the USNO-B
reference catalog.  We removed linear flaws resulting from diffraction
spikes \citep{barroncleaning}, but many other linear flaws remain.  An
example is shown in \figs \ref{fig:falseposA} and \ref{fig:falseposB}.

%%% These figs are generated by sdss-tests/mosaic/mosaic.sh

\section{Discussion}

We have described a system that performs astrometric
calibration---determination of imaging pointing, orientation, and
plate scale---blind, or without any prior information beyond the data
in the image pixels.  This system works by using indexed asterisms to
generate hypotheses, followed by quantitative verification of those
hypotheses.  The system makes it possible to vet or restore the
astrometric calibration information for astronomical images of unknown
provenance, and images for which the astrometric calibration is lost,
unknown, or untrustworthy.

There are several sources of astronomical imagery that could be very
useful to \linebreak[4] researchers---especially those studying the
time domain---if they were consistently and correctly calibrated.
These include photographic archives, amateur astronomers, and a
significant fraction of professional imagery which has incorrect or no
astrometric calibration \metadata.  The photographic archives extend
the time baseline up to a century, while amateur astronomers---some of
whom are highly-skilled and well-equipped---can provide a dense
sampling of the time domain and can dedicate a large amount of
observing time to individual targets.  Our system allows these sources
of data to be made available to researchers by creating trustworthy
astometric calibration \metadata.

The issue of trust is key to the success of efforts such as the
Virtual Observatory to publish large, heterogeneous collections of data
produced by many groups and individuals.  Without trusted \metadata,
data are useless for most purposes.  Our system allows existing
astrometric calibration \metadata to be verified, or new
trustworthy \metadata to be created, from the data.  Furthermore,
applying a principled and consistent calibration procedure to
heterogeneous collections of images enables the kinds of large-scale
statistical studies that are made possible by the Virtual Observatory.

Our experiments on Sloan Digital Sky Survey, Galaxy Evolution
Explorer, and Hubble Space Telescope images have demonstrated the
capabilities and limitations of our system.  The best performance, in
terms of the fraction of images that are recognized and the
computation effort required, is achieved when the index of known
asterisms is well-matched to the images to be recognized.  Differences
in the bandpasses of the index and images lead to a small drop in
performance across the near-infrared to near-ultraviolet range.  By
creating multiple indices across the spectrum we could overcome this
limitation, if suitable reference catalogs were available.  The image
quality has some effect on the performance, though our experiments
using the quality ratings assigned by the SDSS image-processing
pipeline do not fully explore this space since the quality ratings are
in terms of the very high standards of the survey: even the ``bad''
images are reasonable and over $99~\percent$ of them can be recognized
by our system.  More experiments on images with poorly-localized
sources would better characterize the behavior of the system on
low-quality images.

In order to handle images across a wide range of scales, we build a
series of indices, each of which is specialized to a narrow range of
scales.  Each index is able to recognize images within, and extending
somewhat outside, the range to which it is tuned, but the drop-off in
performance is quite fast: the index we used in the majority of our
experiments works very well on $13\times9~\arcmin^2$ SDSS images and
sub-images down to $8\times8~\arcmin^2$ but suffers a serious drop in
performance on $7\times7~\arcmin^2$ images.  Similarly, the
computational effort required is sharply reduced if the system is
given hints about the scale of the images to be recognized.  This is
driven by three main factors.  First, any index that contains quads
that cannot possibly be found in an image of the given range of scales
need not be examined.  Second, only quad features of a given range of
scales in the image need be tested.  Finally, every quad in the image
that is matched to a quad in the index implies an image scale, and any
scale outside the allowed range can be rejected without running the
verification procedure.

Using a set of indices, each of which is tuned to a range of scales,
is related to the idea of \emph{co-visibility constraints} in computer
vision systems: ``closely located objects are likely to be seen
simultaneously more often than distant objects'' \citep{yairi01}.  Each
of our indices contains the brightest stars within grid cells of a
particular size, and contains quad features at a similar scale, so
quads of large angular extent can only be built from bright stars,
while small quads can be built from quite faint stars.  This captures
the practical fact that the angular scale of an image largely
determines the brightnesses of the stars it contains.  Distant pairs
of faint stars are very unlikely to appear in the same image, and we
take advantage of this constraint by only using faint stars to build
quads of small angular size.

The index used in most of our experiments covers the sky in a dense
blanket of quads.  This means that in any image, we have many chances
of finding a matching quad, even if some stars are missing from the
image or index.  This comes at the cost of increasing the number of
features packed into our code feature space, and therefore the number
of false matches that are found for any given quad in a test image.
Reducing the number of quads means that each query will be faster, but
more queries will typically be required before a match is found.

Our experiment using indices built from triangles and quintuples of
stars shows that, for SDSS images, our geometric features built from
quadruples of stars make a good tradeoff between being distinctive
enough that the feature space is not packed too tightly, yet having
few enough stars that the probability of finding all four stars in
both the image and index is high.  We expect that for images much
larger in angular size than SDSS images, a triangle-based index might
perform better, and for images smaller than SDSS images, a quint-based
index might be superior.

Similarly, we found that using a voting scheme---requiring two or more
hypotheses to agree before running the relatively expensive
verification step---was slower than simply running the verification
process on each hypothesis, when using a quad-based index and
SDSS-sized images.  In other domains (such as triangle-based indices),
a voting scheme could be beneficial.

Although we have focused on the idea of a system that can recognize
images of any scale from any part of the sky, our experiments on
Hubble Space Telescope images demonstrate that by building a
specialized index that covers only a tiny part of the sky, we can
recognize tiny images that contain only a few stars that appear in
even the deepest all-sky reference catalogs.

In principle, the geometric feature we are using requires that the
images be \emph{conformal}.  That is, the images must be scaled,
rotated versions of a tangent-plane project of the celestial sphere.
Images with non-isotropic scaling (\ie, rectangular pixels) or shear
in general cannot be recognized.  It is possible to extend the system
by using a different geometric feature that is invariant to these
transformations; this requires adding at least one star to the set of
stars used to define the local reference frame.  In practice, we have
found that although many images have some shear (indeed, the SDSS
images that we have used as our primary test set have this property),
the magnitude of the shear is typically small enough that it does not
distort the geometric features very much.  Even in images with
significant shear or optical distortions, there are
often \emph{regions} of the image that are nearly conformal, and we
are often able to recognize such images by finding a matching feature
within a conformal region.

Our system, built on the idea of geometric hashing---generating
promising hypotheses using a small number of stars and checking the
hypotheses using all the stars---allows fast and robust recognition
and astrometric calibration of a wide variety of astronomical images.
The recognition rate is above $99.9~\percent$ for high-quality images,
with no false positives.  Other researchers have begun using \an to
bring otherwise ``hidden'' data to light, and we hope to continue our
mission ``to help organize, annotate and make searchable all the
world's astronomical information.''

All of the code for the \an system is available under an open-source
license, and we are also operating a web service.  See
\mbox{\texttt{http://astrometry.net}} for details.

\acknowledgments We thank Jon Barron, Doug Finkbeiner, Chris Kochanek,
Robert Lupton, Phil Marshall, John Moustakas, Peter
Nugent, and Christopher Stumm for comments on and contributions to the
prototype version of the online service \an.  It is a pleasure to
thank also our large alpha-testing team.  We thank Leslie Groer and
the University of Toronto Physics Department for use of their Bigmac
computer cluster, and Dave Monet for maintaining and helping us with
the USNO-B Catalog.  Lang, Mierle and Roweis were funded in part by
NSERC and CRC.  Hogg was funded in part by the National Aeronautics
and Space Administration (NASA grants NAG5-11669 and NNX08AJ48G), the
National Science Foundation (grants AST-0428465 and AST-0908357),
and the Alexander
von~Humboldt Foundation.  Hogg and Blanton were funded in part by the NASA
\textit{Spitzer Space Telescope} (grants 30842 and 50568).

This project made use of public SDSS data.  Funding for the SDSS and
SDSS-II has been provided by the Alfred P. Sloan Foundation, the
Participating Institutions, the National Science Foundation, the
U.S. Department of Energy, the National Aeronautics and Space
Administration, the Japanese Monbukagakusho, the Max Planck Society,
and the Higher Education Funding Council for England. The SDSS Web
Site is http://www.sdss.org/.

The SDSS is managed by the Astrophysical Research Consortium for the
Participating Institutions. The Participating Institutions are the
American Museum of Natural History, Astrophysical Institute Potsdam,
University of Basel, University of Cambridge, Case Western Reserve
University, University of Chicago, Drexel University, Fermilab, the
Institute for Advanced Study, the Japan Participation Group, Johns
Hopkins University, the Joint Institute for Nuclear Astrophysics, the
Kavli Institute for Particle Astrophysics and Cosmology, the Korean
Scientist Group, the Chinese Academy of Sciences (LAMOST), Los Alamos
National Laboratory, the Max-Planck-Institute for Astronomy (MPIA),
the Max-Planck-Institute for Astrophysics (MPA), New Mexico State
University, Ohio State University, University of Pittsburgh,
University of Portsmouth, Princeton University, the United States
Naval Observatory, and the University of Washington.

% http://www.ipac.caltech.edu/2mass/releases/allsky/faq.html#reference
This publication makes use of data products from the Two Micron All
Sky Survey, which is a joint project of the University of
Massachusetts and the Infrared Processing and Analysis
Center/California Institute of Technology, funded by the National
Aeronautics and Space Administration and the National Science
Foundation.

% http://hla.stsci.edu/hla_help.html#acknow
Based on observations made with the NASA/ESA Hubble Space Telescope,
and obtained from the Hubble Legacy Archive, which is a collaboration
between the Space Telescope Science Institute (STScI/NASA), the Space
Telescope European Coordinating Facility (ST-ECF/ESA) and the Canadian
Astronomy Data Centre (CADC/NRC/CSA).

% http://www.nofs.navy.mil/data/fchpix/cfbl.html
\sloppy
This research has made use of the USNOFS Image and Catalogue Archive
operated by the United States Naval Observatory, Flagstaff Station
%\newline
(http://www.nofs.navy.mil/data/fchpix/).

This research made use of the NASA Astrophysics Data System.

\bibliographystyle{apj}
\bibliography{apj-jour,astrometry-dot-net}

\end{document}